\documentclass[1p]{elsarticle}

\journal{RSER}

\biboptions{numbers,sort&compress}

\usepackage{libertine}
\usepackage{libertinust1math}

\usepackage{amsmath}
\usepackage{graphicx}
\usepackage{eurosym}
\usepackage{mathtools}
\usepackage{url}
\usepackage{booktabs}
\usepackage{epstopdf}
\usepackage{xfrac}
\usepackage[colorlinks]{hyperref}

\usepackage[
	type={CC},
	modifier={by},
	version={4.0},
]{doclicense}

\newcommand{\abs}[1]{\left|#1\right|}
\def\cL{\mathcal{L}}
\def\cN{\mathcal{N}}
\def\cT{\mathcal{T}}

\begin{document}

\begin{frontmatter}

	\title{Approximating Power Flow and Transmission Losses\\in Coordinated Capacity Expansion Problems}
	
	\author[kitaddress]{Fabian Neumann\corref{correspondingauthor}}
	\ead{fabian.neumann@kit.edu}
	\author[kitaddress]{Veit Hagenmeyer}
	\author[kitaddress]{Tom Brown}
	\cortext[correspondingauthor]{Corresponding author}
	\address[kitaddress]{Institute for Automation and Applied Informatics (IAI), Karlsruhe Institute of Technology (KIT), Hermann-von-Helmholtz-Platz 1, 76344, Eggenstein-Leopoldshafen, Germany}

	\begin{abstract}
		With rising shares of renewables and the need to properly
assess trade-offs between transmission, storage and sectoral integration as balancing options,
building a bridge between energy system models and detailed power flow
studies becomes increasingly important, but is computationally challenging.

In this paper, we compare both common and improved
approximations for two nonlinear phenomena,
power flow and transmission losses, in linear capacity expansion problems
that co-optimise investments in generation, storage and transmission infrastructure.
We evaluate different flow representations discussing differences in investment decisions,
nodal prices, the deviation of optimised flows and losses
from simulated AC power flows, and the computational performance.
By using the open European power system model \mbox{PyPSA-Eur}, 
that combines high spatial and temporal resolution,
we obtain detailed and reproducible results aiming at
facilitating the selection of a suitable power flow model.

Given the differences in complexity, the optimal choice
depends on the application, the user's available computational resources,
and the level of spatial detail considered.
Although the commonly used transport model can already identify key features
of a cost-efficient system while being computationally performant,
deficiencies under high loading conditions arise
due to the lack of a physical grid representation.
Moreover, disregarding transmission losses overestimates optimal grid expansion by 20\%.
Adding a convex relaxation of quadratic losses with two or three tangents to the linearised
power flow equations and accounting for changing line impedances as the network is reinforced
suffices to represent power flows and losses adequately in design studies.
We show that the obtained investment and dispatch decisions are then sufficiently physical
to be used in more detailed nonlinear simulations
of AC power flow in order to better assess their technical feasibility.
This includes determining reactive power flows and voltages,
which the initial linear model neglects.
Simpler approximations are less suitable for such ex-post analysis.
 
	\end{abstract}


\end{frontmatter}


\newpage
\begin{small}
\tableofcontents
\end{small}

\newpage

\section{Introduction}
\label{sec:intro}

Energy system models seek to answer what infrastructure a
future energy system requires for given policy goals,
where and when it should be built, and how much it costs.
For systems with high shares of renewable energy, the question to what extent the variability of
weather-dependent wind and solar energy will be balanced in space
with continent-spanning transmission networks and in time with storage and coupling 
to other energy sectors attracts much research.
Because energy system models are frequently used in policy-making,
it becomes crucial to understand their particular limitations.

To find credible answers for highly renewable systems, it has been demonstrated that
models require coordinated expansion planning of generation,
storage and transmission infrastructure
because they strongly interact \cite{synergies,krishnan};
high temporal resolution and scope to 
account for extreme weather events, storage operation, and
investments shaped by the characteristic daily, synoptic and seasonal
patterns of renewables and load
\cite{kotzur_tsa_2018, kotzur_storage_2018};
high spatial resolution and scope to also
capture the spatio-temporal patterns,
such as correlations of wind speeds across the continent,
and to represent transmission constraints
\cite{hoersch_spatial_2017, schlachtberger_benefits_2017}.
As higher shares of renewables increase the frequency of transmission bottlenecks,
more detailed grid modelling is needed that looks beyond import and export capacities
but accounts for physical conditions such as loop flows, transmission losses,
and curtailment due to otherwise overloaded lines
\cite{nolden_network_2013, Lumbreras2017}.

Especially for planning problems with both static investment and time-dependent
dispatch variables spanning across thousands of operational conditions,
a tractable yet sufficiently trustworthy representation of power flows is essential.
Ideally, outputs are detailed enough to be used
as inputs for more accurate analyses,
bridging the granularity gap between coarsely-resolved planning models
and more detailed engineering models.
Yet, even in strategic problems speed matters for
performing pivotal sensitivity analyses; e.g. regarding uncertain cost parameters,
reference weather years, technology choices, and resource boundaries.
Unfortunately, the first-choice AC power flow equations
are nonlinear and nonconvex, which makes the embedded AC optimal power flow problem NP-hard 
\cite{Bienstock2019,Lehmann2016,Sojoudi2012}.

Even for a linear representation of power flows, considerations of
transmission expansion planning result in a bilinear problem because line impedances change
as line capacities are increased. While we can deal with this challenge
through iterative impedance updates in sequential linear programming \cite{slp2019},
the problem would become even more complex
if a discrete set of transmission expansion plans were considered,
rather than continuous line expansion.
More generally, we can approach computational challenges from multiple angles:
by improving solving algorithms or by figuring out what model details
can be simplified while retaining accuracy \cite{cao_classification_2019}.
Examples include the level of spatial aggregation,
temporal aggregation, technology detail and diversity, or
finally the approximation of power flow.

The transport model, that takes account only of power transfer capacities while ignoring impedances,
and the linearised power flow model, which includes impedances to consider both Kirchhoff laws but no losses, are
commonly used in energy system models.
Among the models reviewed by Ringkj{\o}b et al.~\cite{ringkjob_review_2018}
around four in five models use a transport model if flows are represented,
whereas only one in five uses a linearised power flow model.
Previous work has compared these two major variants
\cite{nijs_2015,haller_bridging_2012,gunkel_modelling_2020,schaber_integration_2013},
and some performed simulations of AC power flow after optimisation
\cite{schaber_integration_2013, ackermann_optimising_2016}.
The comparisons indicate little discrepancy regarding
total system cost and cross-border transmission,
but also differences in nodal prices and overlooked line overloadings
when checked against AC power flow calculations.
However, the cogency of existing comparisons is limited by the use of
low spatial resolution models with fewer than 25 nodes.
Furthermore, the consideration of losses is underrepresented in design studies,
but alongside characteristic weather patterns
shapes the tradeoffs regarding the volume of transmitted energy
because losses increase as more power is transported
\cite{Fitiwi2016, Stott2009}.

In the present contribution, we offer a comprehensive comparison
of linear representations of power flow and losses in theory and practice.
We outline their characteristic benefits and shortcomings
in the context of coordinated capacity expansion problems,
where generation, transmission and storage infrastructure is jointly planned.
Given the multitude of modelling uncertainties, we assess
under which circumstances it is worth embedding more
elaborate flow models than a simple transport model.
We further extend beyond previous research by  
introducing a computationally inexpensive loss approximation
that incorporates an efficient reformulation of the linearised power flow
equations based on a cycle decomposition of the network graph.
By using an open model of the European power system, PyPSA-Eur \cite{pypsaeur},
spanning the whole continent with hundreds of nodes
and hourly temporal coupling due to the consideration of storage units,
we achieve advanced and reproducible comparisons
in systems with high shares of renewables.

The content complements the best practices of energy system modelling,
characterised in Pfenniger et al.~\cite{Pfenninger2014}
and DeCarolis et al.~\cite{DeCarolis2017},
regarding the choice of suitable power flow models.
While we take an investment planning perspective in this paper,
we underline that the way that the transmission of power is represented is relevant
beyond system planning. For instance, it plays a role in the design
of future electricity markets with multiple bidding zones
and flow-based market coupling
\cite{hobbs_improved_2008,Hagspiel2014,gunkel_modelling_2020}.

We structured this contribution as follows. 
We begin with an introduction to the basic long-term power system
planning problem in Section \ref{sec:problem} and briefly review the physics of power flow in Section \ref{sec:powerflow}.
We continue with the different linear power flow representations in Section \ref{sec:flowmodels}.
Section \ref{sec:setup} presents the experimental setup, the results of which
are discussed and critically appraised in Section \ref{sec:results}.
Section \ref{sec:conclusion} concludes this paper
with a summary and recommendations.

\section{Power System Planning Problem}
\label{sec:problem}

This section presents the full long-term power system planning problem.
We confine the formulation to the power system, but it can also serve
to represent the power system embedded within the full energy system.
The representation of power flows is one decisive constituent component
and its variants are later introduced in the context of this problem
in Section \ref{sec:flowmodels}.

The objective is to minimize the total
annual system costs, comprising annualised\footnote{The annuity factor $\frac{1-(1+\tau)^{-n}}{\tau}
  $converts the overnight investment of an asset to annual payments considering its
  lifetime $n$ and cost of capital $\tau$.} capital costs $c_*$ for investments at locations $i$
in generator capacity $G_{i,r}$ of technology $r$, storage power capacity $H_{i,s}$ of technology $s$, and transmission line capacities
$P_{\ell}$, as well as the variable operating costs $o_*$ for generator dispatch $g_{i,r,t}$:
\begin{equation}
  \min_{G,H,F,g} \quad \left[\sum_{i,r} c_{i,r}\cdot G_{i,r} + \sum_{i,s} c_{i,s}\cdot H_{i,s} + \sum_{\ell}c_{\ell}\cdot P_{\ell}+\sum_{i,r,t}w_t\cdot o_{i,r} \cdot g_{i,r,t} \right]
\end{equation}
where representative time snapshots $t$ are weighted by the time span $w_t$ such that their total duration
adds up to one year; \mbox{$\sum_{t\in \cT} w_t=365\cdot 24\text{h}=8760\text{h}$}.
The objective function is subject to a set of linear constraints,
including multi-period linear optimal power flow (LOPF) equations, resulting in a linear programme (LP).

The capacities of generation, storage and transmission infrastructure are
constrained above by their installable potentials and below by any existing components:
\begin{align}
  \underline{G}_{i,r}  &  & \leq &  & G_{i,r}  &  & \leq &  & \overline{G}_{i,r}  & \qquad\forall i, r \\
  \underline{H}_{i,s}  &  & \leq &  & H_{i,s}  &  & \leq &  & \overline{H}_{i,s}  & \qquad\forall i, s \\
  \underline{P}_{\ell} &  & \leq &  & P_{\ell} &  & \leq &  & \overline{P}_{\ell} & \qquad\forall \ell
\end{align}

The dispatch of a generator may not only be constrained by its rated capacity but also by the availability of
variable renewable energy, which may be derived from reanalysis weather data.
This can be expressed as a time- and location-dependent availability
factor $\overline{g}_{i,r,t}$, given per unit of the generator's capacity:
\begin{align}
  0 &  & \leq &  & g_{i,r,t} &  & \leq &  & \overline{g}_{i,r,t} G_{i,r} & \qquad\forall i, r, t
\end{align}

The dispatch of storage units is split into two positive variables;
one each for charging $h_{i,s,t}^+$ and discharging $h_{i,s,t}^-$.
Both are limited by the power rating $H_{i,s}$ of the storage units.
\begin{align}
  0 &  & \leq &  & h_{i,s,t}^+ &  & \leq &  & H_{i,s} & \qquad\forall i, s, t \label{eq:sto1} \\
  0 &  & \leq &  & h_{i,s,t}^- &  & \leq &  & H_{i,s} & \qquad\forall i, s, t \label{eq:sto2}
\end{align}
The energy levels $e_{i,s,t}$ of all storage units have to be consistent with the dispatch in all hours.
\begin{align}
  e_{i,s,t} =\: & \eta_{i,s,0}^{w_t} \cdot e_{i,s,t-1} + w_t \cdot h_{i,s,t}^\text{inflow} - w_t \cdot h_{i,s,t}^\text{spillage} & \quad\forall i, s, t \nonumber \\
                & + \eta_{i,s,+} \cdot w_t \cdot h_{i,s,t}^+ - \eta_{i,s,-}^{-1} \cdot w_t \cdot h_{i,s,t}^- \label{eq:stoe}
\end{align}
Storage units can have a standing loss $\eta_{i,s,0}$, a charging efficiency $\eta_{i,s,+}$, a discharging efficiency $\eta_{i,s,-}$,
natural inflow $h_{i,s,t}^\text{inflow}$ and spillage $h_{i,s,t}^\text{spillage}$.
The storage energy levels are assumed to be cyclic
\begin{align}
  e_{i,s,0} = e_{i,s,\abs{\mathcal{T}}} \qquad\forall i, s
\end{align}
and are constrained by their energy capacity
\begin{align}
  0 &  & \leq &  & e_{i,s,t} &  & \leq &  & \overline{T}_s \cdot H_{i,s} & \qquad\forall i, s, t.
\end{align}
To reduce the number of decisison variables, we link the energy storage volume to
power ratings using a technology-specific parameter $\overline{T}_s$
that describes the maximum duration a storage unit can discharge at full power rating.

Total CO$_2$ emissions may not exceed a target level $\Gamma_{\text{CO}_2}$.
The emissions are determined from the time-weighted generator dispatch $ w_t \cdot g_{i,r,t}$ using the specific emissions $\rho_r$ of fuel $r$
and the generator efficiencies $\eta_{i,r}$
\begin{align}
	\sum_{i,r,t}  \rho_r \cdot \eta_{i,r}^{-1} \cdot w_t \cdot g_{i,r,t} \leq \Gamma_{\text{CO}_2}.
\end{align}

All power flows $p_{\ell,t}$ are also limited by their capacities $P_\ell$
\begin{align}
	|p_{\ell,t}| \leq \overline{p}_{\ell} P_{\ell} & \qquad\forall \ell, t,
	\label{eq:cap}
\end{align}
where $\overline{p}_\ell$ acts as a per-unit security margin on the line capacity
to allow a buffer for the failure of single circuits ($N-1$ condition) and reactive power flows.

Ultimately, we need constraints that define the power flows $p_{\ell,t}$ in the network.
In the next Section \ref{sec:powerflow}, we briefly set foundations for nonlinear power flow and losses.
The various alternative flow models are then presented in Section \ref{sec:flowmodels}.
The subsequent descriptions will omit the time index $t$ for notational simplicity.

\section{Nonlinear Power Flow and Losses}
\label{sec:powerflow}

This section briefly revises
the nonlinear AC power flow equations,
important electrical parameters of transmission lines,
and how to calculate active power losses on a line.
We do this to set the foundations for derivations of
the covered flow models.

\subsection{Nonlinear AC Power Flow}
\label{sec:ac}

The active power flow $p_{\ell(i,j)}$ of a line $\ell\equiv\ell(i,j)$ from bus $i$ to bus $j$ can be described in voltage-polar coordinates by
\begin{equation}
	p_{\ell(i,j)} = g_\ell \abs{V_i}^2 + \abs{V_i}\abs{V_j} \left[g_\ell \cos(\theta_i-\theta_j) - b_\ell \sin(\theta_i-\theta_j)\right]
	\label{eq:pf}
\end{equation}
and, analogously, the reactive power flow $q_{\ell(i,j)}$ is given by
\begin{equation}
	q_{\ell(i,j)} = b_\ell \abs{V_i}^2 + \abs{V_i}\abs{V_j} \left[g_\ell \sin(\theta_i-\theta_j) - b_\ell \cos(\theta_i-\theta_j)\right],
	\label{eq:qf}
\end{equation}
where
$\abs{V_i}$ is the per-unit bus voltage magnitude,
$\theta_i$ is the bus voltage angle,
$g_\ell$ is the line conductance, and
$b_\ell$ is the line susceptance \cite{Taylor2015a}.

\subsection{Nonlinear Active Power Losses}
\label{sec:losses}

To derive an expression for the active power losses in a transmission line,
we apply the convention that departing power flows are positive
and arriving power flows are negative.
Consequently, if power flows from bus $i$ to $j$, $p_{\ell(i,j)}>0$ and $p_{\ell(j,i)}<0$.
The losses $\psi_\ell$ are the difference between power sent and power received \cite{Taylor2015a}, therefore
\begin{equation}
	\psi_\ell = p_{\ell(i,j)} + p_{\ell(j,i)}.
	\label{eq:losses_basic}
\end{equation}
Substituting equation \eqref{eq:pf} into equation \eqref{eq:losses_basic} yields
\begin{align}
	\psi_\ell & = &        & g_\ell \abs{V_i}^2 + \abs{V_i}\abs{V_j} \left[g_\ell \cos(\theta_i-\theta_j) - b_\ell \sin(\theta_i-\theta_j)\right] \\
	           &   & +\quad & g_\ell \abs{V_j}^2 + \abs{V_j}\abs{V_i} \left[g_\ell \cos(\theta_j-\theta_i) - b_\ell \sin(\theta_j-\theta_i)\right]
\end{align}
and using the trigonometric identities $\cos(-\alpha) = \cos(\alpha)$ and $\sin(-\alpha) = -\sin(\alpha)$
translates to
\begin{align}
	\psi_\ell & = &        & g_\ell \abs{V_i}^2 + \abs{V_i}\abs{V_j} \left[g_\ell \cos(\theta_i-\theta_j) - b_\ell \sin(\theta_i-\theta_j)\right] \\
	           &   & +\quad & g_\ell \abs{V_j}^2 + \abs{V_j}\abs{V_i} \left[g_\ell \cos(\theta_i-\theta_j) + b_\ell \sin(\theta_i-\theta_j)\right].
\end{align}
We can further simplify this expression to the loss formula
\begin{equation}
	\psi_\ell = g_\ell \left(\abs{V_i}^2+\abs{V_j}^2\right) - 2\abs{V_i}\abs{V_j}g_\ell \cos(\theta_i - \theta_j).
	\label{eq:losses}
\end{equation}
We will use this formula in Section \ref{sec:mod-lossy} to derive a linear approximation for losses.

\subsection{Electrical Parameters of Transmission Lines}
\label{sec:impedances}

The complex per-unit impedance $z_\ell = r_\ell + i x_\ell$ is composed of resistance $r_\ell$ and reactance $x_\ell$.
Likewise, the admittance $y_\ell = g_\ell + i b_\ell$ is composed of conductance $g_\ell$ and susceptance $b_\ell$.
Impedance and admittance are reciprocals ($y_\ell=z_\ell^{-1}$).
Hence, if we assume a dominance of reactance over ohmic resistance ($r_\ell \ll x_\ell$),
as applies for high voltage overhead transmission lines, we obtain the approximations
\begin{align}
	g_\ell & \approx \frac{r_\ell}{x_\ell^2} \label{eq:g-approx}, \\
	b_\ell & \approx \frac{1}{x_\ell} \label{eq:b-approx}.
\end{align}
For a derivation, see \ref{sec:app-impedance}.
We will use these relations in Section \ref{sec:mod-lossless} and Section \ref{sec:mod-lossy}.
In view of the approximation of line losses in later sections,
note that although we assume that resistance is dominated by reactance, 
we do not assume resistance to be zero (cf.~\cite{doerfler2013,Coffrin2012}).

\section{Linear Power Flow Models}
\label{sec:flowmodels}

The AC power flow equations \eqref{eq:pf} are nonlinear and nonconvex.
This is challenging because multiple local minima exist due to the trigonometric expressions and when directly incorporated
in the optimisation problem they would make the problem NP-hard
\cite{Bienstock2019,Lehmann2016,Sojoudi2012}.
To be able to run large optimisation problems of the continental power system at
sufficient spatial and temporal resolution
it is hence inevitable to retain a convex problem that can be solved in
polynomial time and does not possess local minima.

\begin{figure}
	\centering
	\includegraphics[width=\textwidth]{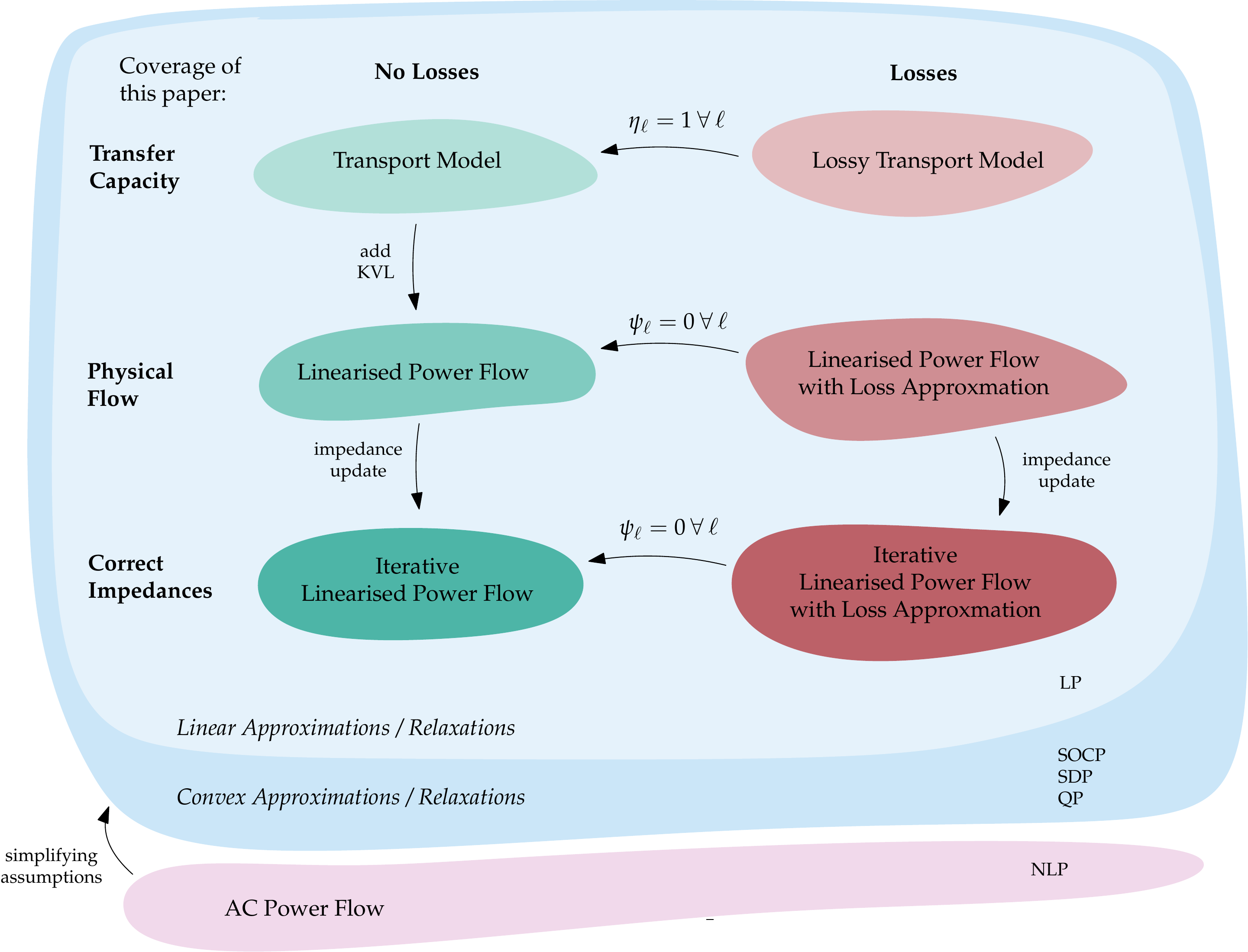}
	\label{fig:overview}
	\caption{Illustration of the scope of the present paper and its context.
	It shows the connections between the covered linear power flow models,
	their main features, and how they are related to other
	(convexified) nonlinear formulations.}
\end{figure}

\begin{table}[]
	\centering
	\begin{tabular}{@{}llrrr@{}}
	\toprule
	Model & Section & Variables & Eq.~Constraints & Ineq.~Constraints \\ \midrule
	Transport & \ref{sec:mod-transport} &  $\abs{\cL}$ & $\abs{\cN}-1$ & 0 \\ 
	Lossy Transport & \ref{sec:mod-lossytransport} &  $2\abs{\cL}$& $\abs{\cN}-1$ & 0 \\
	Linearised Power Flow & \ref{sec:mod-lossless} & $\abs{\cL}$ & $\abs{\cL}$ & 0 \\
	Linearised Power Flow & \ref{sec:mod-lossy} & $2\abs{\cL}$ & $\abs{\cL}$ & $2n\abs{\cL}$ \\
	with Loss Approximation & &  &  &  \\ \bottomrule
	\end{tabular}
	\label{tab:numvarcon}
	\caption{Comparison of the number of variables and equality/inequality constraints
	related to flow models per snapshot $t\in \cT$.
	The constraint count excludes variable bounds.
	$\abs{\cL}$ is the number of lines,
	$\abs{\cN}$ is the number of nodes,
	and $2n$ represents the number of tangents used for the loss approximation.
	}
\end{table}

In this section we describe and develop various
linear representations of power flow.
These are introduced in the order from least to most accurate,
progressively increasing the complexity; namely
(i)  the common transport model,
(ii) a lossy transport model,
(iii) the lossless linearised power flow, and
(iv) a lossy linearised power flow model.
Figure \ref{fig:overview} shows the relations between the
formulations and Table \ref{tab:numvarcon} documents
differences in the number of variables and constraints.
The scope of this work is deliberately constrained to:

\begin{itemize}
	\item \textit{only linear problems}:
	  To avail of powerful, scalable and fast interior-point solvers,
	  and to guarantee an optimal solution,
	  we only include formulations that entail linear problems.
	  However, there exist promising second-order cone
	  or semidefinite convex relaxations of the power flow equations.
	  For excellent theoretical reviews of convex relaxations and
	  approximations of power flow
	  see Molzahn et al.~\cite{molzahn_2019}, Taylor~\cite{Taylor2015a},
	  and Coffrin et al.~\cite{coffrin_convex_2018}.
	\item \textit{only active power}:
	  We furthermore confine our analysis to formulations that do not capture
	  reactive power flows or information on bus voltages.
	  Nonetheless, linear problems that capture selected aspects of this
	  are under active research;
	  see e.g.~Coffrin et al.~\cite{coffrin_lp_2014}.
	\item \textit{only comparison of different feasible spaces}:
	  We compare different linear flow models that define different feasible spaces.
	  We do not compare equivalent reformulations of identical models, since this has
	  been analysed in Hörsch et al.~\cite{cycleflows}.
	\item \textit{no copper plate model}:
	  Although occasionally encountered in generation and storage capacity expansion models,
	  we do not include the copper plate relaxation in our comparisons
	  because it does not capture information on power flows in transmission networks.
	  The copper plate model removes all lines and aggregates
	  all components to a single node.
	  It is a relaxation because any transmission of power
	  becomes unconstrained and incurs no cost.
	  For the impact of spatial clustering -- of which the copper plate model
	  is the extreme -- on optimal investments we refer to Hörsch et al.~\cite{hoersch_spatial_2017}.
\end{itemize}

\subsection{Transport Model}
\label{sec:mod-transport}

The transport model is also known as a
network flow model, trade model, transshipment model
or net transfer capacity (NTC) model \cite{Taylor2015a}.
It ignores the effect of impedances on flows (including losses) and,
besides the capacity constraints of lines,
only requires nodal power balance
according to Kirchhoff's Current Law (KCL); i.e.
the power injected at each bus must equal the power withdrawn
by attached lines
\begin{equation}
	p_i = \sum_\ell K_{i\ell} p_\ell \qquad \forall i \in \cN,
	\label{eq:kcl}
\end{equation}
where $p_i$ is the active power injected or consumed at node $i$ and
$K$ is the incidence matrix of the network graph
which has non-zero values $+1$ if line $\ell$ starts at bus 
$i$ and $-1$ if line $\ell$ ends at bus $i$.

Because the columns of the incidence matrix each sum up to zero,
KCL yields $\abs{\cN}-1$ linearly independent constraints.
These are not sufficient to uniquely determine the $\abs{\cL}$ unknown flows.
The transport model allows arbitrary flows
as long as flow conservation is fulfilled, also because
it is free and lossless to transmit power.
This makes the transport model degenerate, which
can be detrimental to the solving speed.
Also, of course, this does not adequately reflect
the physical behaviour of power flows in the transmission network.

Despite its drawbacks, the transport model is very popular.
In the comprehensive review by Ringkj{\o}b et al.~\cite{ringkjob_review_2018},
it is applied in a majority of models.
This minimalistic representation of flows is useful to develop an
understanding for the potential of increased transfer capacity between regions,
rather than assessing specific transmission bottlenecks and reinforcement needs.
It is often applied in investment models where the grid is highly aggregated to a few nodes
(e.g.~one node per country in Europe or federal state in the United States)
or analyses of energy markets across multiple bidding zones.
Its main advantages are ease of implementation and fast solving speed.
For pure dispatch problems without investment decisions one
can even utilise specialised network flow algorithms;
for instance the minimum cost flow algorithm \cite{ahuja_1993}.

\subsection{Transport Model with Loss Approximation}
\label{sec:mod-lossytransport}

Part of the drawbacks and degeneracy of the transport model stems from
the disregard of transmission losses.
As partial remedy, we can amend the transport model with a simple loss approximation
which assumes lines to have a constant transmission efficiency $\eta_\ell$
depending on their length.
In this case, the power arriving at the receiving bus
is lower than the power injected at the sending bus.
To differentiate between sending bus and receiving bus, we need to split
the bidirectional power flow variable $p_\ell$ into forward flows
$p_\ell^+$ and backward flows $p_\ell^-$ with capacity limits
\begin{align}
	0 \leq p_\ell^+ \leq  \overline{p}_{\ell} P_{\ell}  \qquad \forall \ell \in \cL \\
	0 \leq p_\ell^- \leq  \overline{p}_{\ell} P_{\ell}  \qquad \forall \ell \in \cL 
\end{align}
which substitute the variables $p_\ell$ and their bounds given in equation \eqref{eq:cap}.
Furthermore, we need to adjust the nodal balance constraints \eqref{eq:kcl} to reflect
the transmission losses and separated power flow variables to
\begin{equation}
	p_i = \sum_\ell K_{i\ell}^+ p_\ell^+ - \sum_\ell K_{i\ell}^- p_\ell^- \qquad \forall i \in \cN,
	\label{eq:kcl-mod-lossytransport}
\end{equation}
where $K^+$ is the lossy incidence matrix of the network 
graph regarding forward flows $p_\ell^+$
which has non-zero values $+1$ if line $\ell$ starts at
bus $i$ and $-\eta_\ell$ if line $\ell$ ends at bus $i$.
Analogously, $K^-$ regards backward flows $p_\ell^-$
with non-zero values $\eta_\ell$ if line $\ell$ starts
at bus $i$ and $-1$ if line $\ell$ ends at bus $i$.

The transmission losses alleviate some degeneracy
of the transport model since considering losses yields an
incentive to minimise power flows rather than to
distribute them arbitrarily.
However, this is paid for with a doubling in the number of flow variables.
Additionally, while the use of a constant transmission efficiency
is an improvement from the plain transport model,
it still ignores the quadratic relationship between power flow and losses \cite{Fitiwi2016}.
Note, that if all lines have no losses ($\eta_\ell=1$),
the lossy transport model is equivalent to the regular transport model.

\subsection{Linearised Power Flow}
\label{sec:mod-lossless}

The linearised power flow model,
which is also known as linearised load flow, DC power flow or B$\Theta$ model,
extends the lossless transport model.
In addition to the nodal power balance constraints \eqref{eq:kcl} from KCL 
and capacity limits \eqref{eq:cap}, 
linear constraints for Kirchhoff's Voltage Law (KVL) are included,
which define how power flows split in parallel paths.
We derive these by simplifying the nonlinear power flow
equations \eqref{eq:pf} and \eqref{eq:qf}.
Assuming
\begin{itemize}
	\item all per-unit voltage magnitudes are close to one
	      ($\abs{V_i}\approx 1$),
	\item conductances $g_\ell$ are negligible relative to susceptances $b_\ell$
	      ($b_\ell \gg g_\ell$),
	\item voltage angle differences are small enough
	      ($\sin(\theta_i - \theta_j) \approx \theta_i - \theta_j$ and $\cos(\theta_i - \theta_j) \approx 0$),
	\item reactive power flows $q_\ell$ are negligible compared to real power flows $p_\ell$
		  ($q_\ell \approx 0$),
\end{itemize}
leads to
\begin{equation}
	p_\ell = b_\ell (\theta_i - \theta_j),
	\label{eq:angle-kvl-b}
\end{equation}
and when we further assume $r_\ell \ll x_\ell$, by substituting \eqref{eq:b-approx} we obtain
\begin{equation}
	p_\ell = \frac{\theta_i - \theta_j}{x_\ell}.
	\label{eq:angle-kvl-x}
\end{equation}
This angle-based formulation is the most common linear formulation of KVL \cite{Taylor2015a}.
But it is possible to avoid the auxiliary voltage angle variables and reduce
the required number of constraints by using a cycle basis of the network graph  \cite{cycleflows}.
Namely, KVL states that the sum of voltage angle differences across lines around
all cycles in the network must sum up to zero.
Considering a set of independent cycles $c$ of the network forming a cycle basis,
which are expressed as a directed linear
combination of the lines $\ell$ in a cycle incidence matrix
\begin{equation}
	C_{\ell c} = \begin{cases}
		1  & \text{if edge $\ell$ is element of cycle $c$,}          \\
		-1 & \text{if reversed edge $\ell$ is element of cycle $c$,} \\
		0  & \text{otherwise,}
	\end{cases}
\end{equation}
KVL is formulated by
\begin{equation}
	\sum_\ell C_{\ell c} (\theta_i - \theta_j) = 0 \qquad \forall c \in \mathcal{C}.
\end{equation}
Using equation \eqref{eq:angle-kvl-x}, we can express KVL directly in terms of the power flows
and circumvent the auxiliary voltage angle variables
\begin{equation}
	\sum_\ell C_{\ell c} p_\ell x_\ell = 0 \qquad \forall c  \in \mathcal{C}.
	\label{eq:cycle-kvl-x}
\end{equation}
Although less common, this cycle-based formulation \eqref{eq:cycle-kvl-x}
has been shown to significantly outperform
the angle-based formulation \eqref{eq:angle-kvl-x} \cite{cycleflows,cycletep_2020}.
There are even further equivalent reformulations of the linearised power flow \cite{cycleflows};
for example the Power Transfer Distribution Factor (PTDF) formulation, which
directly relates nodal power injections to line flows.
Because our focus lies on the comparison of different flow models,
not their diverse reformulations,
we only evaluate the computationally performant cycle-based
formulation in the present contribution.

With the cycle-based formulation one can clearly see that the transport
model is equivalent to the linearised power flow in radial networks;
i.e. when the network has no cycles.
Also, the absence of auxiliary voltage angle variables
facilitates the insight that the transport model is a
relaxation of the linearised power flow because the latter only adds constraints
in the same variable space.

The linearised power flow model is claimed to be accurate when
reactance dominates \mbox{($x_\ell \gg r_\ell$)} and when parallel lines
have similar ratios \cite{Simpson-Porco2018},
but very long lines in highly aggregated networks 
can deteriorate the quality of the approximation (see Section \ref{sec:res-flowcomparison}).
An advantage of this model over the transport model is that it
captures some meaningful physical characteristics observed in the operation of electrical grids.
Namely, it is capable of revealing loop flows in meshed networks;
for instance recurring spillover effects between Germany and the Czech Republic.
Nevertheless, it still disregards losses.

If we would consider that lines can be built between buses
where there are currently none,
another variant is the so-called hybrid model. 
This version formulates linearised power flow constraints for existing
lines and employs a transport model for candidate lines.

\subsection{Linearised Power Flow with Loss Approximation}
\label{sec:mod-lossy}

Neglecting resistive losses is considered to be among the largest sources
of error in the linearised power flow formulation,
particularly in large networks \cite{Simpson-Porco2018}.
The following extension of the lossless linearised power flow
(Section \ref{sec:mod-lossless}) 
is a mixture of similar variants encountered in the literature
with a focus on computational efficiency.
We reference where we follow or deviate from previous work below.
This or similar formulations have rarely been applied in the
co-optimisation of transmission, storage and generation capacities,
but rather in detailed operational optimal power flow (OPF)
or transmission expansion planning (TEP) problems;
see overview in \cite{nolden_network_2013}.

We start by adding a loss variable $\psi_\ell$ for each line.
Losses reduce the effective transmission capacity of a line
\begin{equation}
	\abs{p_\ell} \leq \overline{p}_\ell P_\ell - \psi_\ell
	\label{eq:cap-mod}
\end{equation}
and must be accounted for in the nodal balance equation \eqref{eq:kcl}
\begin{equation}
	p_i = \sum_\ell K_{i\ell} p_\ell + \frac{\abs{K_{i\ell}}}{2} \psi_\ell\qquad \forall i \in \cN.
	\label{eq:kcl-mod-ri2}
\end{equation}
We split the losses $\psi_\ell$ equally between both buses (like in \cite{Zhang2013a, zhang_improved_2013, Ramos})
and do not allocate them at the sending bus exclusively (like in \cite{Coffrin2012,farahmand_impact_2011}).
The latter could be modelled with an absolute value function in the linear problem. However,
this would involve splitting flow and loss variables each into positive and negative segments.
Because this adds many auxiliary decision variables, we decided in favor of distributing
the losses evenly.
This choice is paid for with the possibility of overestimating losses due to 
an extensive convex relaxation.

Assuming close to nominal per-unit voltage magnitudes $\abs{V_i}\approx 1$ the loss formula given in equation
\eqref{eq:losses} becomes
\begin{equation}
	\psi_\ell = 2g_\ell \left[1 - \cos(\theta_i - \theta_j)\right].
	\label{eq:cosine-loss}
\end{equation}
This is the basis for the linearised loss formulation in \cite{Ramos}.
We can also express this in terms of active power flows $p_\ell$
by substituting equation \eqref{eq:angle-kvl-x} into equation \eqref{eq:cosine-loss}
\begin{equation}
	\psi_\ell = 2g_\ell \left[1 - \cos(p_\ell x_\ell)\right].
	\label{eq:cosine-loss-power}
\end{equation}
This makes the loss formulation independent from the voltage angle variables and
we can therefore avail of the speed-up obtained by using the cycle-based
formulation \eqref{eq:cycle-kvl-x}.

\begin{figure}
	\centering
	\includegraphics[width=0.5\textwidth]{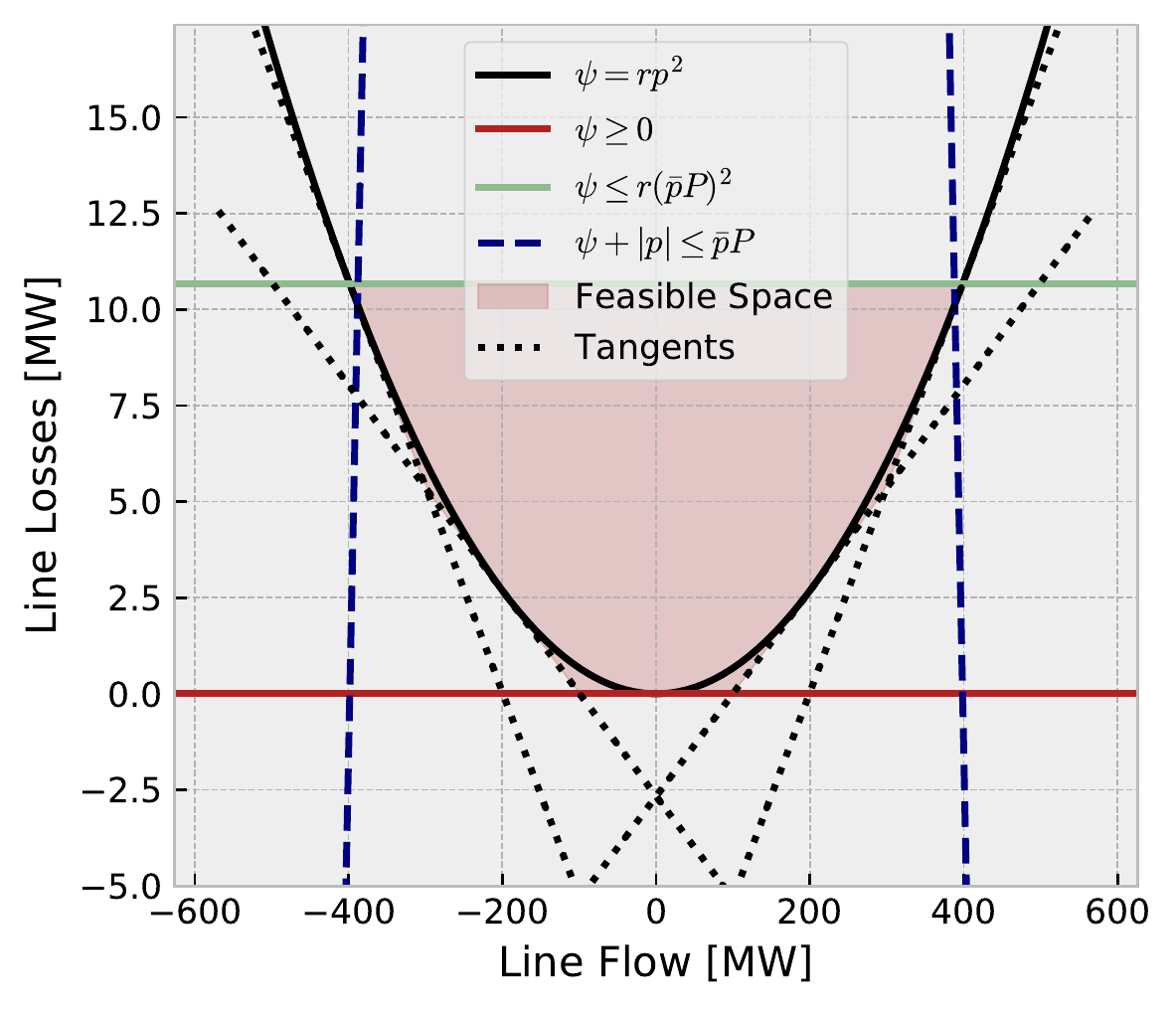}
	\caption{Illustration of the feasible space in the flow--loss ($p_\ell$--$\psi_\ell$) dimensions.}
	\label{fig:feasiblespace}
\end{figure}

Using the small-angle approximation $\cos(\alpha) \approx 1 - \alpha^2 / 2$,
equation \eqref{eq:cosine-loss-power} becomes quadratic
\begin{equation}
	\psi_\ell = 2g_\ell \left[1- \left(1-\frac{(p_\ell x_\ell)^2}{2}\right)\right] = g_\ell (p_\ell x_\ell)^2.
\end{equation}
By inserting equation \eqref{eq:g-approx} we get
\begin{equation}
	\psi_\ell = \frac{r_\ell}{x_\ell^2} (p_\ell x_\ell)^2
\end{equation}
or simply
\begin{equation}
	\psi_\ell = r_\ell p_\ell^2.
	\label{eq:ri2}
\end{equation}
This is the basis for the linearised loss formulation in \cite{Coffrin2012}.
Equation \eqref{eq:ri2} is still a quadratic equality constraint, and therefore nonconvex.
Other works have discussed or applied a piecewise linearisation
of equation \eqref{eq:ri2} \cite{Ramos,molzahn_2019,vce_2019,zhang_improved_2013,dos_santos_dynamic_2011}.
But because the use of integer variables to define the segments would entail a nonconvex
mixed-integer problem (MILP), we choose not to pursue this approach.
Instead, by building a convex envelope around this constraint
from the upper and lower bounds for $\psi_\ell$
as well as a number of tangents as inequality constraints,
we can incorporate transmission losses while retaining a linear optimisation problem.
This is illustrated in Figure \ref{fig:feasiblespace}.
For setting the lower limit, by definition losses are positive
\begin{equation}
	\psi_\ell \geq 0
	\label{eq:sqr-lower}
\end{equation}
and by substituting maximal line flows
\begin{equation}
	p_\ell \leq \overline{p}_\ell P_\ell \leq \overline{p}_\ell \overline{P}_\ell
	\label{eq:line-limit}
\end{equation}
into \eqref{eq:ri2} we obtain the upper limit
\begin{equation}
	\psi_\ell \leq r_\ell (\overline{p}_\ell \overline{P}_\ell)^2.
	\label{eq:sqr-upper}
\end{equation}
Next, we derive $2n$ evenly spaced (like in \cite{molzahn_2019}) mirrored tangents
which approximate equation \eqref{eq:ri2} as inequalities
from below. These have the form
\begin{align}
	\psi_\ell & \geq &   & m_k\cdot p_\ell + a_k & \qquad\forall k=1,\dots,n \label{eq:tangents-pos} \\
	\psi_\ell & \geq & - & m_k\cdot p_\ell + a_k & \qquad\forall k=1,\dots,n \label{eq:tangents-neg}
\end{align}
At segment $k$ we calculate the losses
\begin{equation}
	\psi_\ell(k) = r_\ell \left(\frac{k}{n}\cdot \overline{p}_\ell \overline{P}_\ell\right)^2
\end{equation}
and the corresponding slope
\begin{equation}
	m_k = \frac{\text{d} \psi_\ell(k) }{\text{d}k} = 2 r_\ell \left(\frac{k}{n}\cdot \overline{p}_\ell \overline{P}_\ell\right)
\end{equation}
and the offset
\begin{equation}
	a_k = \psi_\ell(k) - m_k \left(\frac{k}{n}\cdot\overline{p}_\ell \overline{P}_\ell\right).
\end{equation}
Together, equations \eqref{eq:sqr-lower}, \eqref{eq:sqr-upper}, \eqref{eq:tangents-pos}, and \eqref{eq:tangents-neg}
form a convex envelope of equation \eqref{eq:ri2}.

A recurring criticism of this extensive convex relaxation is the possibility of so-called fictitious or artificial losses
\cite{molzahn_2019,Fortenbacher2017,palma-benhke_modelling_2013,Yang2018,zhong_dynamic_2013}.
As illustrated by Figure \ref{fig:feasiblespace},
the model does indeed allow for overestimating losses.
This can be economical if negative locational marginal prices occur.
Overestimating losses is equivalent to dissipating power at a node.
Another component in the problem formulation that already permits
this behaviour are storage units 
(see equations \eqref{eq:sto1}--\eqref{eq:stoe}).
To avoid binary variables, storage units may charge
and discharge at the same time.
Power is then lost by cycling through the conversion efficiencies.
We argue that fictitious losses are not problematic because
(i) negative nodal prices are rare,
(ii) such behaviour could be realised in operation by low-cost resistors and demand response, and
(iii) the loss overestimation is bounded.
We will substantiate this argument with experimental
results in Section \ref{sec:res-lossapproximation}.

\subsection{Iterative Linearised Power Flow (with Loss Approximation)}
\label{sec:mod-iterative}

When using the linearised power flow,
with and without losses
(Sections \ref{sec:mod-lossless} and \ref{sec:mod-lossy}),
the impedances of transmission lines affect the flows and losses.
The relations of reactances $x_\ell$ determine the distribution of flows
(cf.~equation \eqref{eq:cycle-kvl-x}).
The resistances $r_\ell$ set the losses (cf.~equation \eqref{eq:ri2}).
Thus, for reactances we are only interested in relative values,
whereas for the resistances the absolute values are decisive.

Line impedances change as line capacities are increased
($x_\ell \propto 1/P_\ell$ and $r_\ell \propto 1/P_\ell$).
Ignoring this dependency would result in distorted power flows.
Expanded lines would experience less flow than they should.
Losses may also be overestimated as the extension of parallel lines
reduces the effective resistance.

Consequently, the representation of grid physics is improved by
taking account of the relation between line capacities and impedances,
yet also complicates the problem.
If we considered discrete expansion options we would use a
big-$M$ disjunctive relaxation to resolve the nonlinearity \cite{cycletep_2020}.
But since we assume continuous line expansion in view of computational performance,
we instead pursue an iterative heuristic approach.
In previous works, we have shown that this is an acceptable approximation \cite{slp2019}.

We sequentially solve the optimisation problem from Section \ref{sec:problem}
and in each iteration update the line impedances according to their optimised capacities.
We repeat this process until
(i) line expansion choices do not change in subsequent
    iterations and convergence is reached, or
(ii) a predetermined number of iterations are performed.
In the latter case, the final iteration should be run with fixed line capacities
such that impedances do not change anymore.

\section{Simulation Setup}
\label{sec:setup}

Having developed the individual power flow models in theory,
this section outlines the setup we use to test them.
First, we introduce the power system model in Section \ref{sec:pypsa-eur}
for which we optimise investments.
Second, we outline the methodology we use to validate the
resulting approximated line flows in Section \ref{sec:newtonraphson}.

\subsection{Power System Model for Optimisation}
\label{sec:pypsa-eur}

We evaluate the different flow models on the open power system dataset PyPSA-Eur (v0.1.0),
which covers the whole European transmission system \cite{pypsaeur}.
We choose a spatial resolution of 250 nodes and
a temporal resolution of 4380 snapshots,
one for every two hours of a full year.
This reflects the maximum for which all flow models presented
in Section \ref{sec:flowmodels} could be solved.

Targeting an emission reduction of 100\% in the power sector,
we only consider renewable resources \cite{responseto2018}.
Following the problem formulation from Section \ref{sec:problem},
solar PV, onshore and offshore wind capacities are co-optimised with battery storage,
hydrogen storage, and transmission infrastructure in a greenfield planning approach,
subject to spatio-temporal capacity factors and geographic potentials.
Exceptions to greenfield planning are existing transmission infrastructure,
which can only be reinforced but not removed, and
today's run-of-river and hydropower capacities including pumped hydroelectric energy storage,
which are not extendable due to assumed geographical constraints.
HVDC links are assumed to have losses of 3\% per \mbox{1000 km} \cite{siemens_hvdc}
and can be expanded continuously up to 20 GW (each composed of several smaller parallel circuits).
Planned projects from the 2018 Ten Year Network Development Plan (TYNDP) are included \cite{tyndp}.
We only apply link losses to flow models which also account for
losses in HVAC lines.
HVAC line capacity can also be expanded continuously; by the maximum of doubled capacity or additional 5 GW.
When using the lossy transport model,
HVAC lines are assumed to have constant losses in the order of 5\% per \mbox{1000 km} \cite{siemens_hvdc}.
To approximate $N-1$ security, lines may only be used up to 70\% of their nominal rating.
More details are provided in Hörsch et al. \cite{pypsaeur}.

Technically, the optimisation problem is implemented using the free Python modelling framework
PyPSA (v0.17.0) working with the Pyomo interface \cite{pypsa}.
Both optimality and feasibility tolerances are set to a value of 0.1\%,
which is sufficient given the mentioned approximations made in the model.
We use the cycle-based formulation of Kirchhoff's Voltage Law for any model that accounts for it.
The code to reproduce the experiments is openly available at
\href{https://github.com/fneum/power-flow-models}{github.com/fneum/power-flow-models}.

In accordance with descriptions in Section \ref{sec:flowmodels},
the following flow models are evaluated:
\begin{itemize}
	\setlength\itemsep{0em}
	\item lossless transport model as \textsc{Transport},
	\item lossy transport model as \textsc{Lossy Transport},
	\item lossless linearised power flow
	      with no iterations as \textsc{Lossless},
	\item lossless linearised power flow
	      with 3 iterations as \textsc{Iterative Lossless},
	\item lossy linearised power flow 
		  with 6 tangents and no iterations as \textsc{Lossy}, and
	\item lossy linearised power flow
	      with 6 tangents and 3 iterations as \textsc{Iterative Lossy}.
\end{itemize}

\subsection{Nonlinear AC Power Flow Simulation}
\label{sec:newtonraphson}

All presented flow models approximate the AC power flow equations
(Section \ref{sec:powerflow}).
Thus, to identify possibly overlooked line overloading, and
to demonstrate characteristic features of particular flow models,
we use the AC power flow equations to assess the quality
of the respective approximations.

We compare optimised flows to simulated flows which we obtain by
solving the AC power flow equations ex-post 
based on the optimised dispatch of controllable system components.
Specifically, we do not reoptimise dispatch decisions subject to
the AC power flow model due to computational constraints given such large
multi-period problems, but only check their feasibility.
We use the Newton-Raphson method (see e.g.~\cite{grainer1994}) and
distribute the total slack power across all buses 
in proportion to their total generation capacity \cite{Milano2008,Stott2009}
(details are provided in \ref{sec:app-powerflow}).
Moreover, we consider PV
buses\footnote{For PV buses, the nodal active power injections $p_i$
and voltage magnitudes $\abs{V_i}$ are known
(we assume nominal voltage magnitudes $\abs{V_i}=1$).
Bus voltage angles $\theta_i$ and reactive power feedin $q_i$ are to be found.
Conversely, for PQ buses the nodal active power injections $p_i$ and
reactive power injections $q_i$ are known.
Bus voltage magnitudes $\abs{V_i}$ and angles $\theta_i$ are to be found.}
at each node since the reactive power set points are unknown.
Hence, we assume there to be sufficient reactive power control infrastructure
to maintain nominal voltages.
We argue that in systems with high shares of renewables
the PV bus assumption is justified in view of a
growing number of distributed power generation units,
each capable of contributing to voltage control
by reactive power injection or consumption,
and power electronic devices such as
Flexible Alternating Current Transmission Systems (FACTS).
While the linearised power flow approximations neglect the shunt capacitance of lines,
these are taken into account in the subsequent AC power flow simulation
according to the standard equivalent $\Pi$ model \cite{kundur1994}.
Suitable short- to medium-length lines between 25km and 250km make up about 85\% 
of all lines in the model (Figure \ref{fig:linelengths}).
The remaining 15\% of lines, which are longer than 250km, are modelled identically although
more rigorous alternatives exist. These include partitioning long lines into multiple
shorter sections to model series compensation \cite{kundur1994},
or using equations specifically for long lines that include fewer
simplifying approximations of impedances than the $\Pi$ model \cite{machowski2008}.

\section{Results and Discussion}
\label{sec:results}

This section presents and discusses the results from the 
experiments as described in Section \ref{sec:setup}.
As evaluation criteria we consider the
total system costs and the optimal system composition (Section \ref{sec:res-costs}),
the error of optimised losses  (Section \ref{sec:res-lossapproximation}),
the error of optimised flows compared to simulated flows (Section \ref{sec:res-flowcomparison}),
as well as peak memory consumption and solving time (Section \ref{sec:res-benchmark}).

\subsection{Investments, Nodal Prices and Total System Costs}
\label{sec:res-costs}

\begin{table}
	\centering
	\footnotesize
	\begin{tabular}{ll|rrrrrr}
		\toprule
		 &  &  & Lossy &  & Iterative & & Iterative \\
		 Indicator & Unit & Transport & Transport & Lossless & Lossless & Lossy & Lossy \\ \midrule
		 System cost & bn\euro p.a. & 220.2 & 226.2 & 224.9 & 225.7 & 243.8 & 238.5 \\
		  & \euro/MWh & 70.2 & 72.1 & 71.7 & 71.9 & 77.7 & 76.0 \\
		  Energy transmitted & EWhkm & 1.56 & 1.26 & 1.36 & 1.28 & 0.90 & 0.94 \\
		  Network expansion & TWkm & 216 & 214 & 206 & 206 & 160 & 170 \\
		  Transmission losses & \% of load & 0 & 2.3 & 0 & 0 & 5.1 & 3.7 \\
		Curtailment & \% & 2.0 & 1.9 & 2.3 & 2.4 & 2.2 & 2.4 \\
		Share of $\abs{\theta_i-\theta_j} \geq 30^\circ$ & \% & 5.1 & 3.7 & 4.6 & 3.9 & 1.4 & 1.5 \\
		\bottomrule
	\end{tabular}
	\caption{Various statistical indicators compared accross covered flow models.}
	\label{tab:misc}
\end{table}

\begin{figure}
	\includegraphics[width=0.65\textwidth, trim=0cm 0cm 5.7cm 1.5cm, clip]{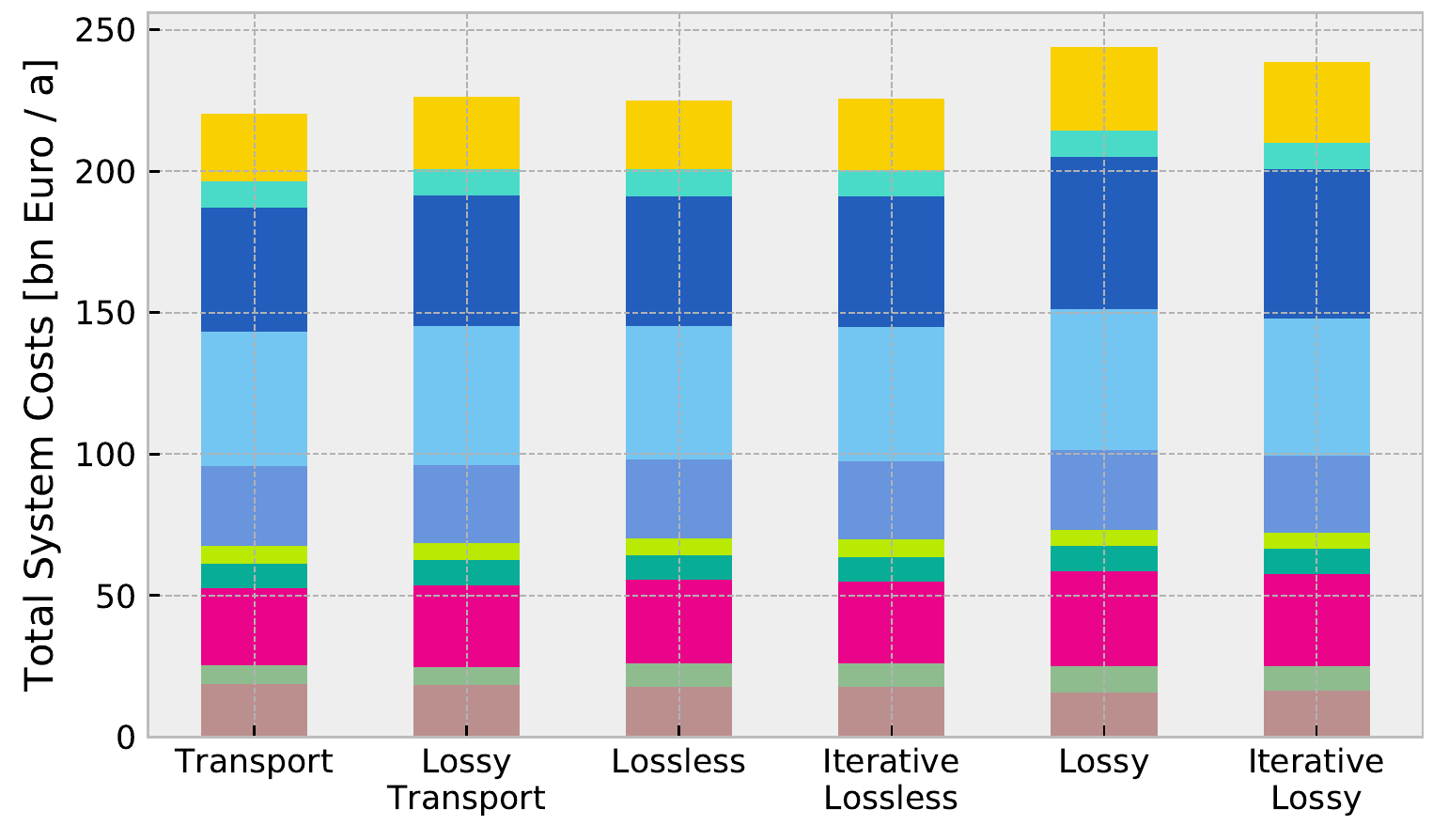}
	\includegraphics[width=0.23\textwidth, trim=15.5cm 1cm 0cm 0cm, clip]{graphics/cost-bar.pdf}
	\caption{Comparison of total annual system costs split by system component for the covered flow models.}
	\label{fig:cost-bar}
\end{figure}

\begin{figure}
	\includegraphics[width=0.245\textwidth]{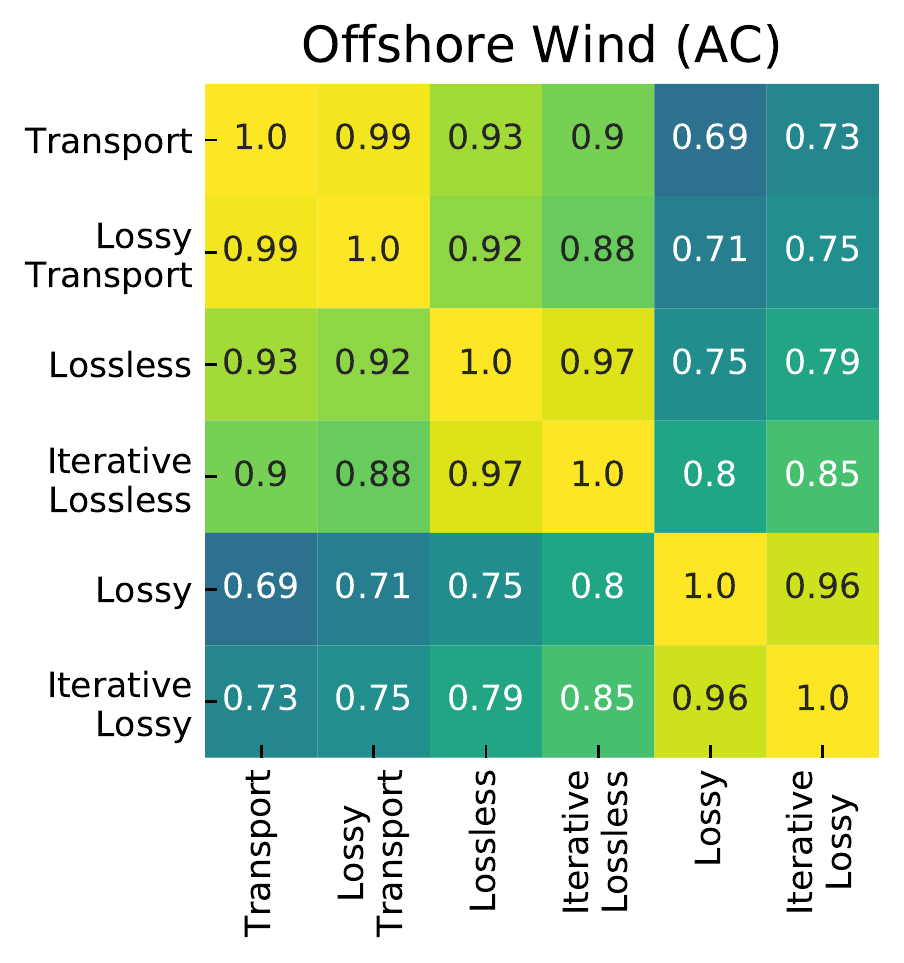}
	\includegraphics[width=0.245\textwidth]{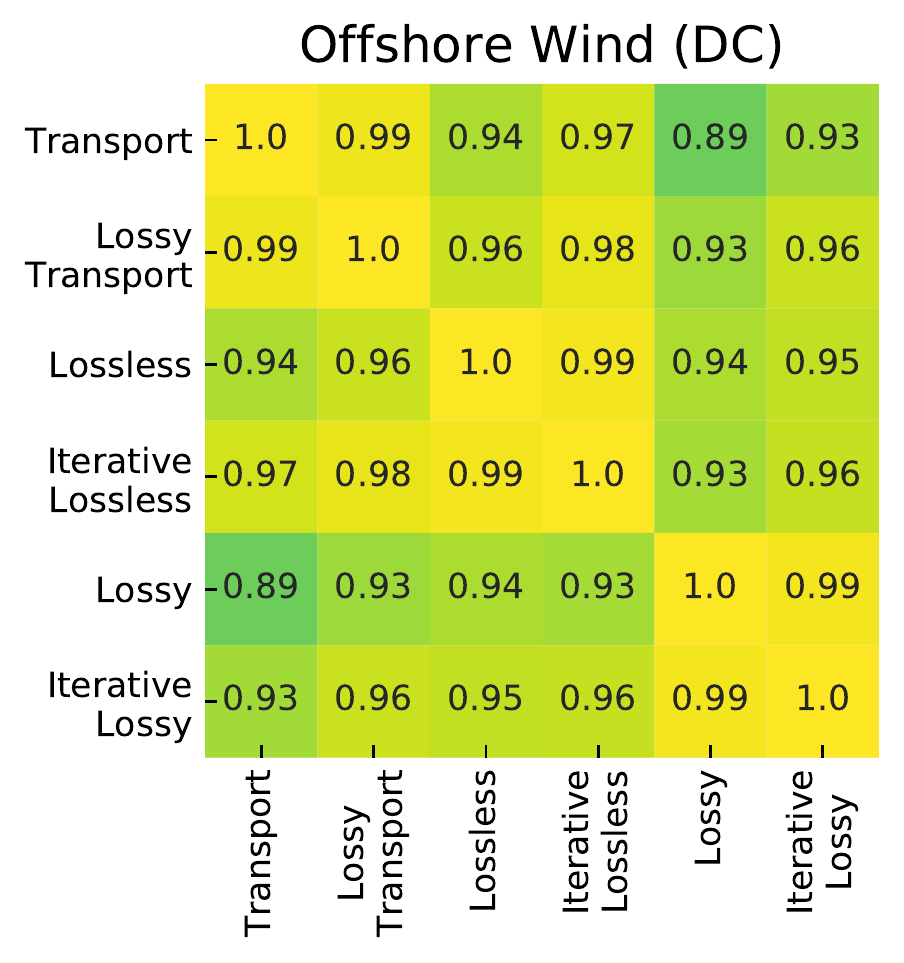}
	\includegraphics[width=0.245\textwidth]{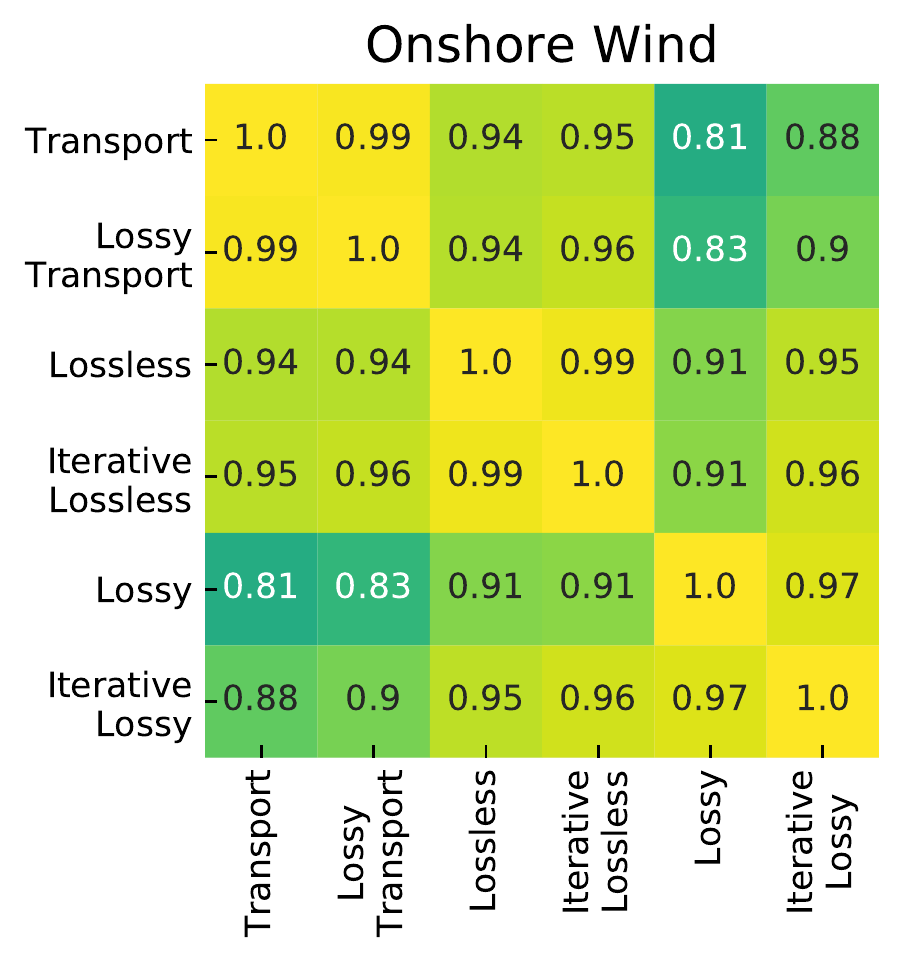}
	\includegraphics[width=0.245\textwidth]{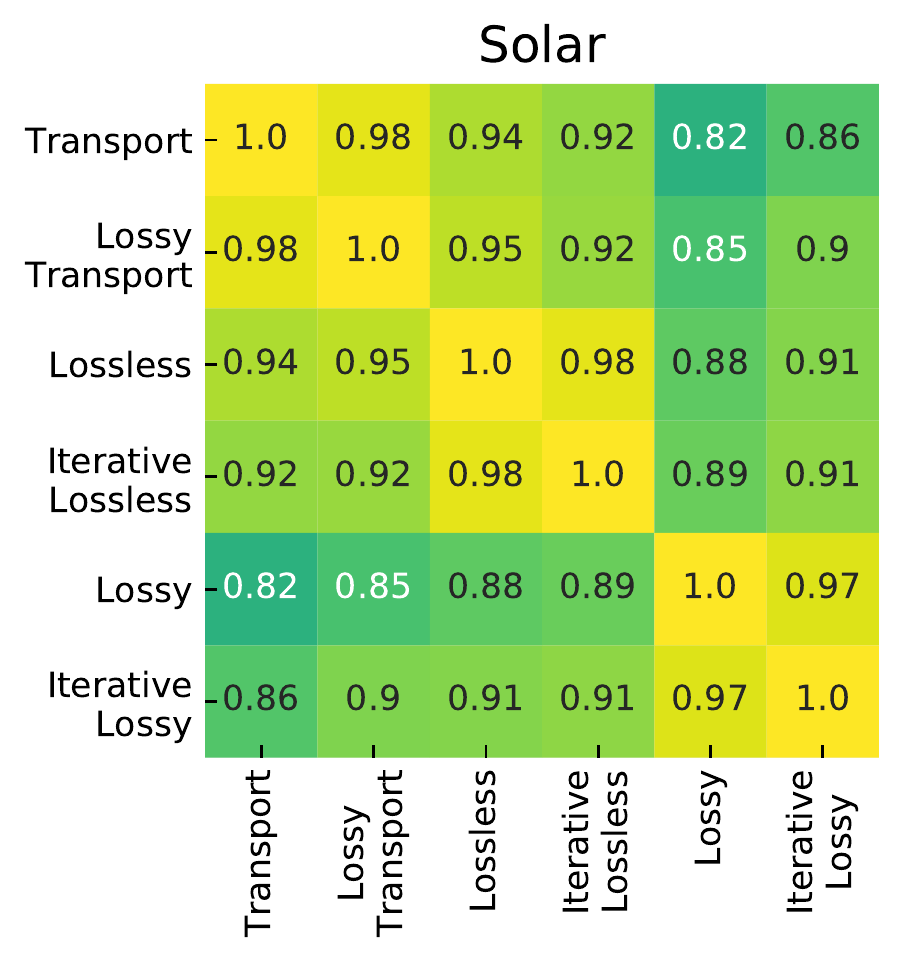} \\
	\includegraphics[width=0.245\textwidth]{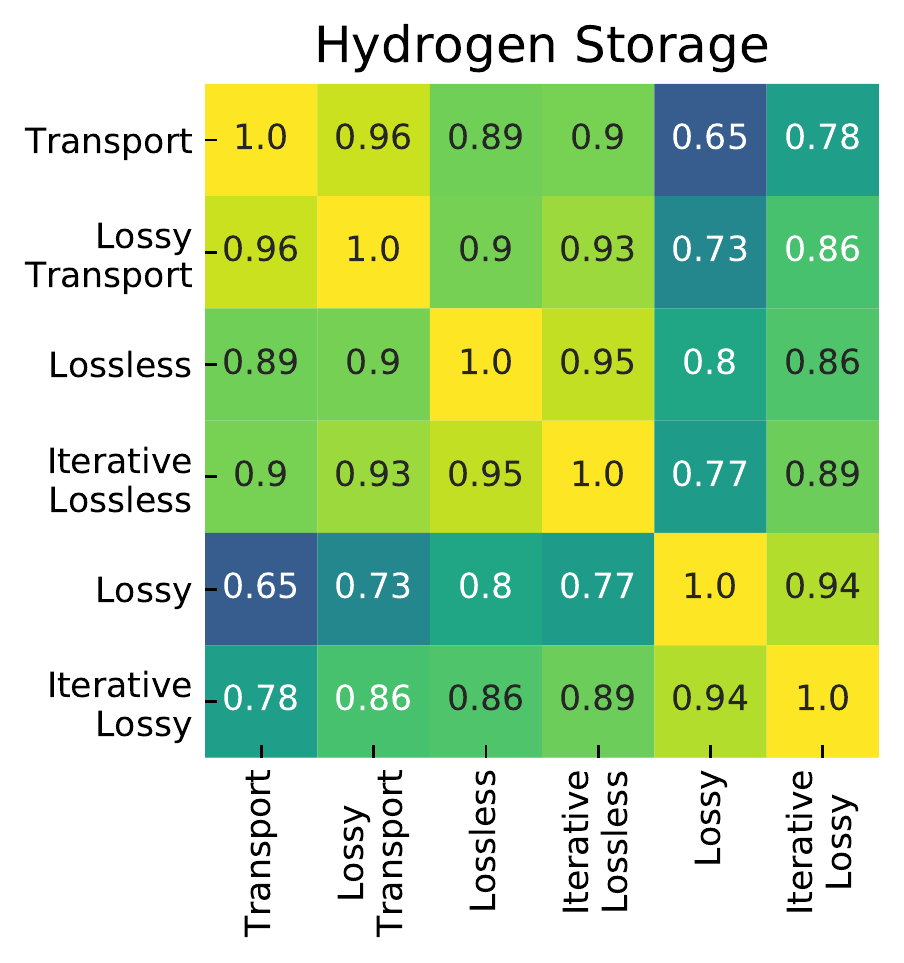}
	\includegraphics[width=0.245\textwidth]{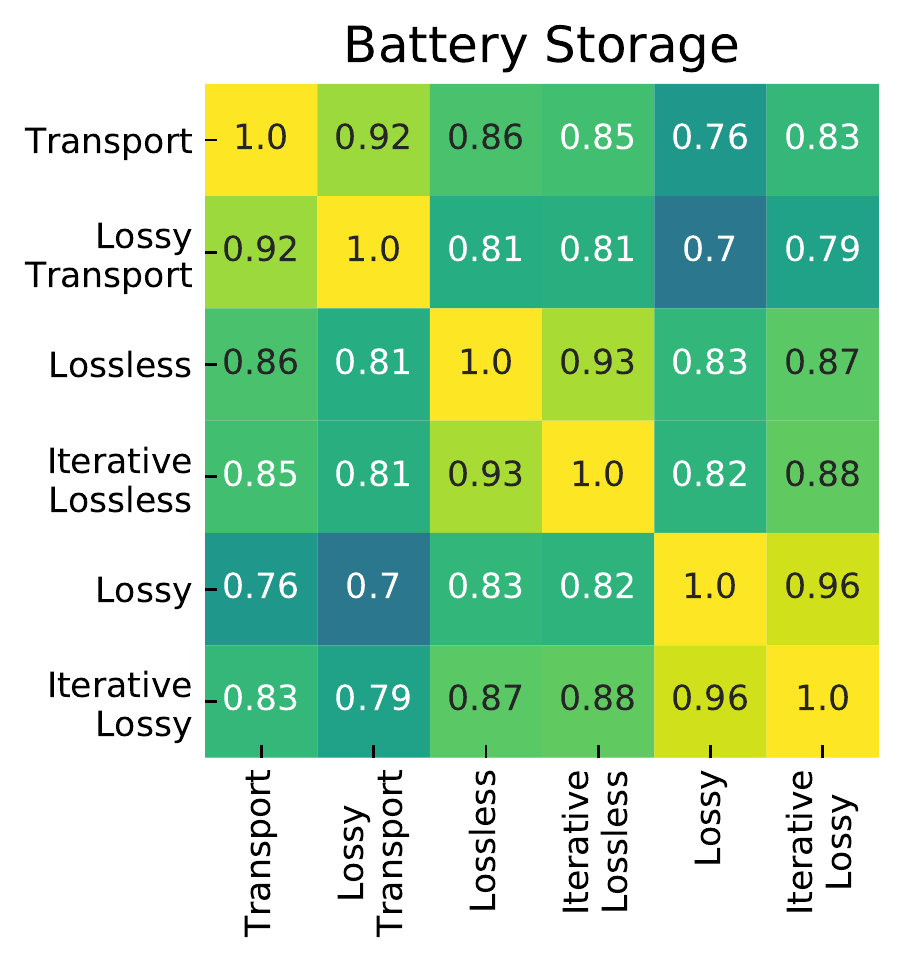}
	\includegraphics[width=0.245\textwidth]{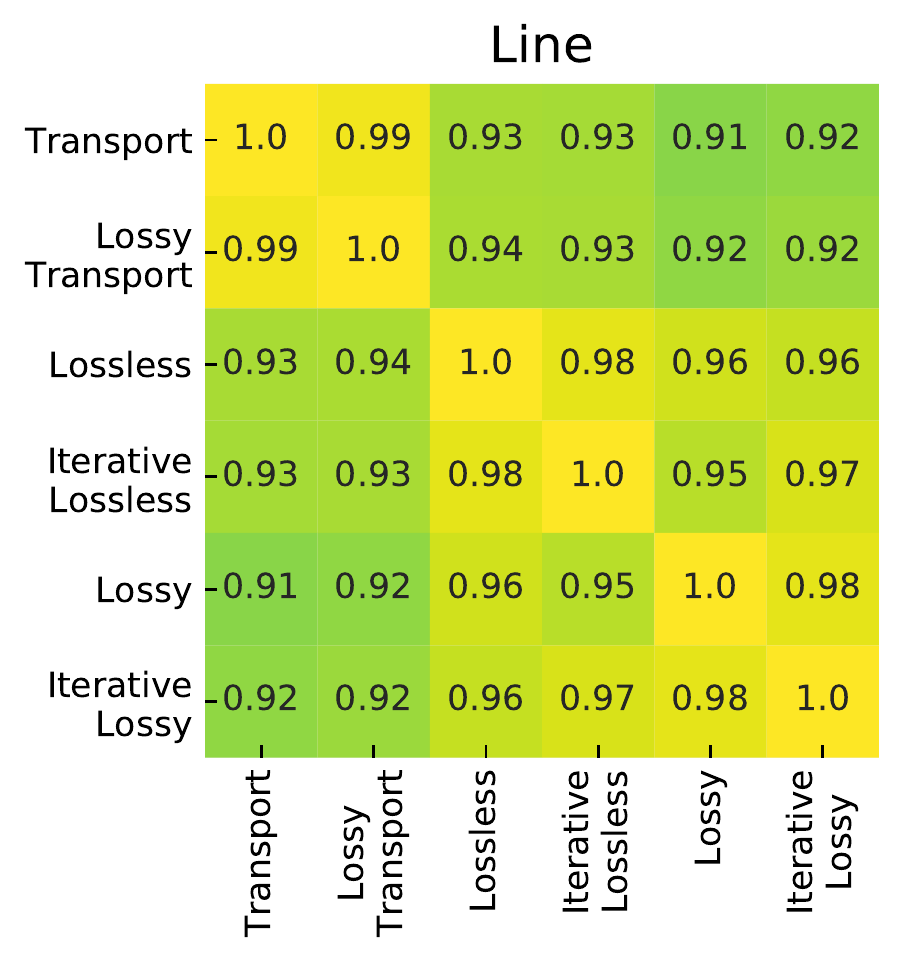}
	\includegraphics[width=0.245\textwidth]{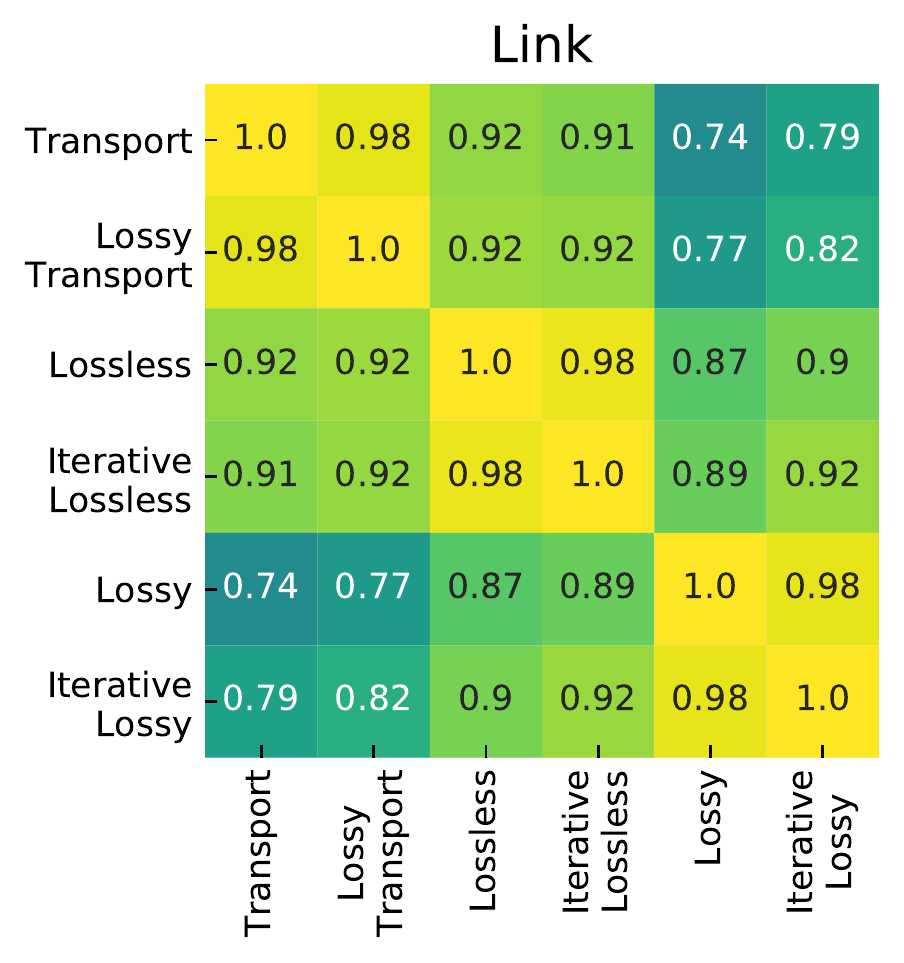}
	\caption{Capacity correlation of optimised nodal investments
	among covered flow models distinguished by technology.}
	\label{fig:capacity-corr}
\end{figure}

\begin{figure}
	\includegraphics[width=\textwidth]{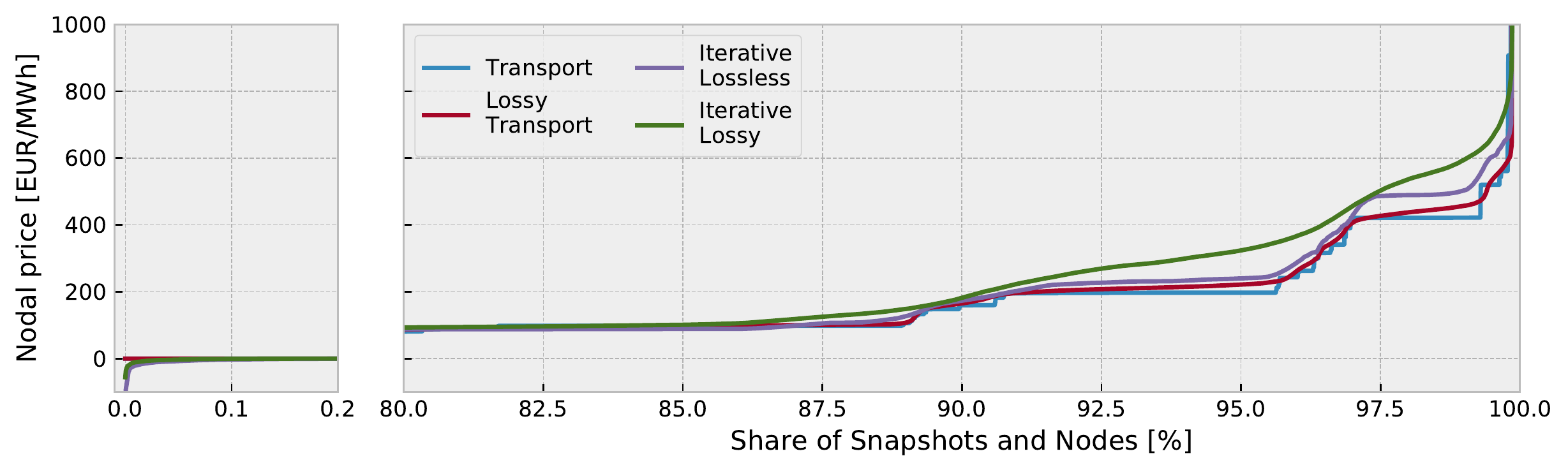}
	\caption{Nodal price duration curves (snapshots and nodes) for selected flow models.
	In the omitted section, prices rise steadily and similarly for all models. Some models allow for negative nodal prices with occurence below 0.2\%.}
	\label{fig:price-duration}
\end{figure}

\begin{figure}
	\centering
	\begin{tabular}{cc}
		Transport & Lossy Transport \\
		\includegraphics[width=0.49\textwidth, clip, trim=0cm 4.5cm 0cm 0cm]{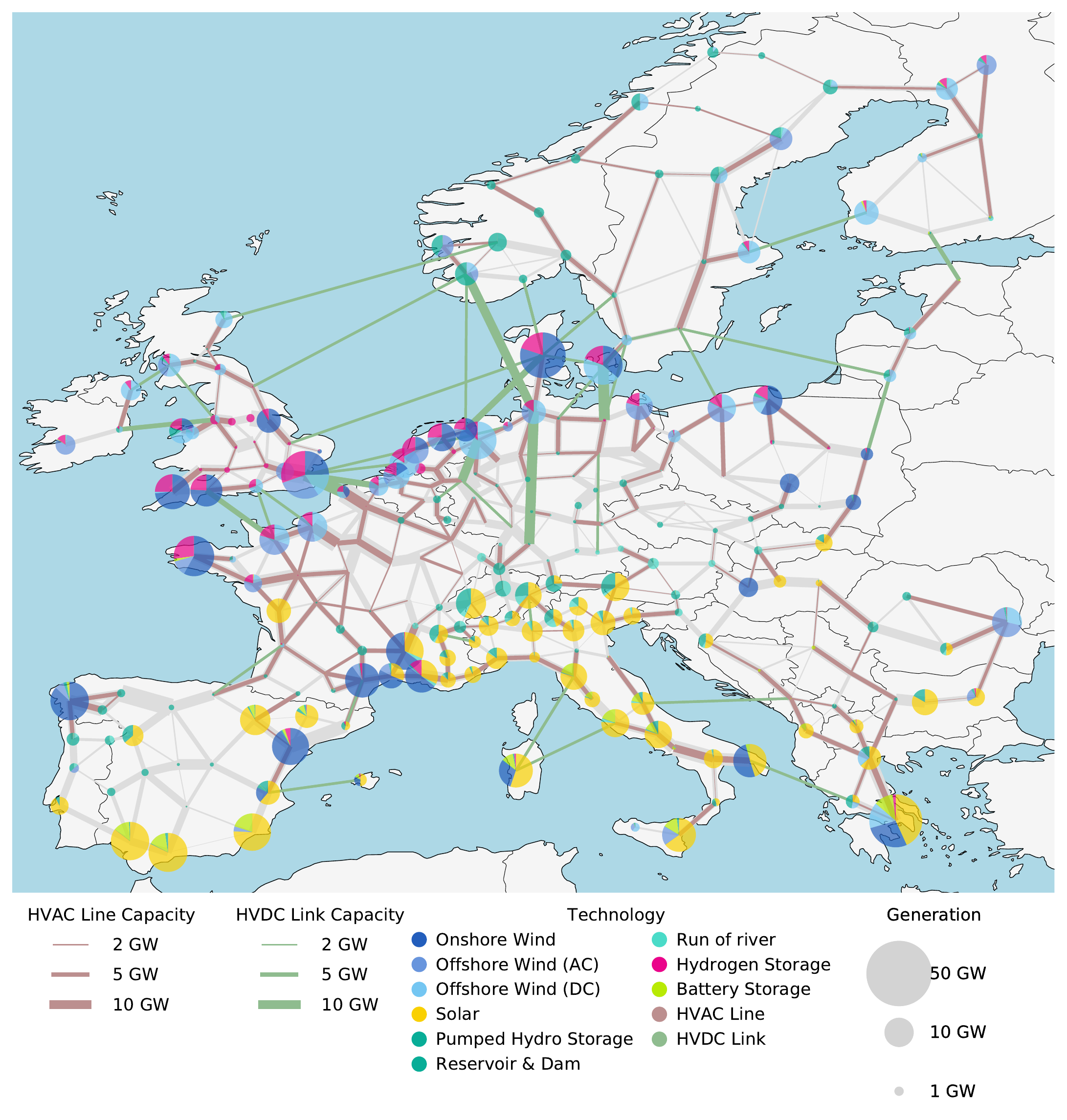} &
		\includegraphics[width=0.49\textwidth, clip, trim=0cm 4.5cm 0cm 0cm]{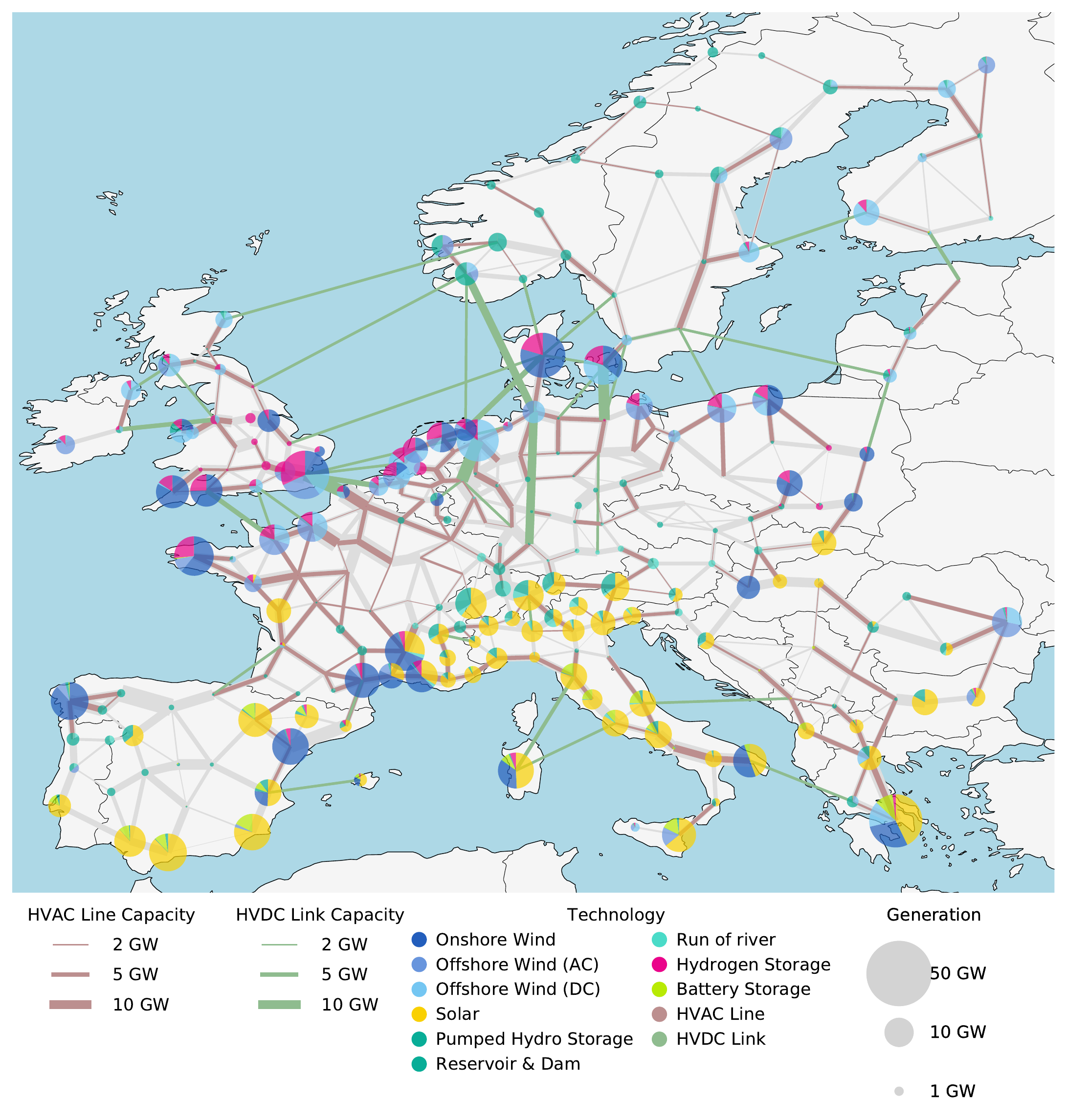} \\
		Lossless & Iterative Lossless \\
		\includegraphics[width=0.49\textwidth, clip, trim=0cm 4.5cm 0cm 0cm]{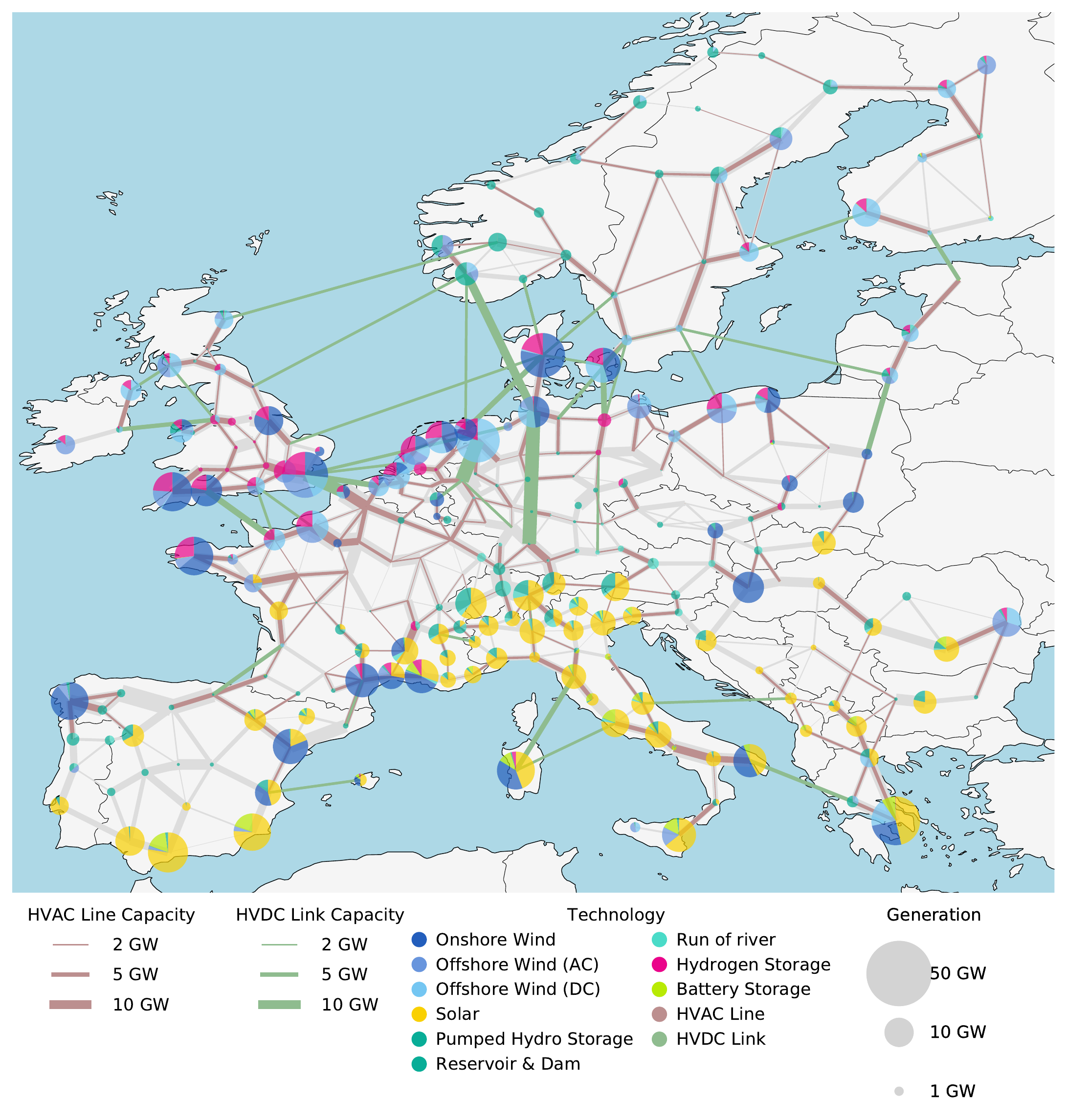} &
		\includegraphics[width=0.49\textwidth, clip, trim=0cm 4.5cm 0cm 0cm]{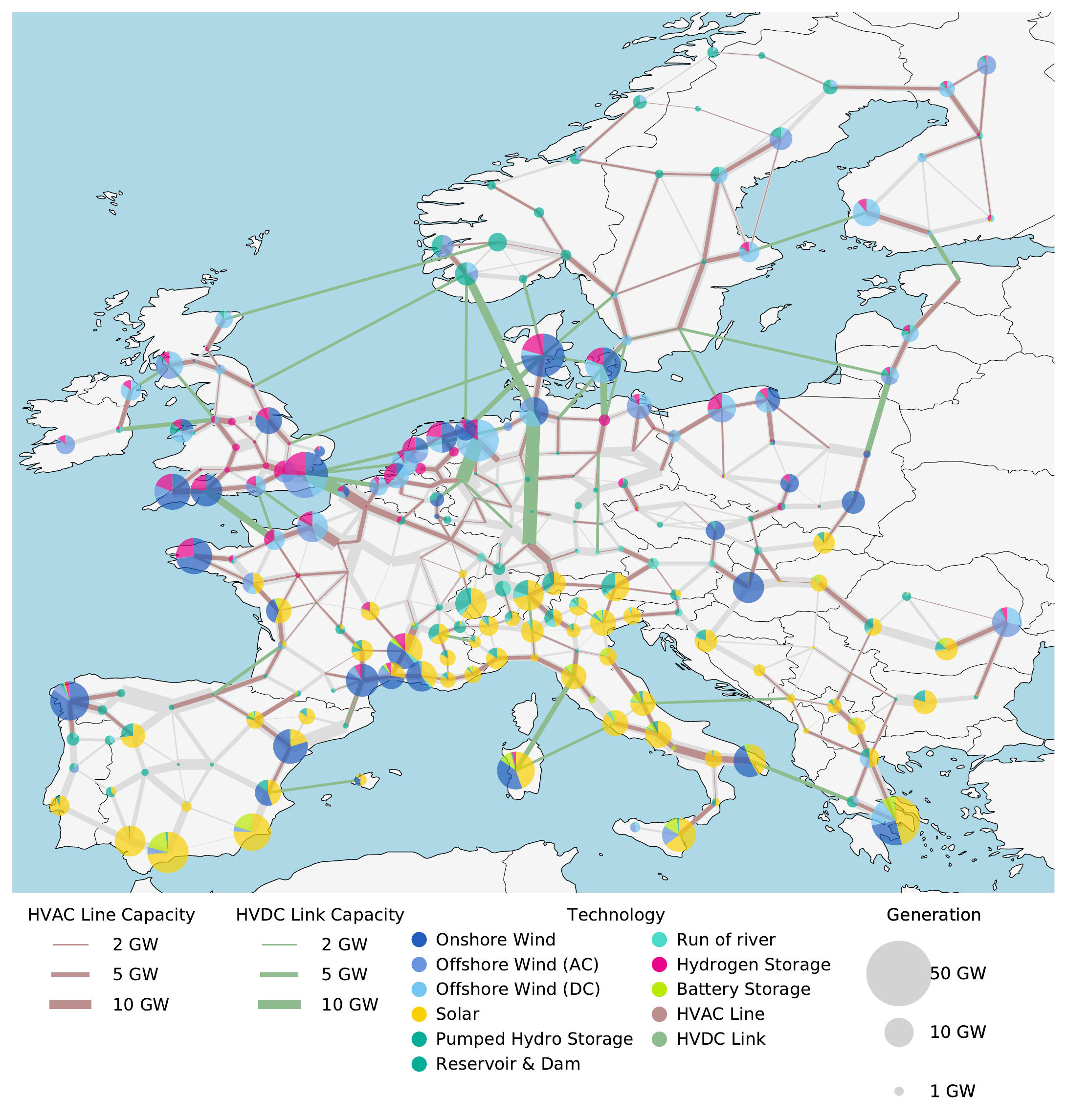} \\
		Lossy & Iterative Lossy \\
		\includegraphics[width=0.49\textwidth, clip, trim=0cm 4.5cm 0cm 0cm]{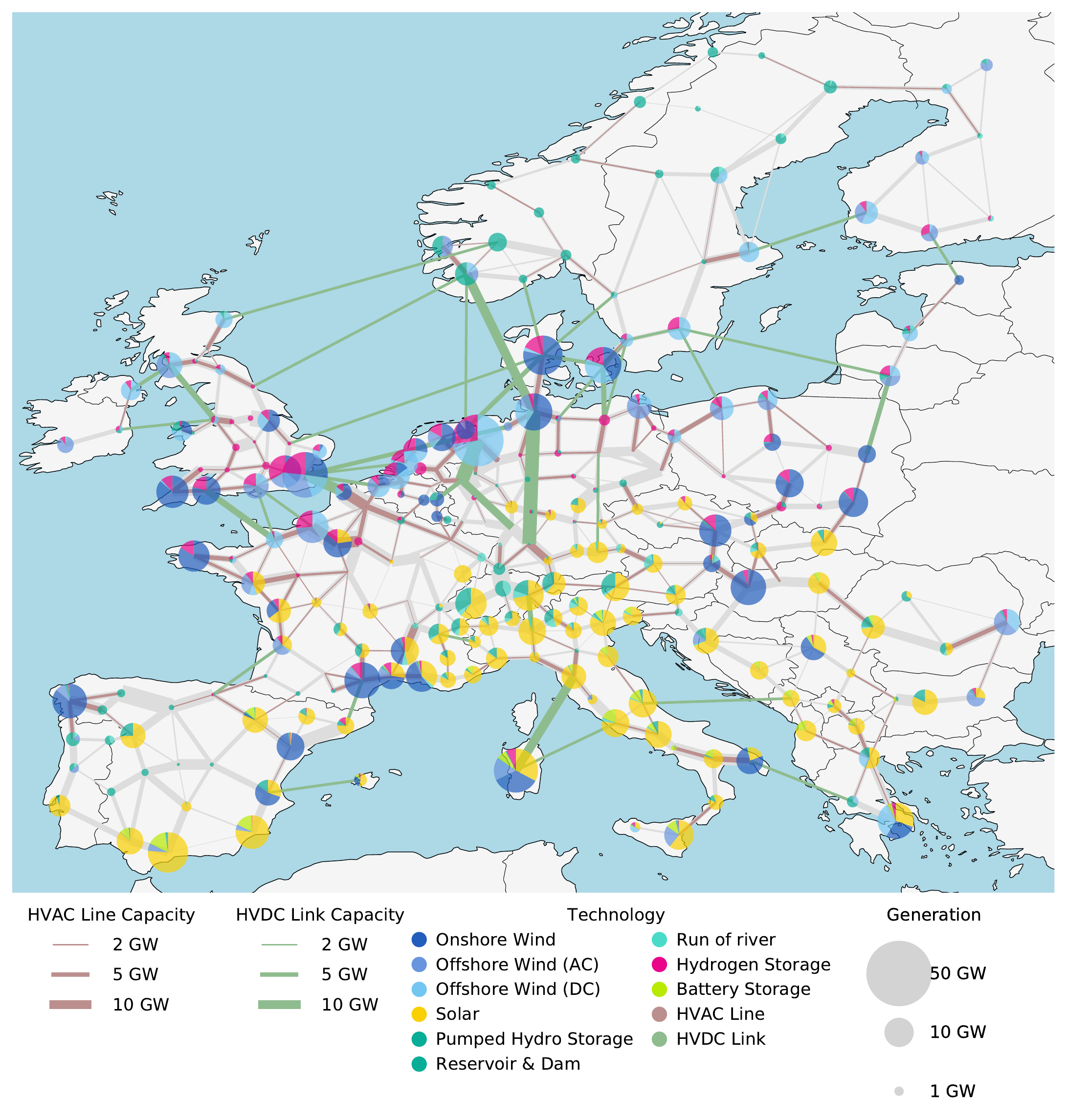} &
		\includegraphics[width=0.49\textwidth, clip, trim=0cm 4.5cm 0cm 0cm]{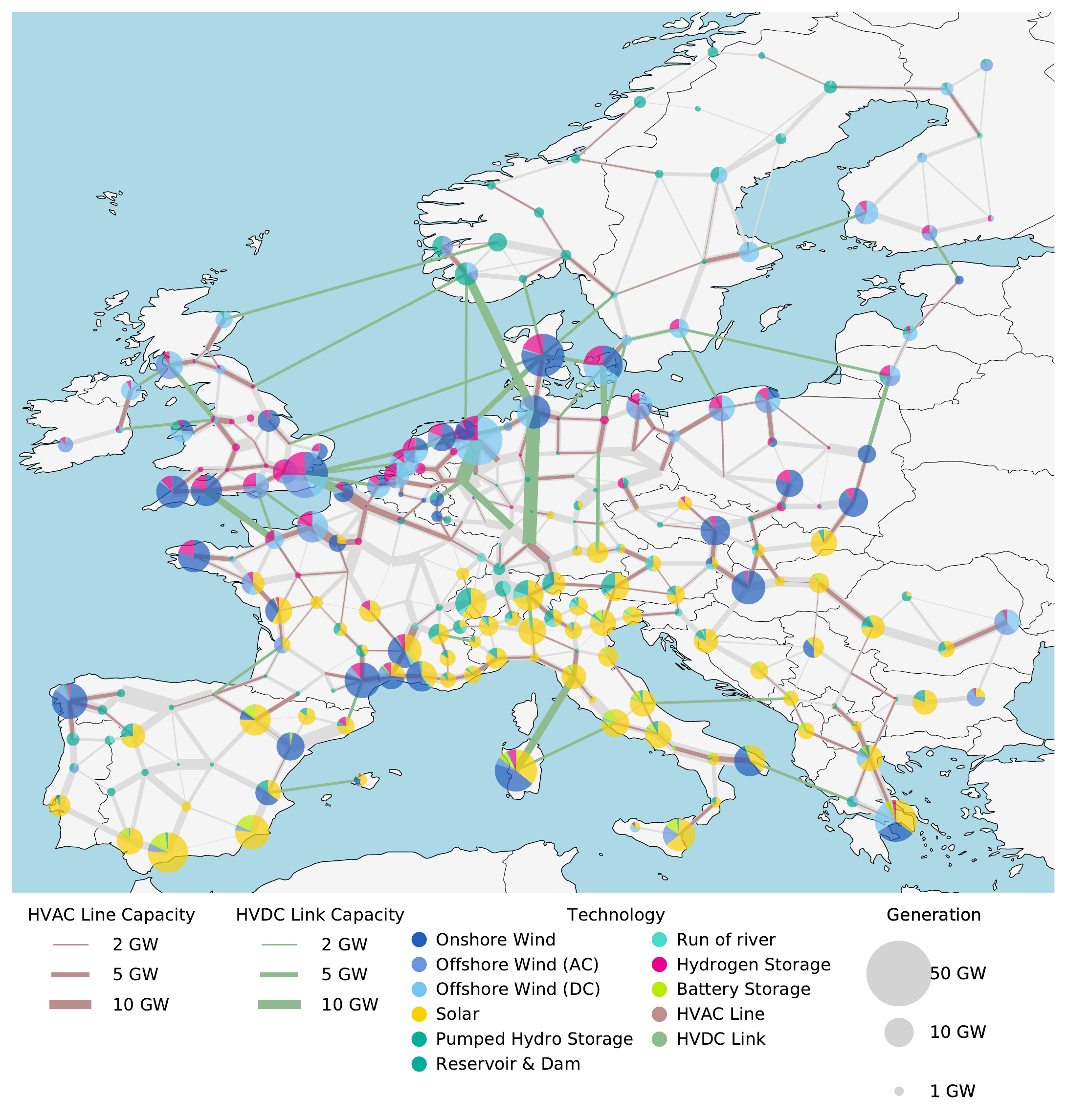}
	\end{tabular}
	\includegraphics[width=0.49\textwidth, clip, trim=0cm 0cm 0cm 19cm]{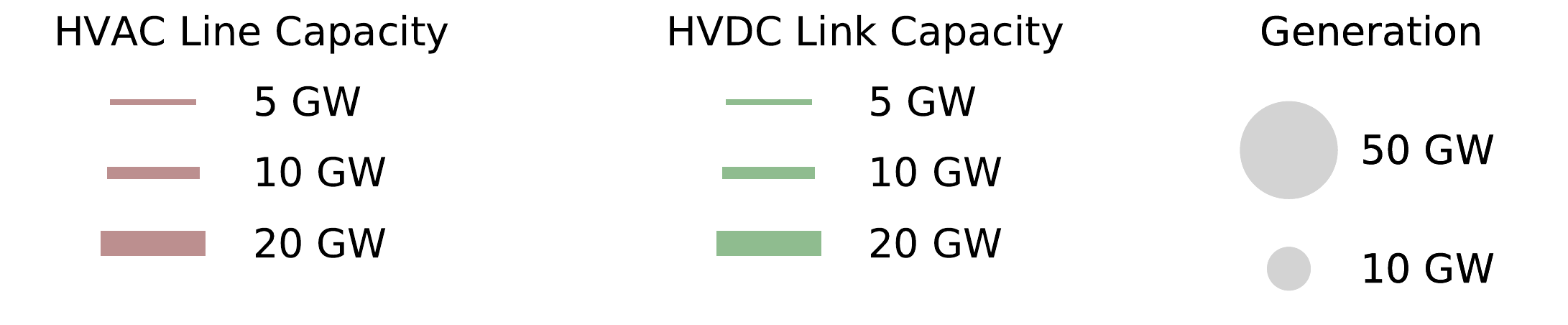}
	\includegraphics[width=0.49\textwidth, clip, trim=0cm 0cm 0cm 19cm]{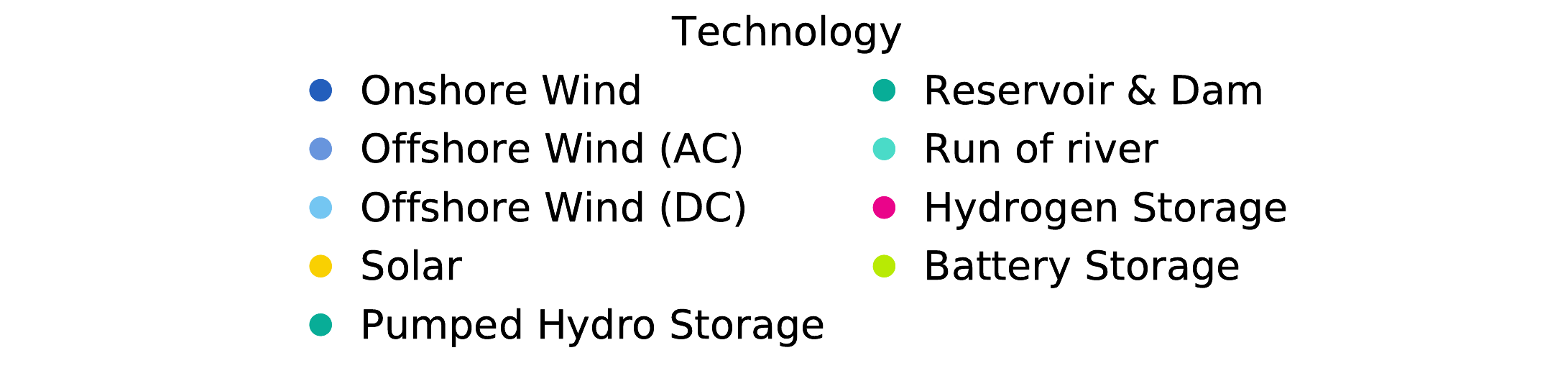}
	\caption{Maps of cost-optimal capacity expansion results for the covered lossless flow models.}
	\label{fig:network-map-lossless}
\end{figure}

Table \ref{tab:misc} presents total transmission losses 
to sum up to around 4\% of the total load when updated impedances according to line expansion are used.
In comparison to the 1.2\% transmission losses reported by the
German Federal Network Agency for the year 2019 \cite{bnetza2019}, this value is
higher owing to the larger volume of power transmission
across the whole continent in scenarios with high shares of renewables.
Skipping the update of impedances overestimates losses (5.5\%)
because additional parallel lines reduce the total impedance.
The lossy transport model underestimates losses (2.5\%) since it
neglects the quadratic relationship between power and losses.
Table \ref{tab:misc} further shows low curtailment at around 2\% across
all flow models due to generous line expansion allowances.

At first sight, the optimised technology mix appears relatively similar across
all flow models, both in terms of cost composition in Figure \ref{fig:cost-bar}
and the map of investments in Figure \ref{fig:network-map-lossless}.
This is further underlined by the high correlations of optimised capacities shown in Figure \ref{fig:capacity-corr}.
Potentially due to some placement degeneracies, lowest correlations are found for battery and hydrogen storage.
Further notable differences concern grid reinforcement.
The lossless and lossy transport models feature many new
transmission lines in France and Scandinavia,
which disappear as more accurate flow models are applied.
The difference adds up to 20\% less network reinforcement.
Likewise, the energy transmitted decreases as more constraints are imposed on power transmission.
In order to avoid grid losses, models that consider transmission losses and KVL
transmitted up to 66\% less energy than the transport model.
The reduced spatial transport of power is then compensated
by a shift towards hydrogen storage and controllable HVDC links (e.g.~in the West of Germany).
Despite the involved conversion losses, balancing renewables
in time through storage becomes more attractive.
Additionally, to offset the energy lost by transmission
but also the reduced amount of power transmission,
lossy models feature more wind and solar generation capacity.
This includes both more localised generation
(e.g.~more solar panels in Southern Germany and more onshore wind
turbines in Eastern Europe) where previously there were few production sites,
and more concentrated generation in the North Sea region to pair
with the appended storage units.
The added capacities raise the system cost.
In total, the annual system costs increase by approximately
5.7\% compared to iterative linearised power flow, or
8\% relative to the transport model.

Besides investments, we also compare electricity prices in an idealised
nodal market by using the dual variables of the nodal balance constraints.
The price duration curves depicted in Figure \ref{fig:price-duration} show that nodal
prices are more evenly distributed in the lossless linear power flow compared compared to the transport model.
The even distribution of prices was also found in Gunkel et al.~\cite{gunkel_modelling_2020}.
The transport model and lossy transport model do not have the properties that would allow negative prices.
Negative nodal prices are a consequence of KVL and occur when increasing demand at a bus relieves a transmission line,
allowing power to be exported from somewhere cheap to somewhere expensive. This lowers the system cost and consequently
results in a negative price at that bus. Other constraints that can generally entail negative prices
are unit commitment constraints, but these are not considered in this contribution.
We find that even for models with KVL and loss approximations, negative prices are rare ($\leq 0.2\%$).
The major differences regarding nodal prices can be observed in the 10\% of highest prices.
The transport model features step-like price profiles, whereas the profiles of the other models are smoother.
The iterative lossy linearised power flow model possesses the highest yet smoothest price duration curve.

\subsection{Validation of Loss Approximation}
\label{sec:res-lossapproximation}

\begin{figure}
	\centering
	\begin{tabular}{cc}
		Lossy Transport & Iterative Lossy \\
		\includegraphics[width=0.49\textwidth]{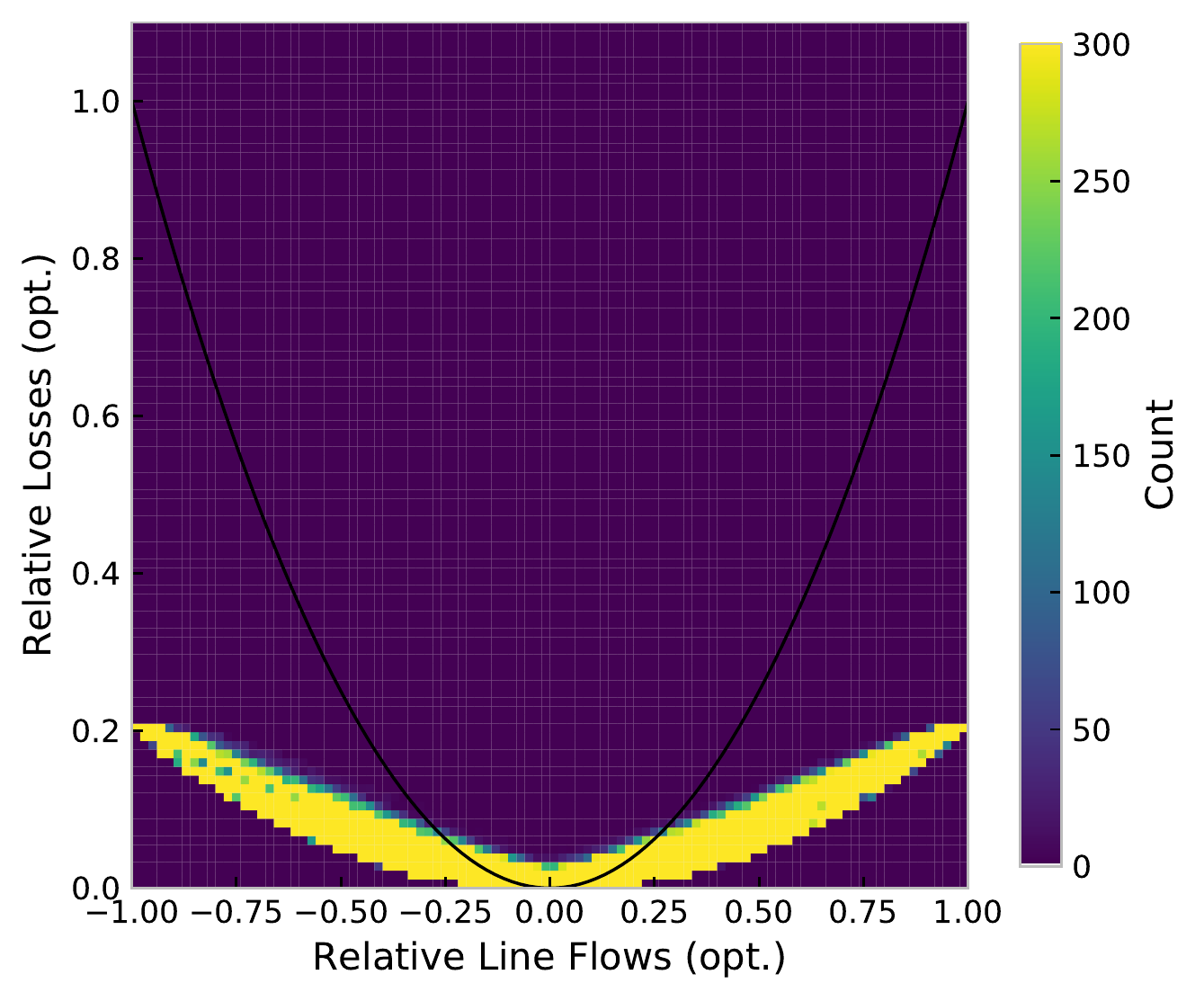} &
		\includegraphics[width=0.49\textwidth]{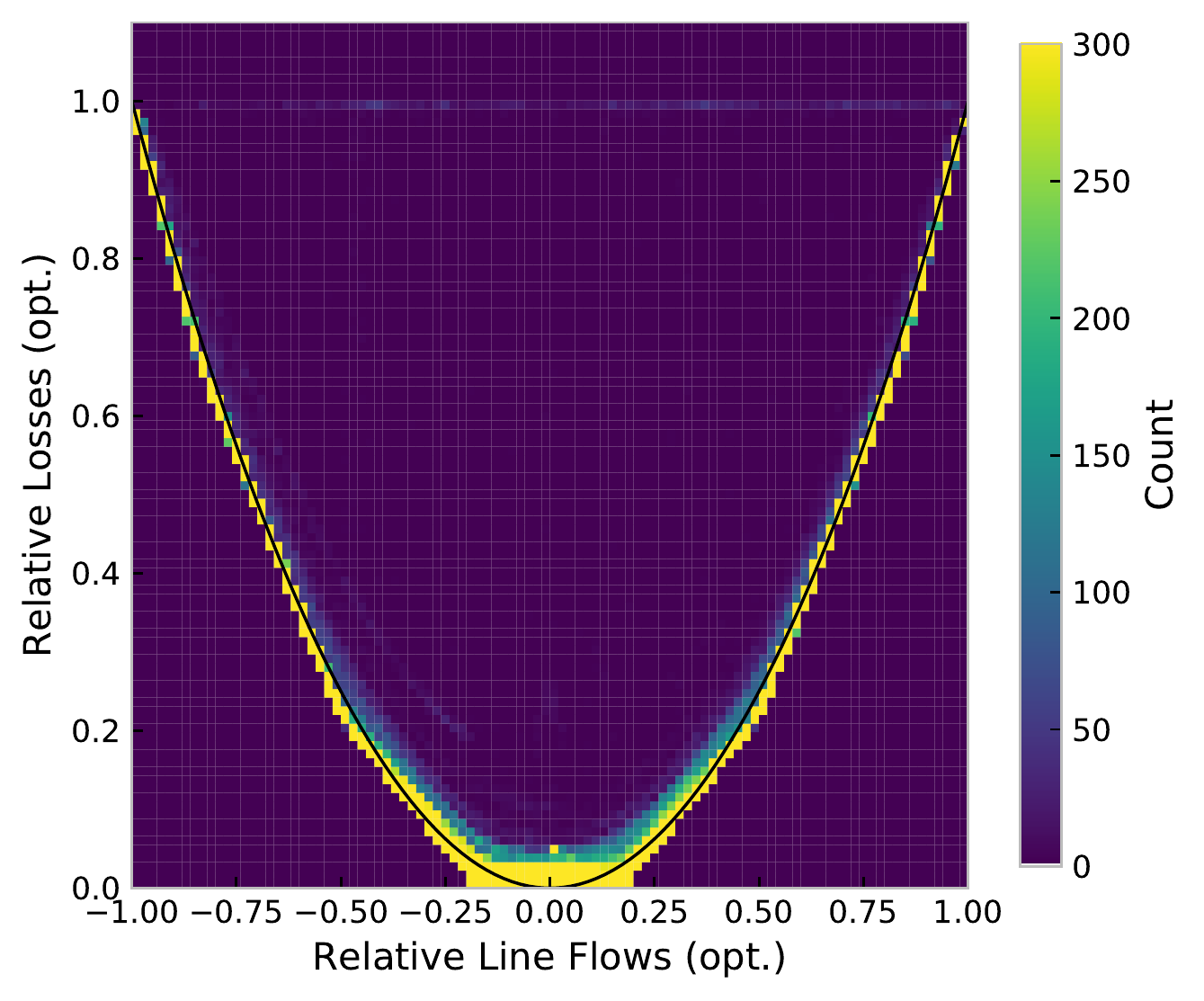}
	\end{tabular}
	\caption{Examination of convex envelope relaxation around
			 loss formula $\psi=r p^2$ 
			 given in equation \eqref{eq:ri2}
			 for lossy transport model and the iterative lossy
			 linearised power flow model
			 in a two-dimensional histogram.
			 The line flows are normalised by their nominal capacity
			 including the $N-1$ security margin ($\sfrac{p_{\ell,t}}{\overline{p}_{\ell} P_{\ell}}$)
			 and maximum losses according to security-constrained line capacity respectively,
			 such that lines with different electrical parameters can be mapped onto the same chart.
			 The count refers to a tuple $(\ell,t)$ of line and snapshot.
			 The black line depicts the normalised quadratic loss formula \eqref{eq:ri2}.}
	\label{fig:flow-vs-loss}
\end{figure}

\begin{figure}
	\centering
	\begin{tabular}{cc}
		Lossy & Iterative Lossy \\
		\includegraphics[width=0.49\textwidth]{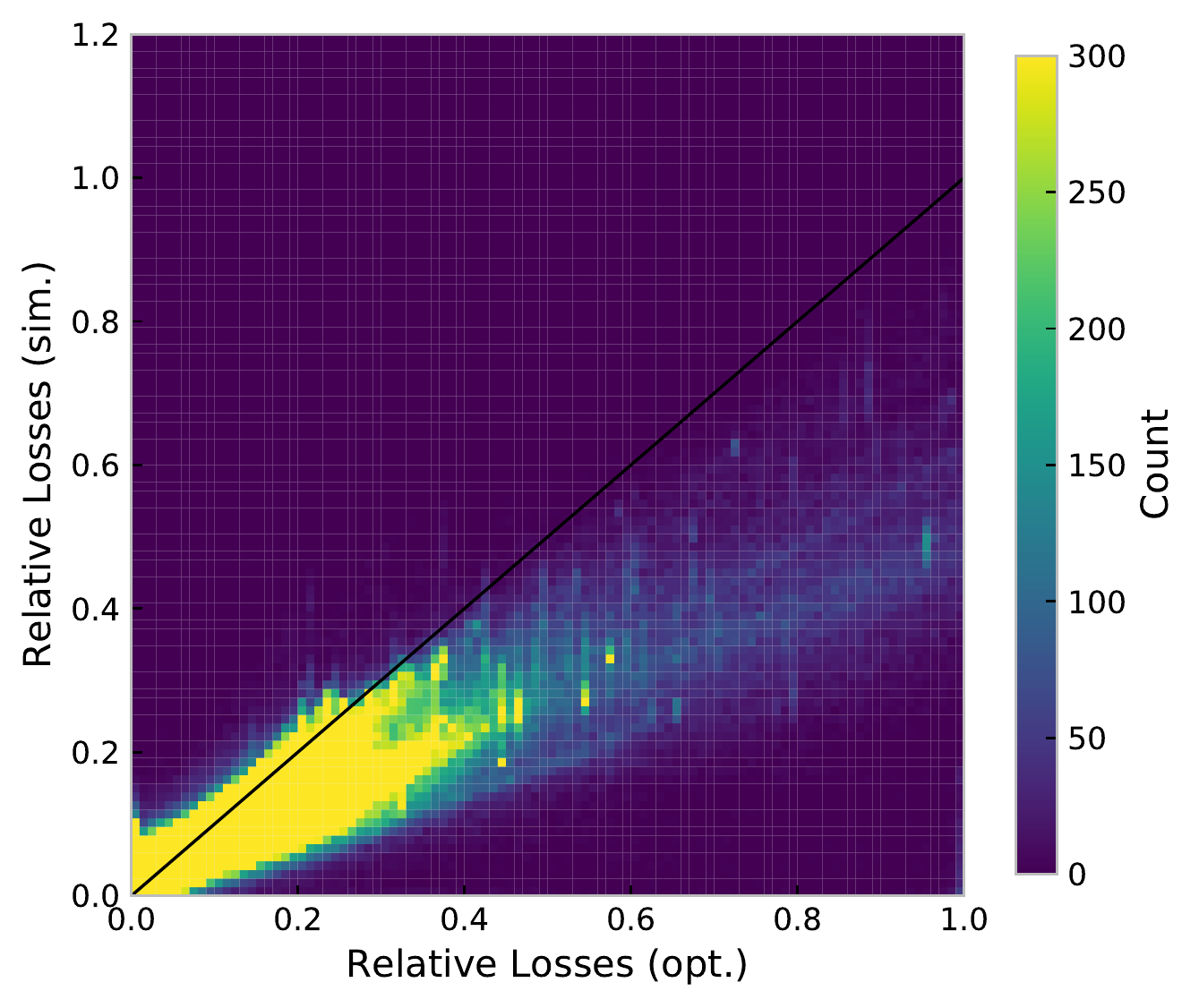} &
		\includegraphics[width=0.49\textwidth]{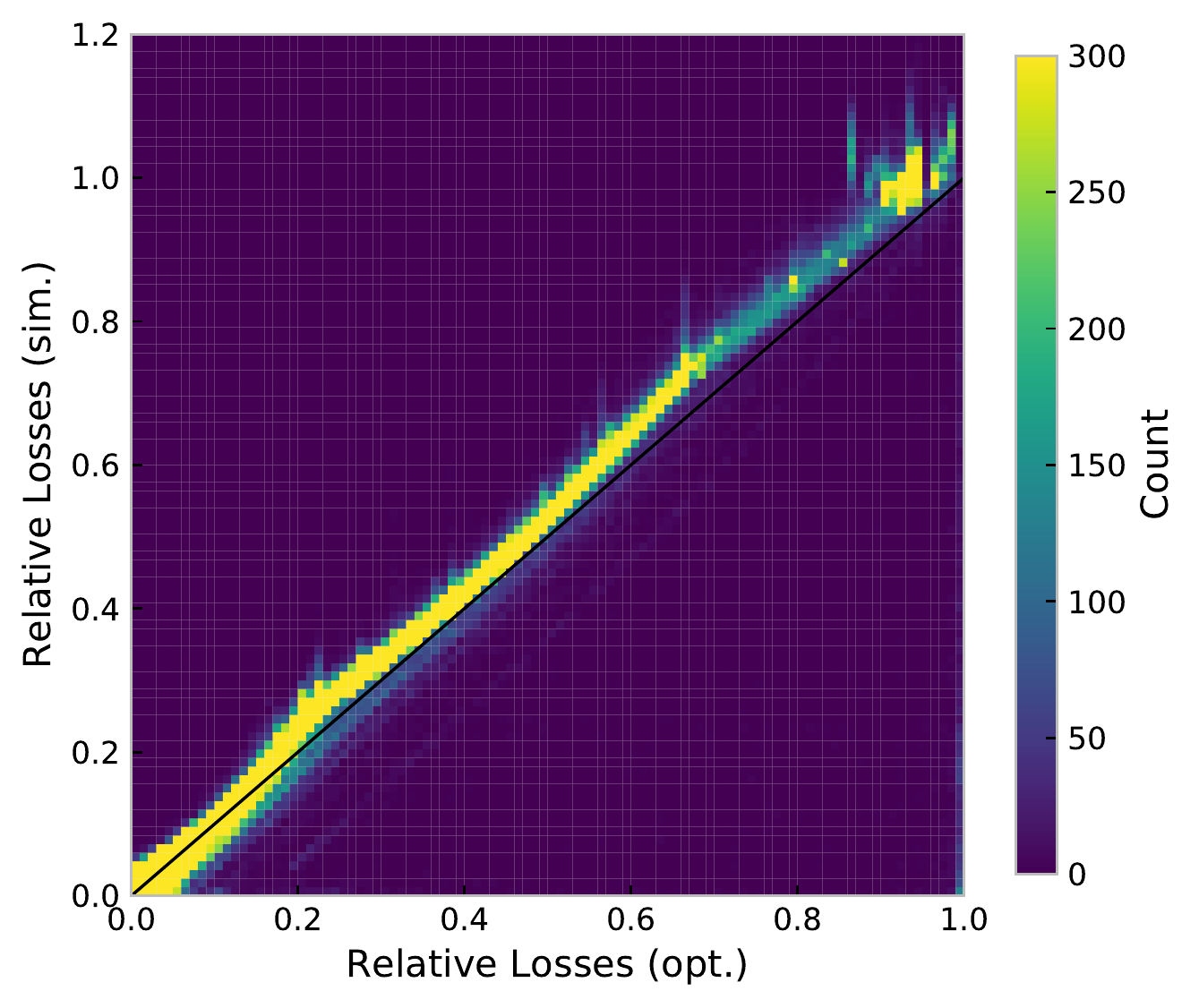}
	\end{tabular}
	\caption{Comparison of simulated losses from AC power flow equations
			 and optimised losses for iterative and non-iterative
			 lossy linearised power flow
			 in a two-dimensional histogram.
			 Relative losses are shown as $\sfrac{\psi_{\ell,t}}{\psi_\ell^{\mathrm{max}}}$
			 according to security-constrained line capacity $\overline{p}_{\ell} P_{\ell}$.
			 The count refers to a tuple $(\ell,t)$ of line and snapshot.
			 The black line indicates perfect alignment of simulated and optimised losses.}
	\label{fig:loss-vs-loss}
\end{figure}

Figure \ref{fig:flow-vs-loss} relates optimised line flows $p_{\ell,t}$ to
optimised losses $\psi_{\ell,t}$ for the lossy transport model and the iterative
lossy linearised power flow model.
The lossy transport model underestimates losses under high loading conditions
depending on the assumed constant loss factor and fails to reflect the
quadratic relationship between losses and flow.
On the contrary, the results also confirm that approximating losses
in linearised optimal power flow
with a convex envelope does not degrade the obtained solutions.
Although the envelope around the loss parabola \eqref{eq:ri2}
(cf.~Section \ref{sec:mod-lossy}, approximates cosine in \eqref{eq:cosine-loss-power})
allows for losses to take values above the parabola,
the cost associated with losses tends to push losses downwards.
Substantial deviations from the parabola to above only occur when there is no cost
(or even a benefit in the case of negative nodal prices)
associated with higher losses;
e.g.~when energy is being curtailed,
or when there is some extra consumption of interest to control power flows
or some other problem degeneracy.
As previously shown in Figure \ref{fig:price-duration}, negative nodal prices
and consequently incentives for loss overestimation are rare ($\leq0.2\%$).
These circumstances cause the generous convex relaxation to function well.
Underestimating losses is also possible, albeit to a much smaller extent,
as a small fraction of the feasible space lies between the loss parabola and
the tangents that form the convex envelope. Recall that the loss parabola \eqref{eq:ri2}
is already an approximation of the cosine terms in equation \eqref{eq:cosine-loss-power}.

Figure \ref{fig:loss-vs-loss} compares transmission losses retrieved
from the optimisation problem to the simulated losses from AC power flow
for the iterating and non-iterating loss approximation.
Like in Figure \ref{fig:flow-vs-loss}, we note that the iterative lossy
formulation manages to sufficiently represent losses observed in the respective AC power flow simulation.
However, when the iteration is skipped and hence line impedances are
not updated according to their optimised capacities, losses are overestimated.

\subsection{Validation of Optimised Line Flows}
\label{sec:res-flowcomparison}

\begin{figure}
	\centering
	\small
	\begin{tabular}{cc}
		Transport & Lossy Transport \\
		\includegraphics[width=0.49\textwidth]{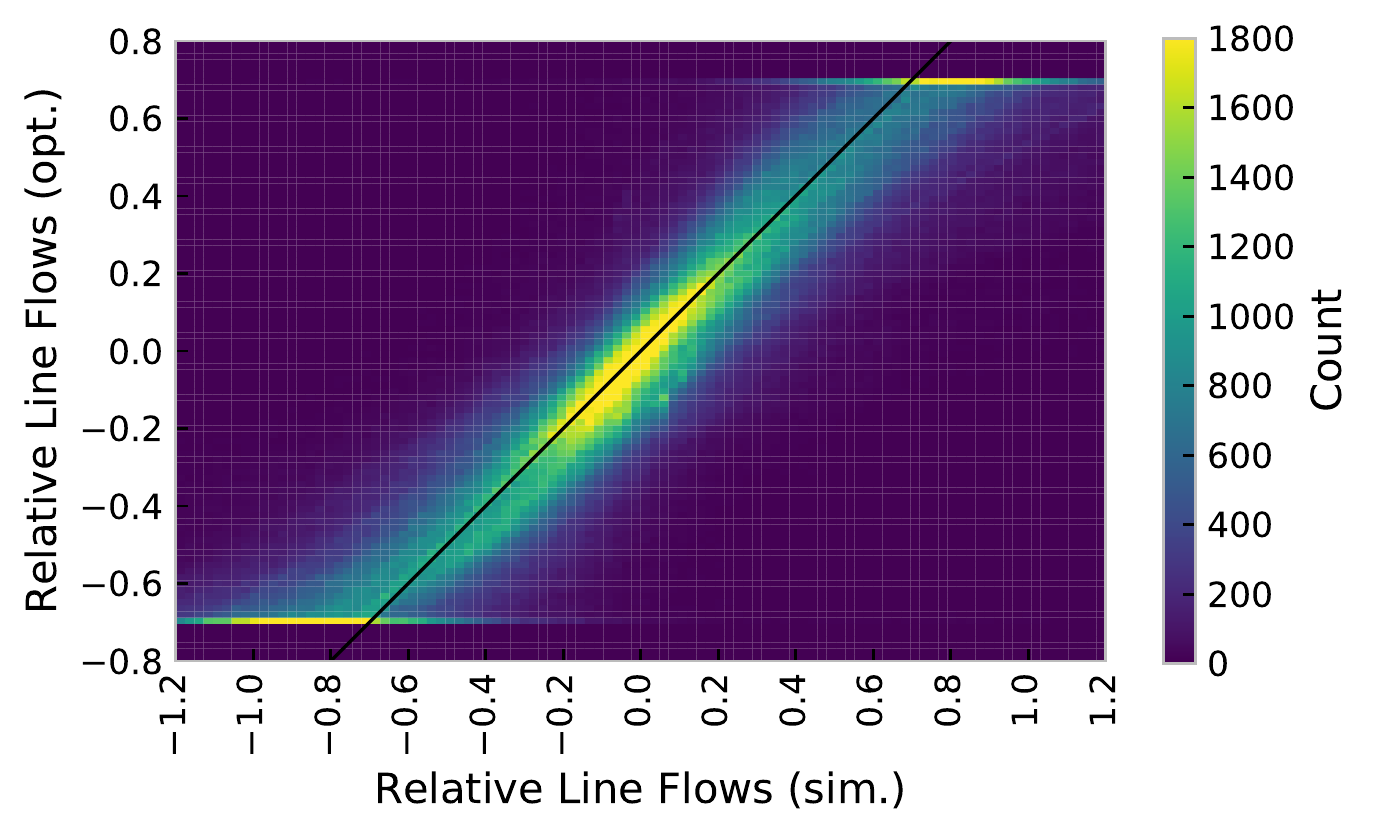} &
		\includegraphics[width=0.49\textwidth]{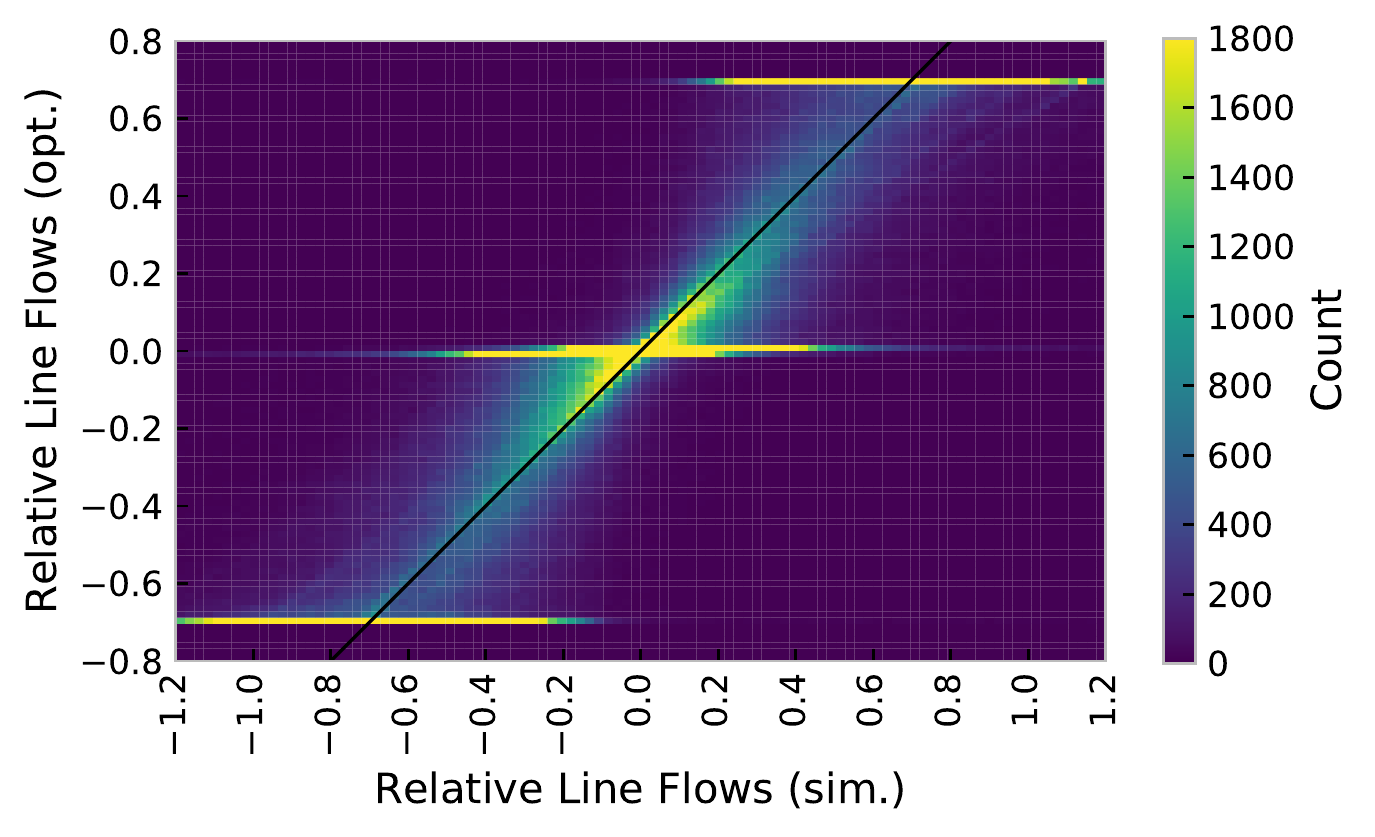} \\
		Lossless & Iterative Lossless \\
		\includegraphics[width=0.49\textwidth]{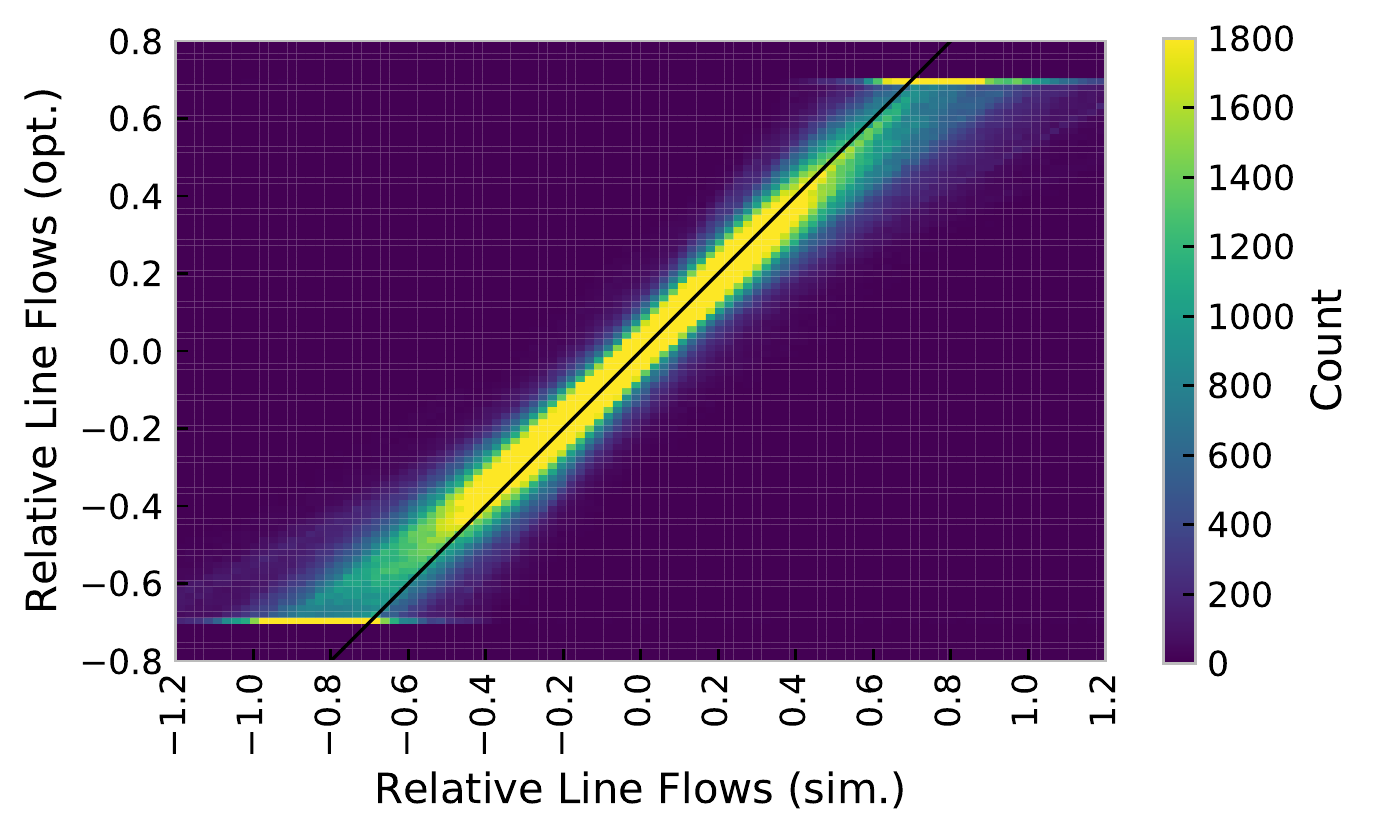} &
		\includegraphics[width=0.49\textwidth]{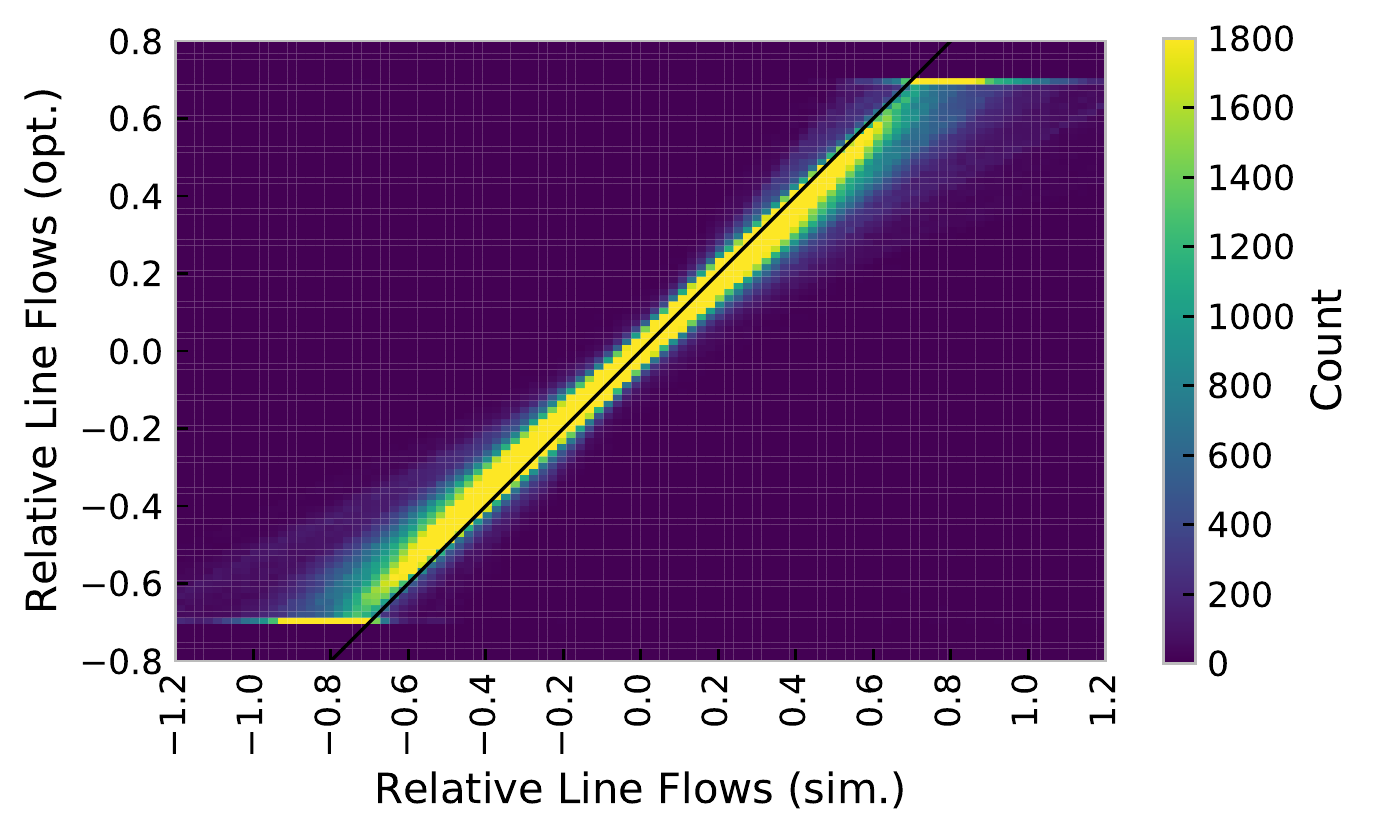} \\
		Lossy & Iterative Lossy \\
		\includegraphics[width=0.49\textwidth]{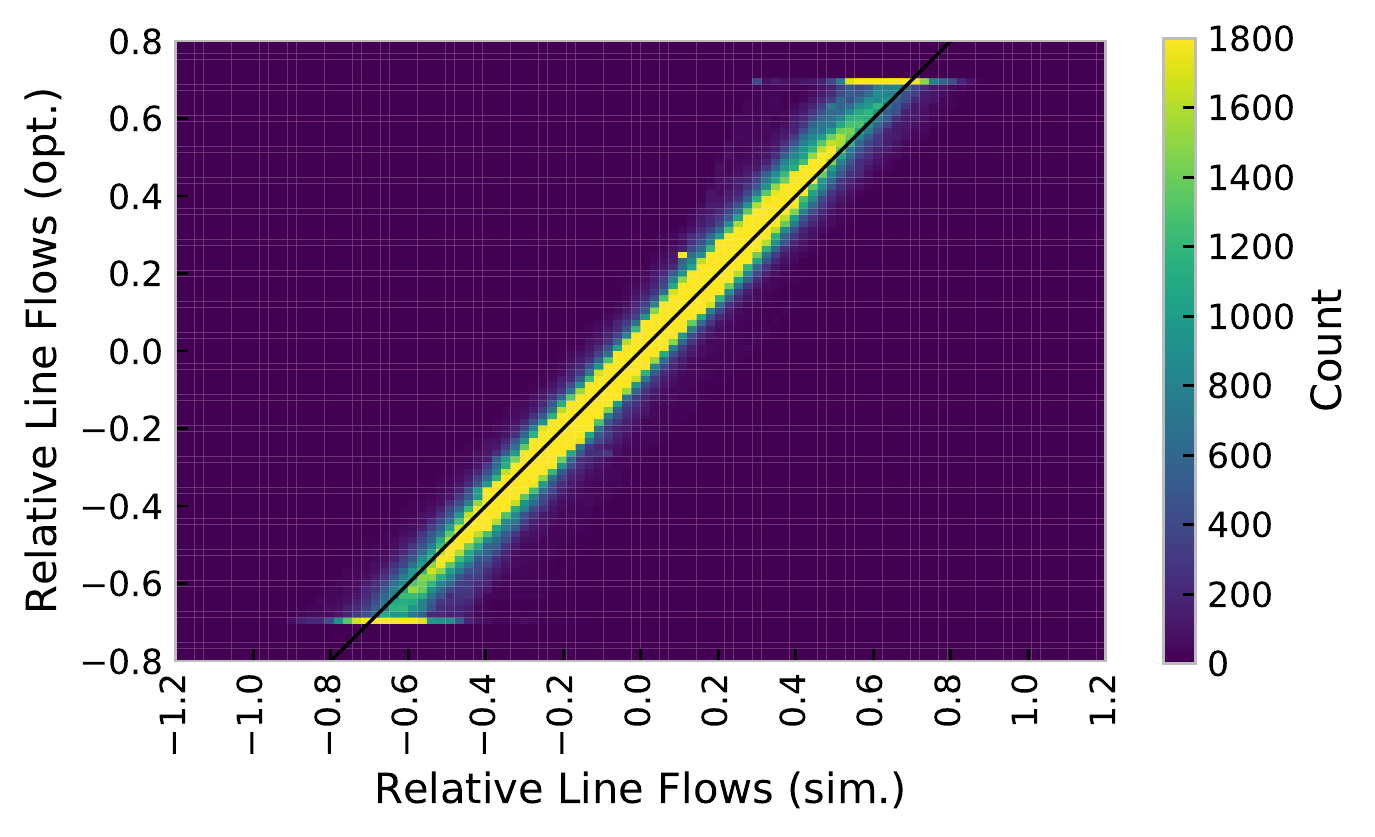} &
		\includegraphics[width=0.49\textwidth]{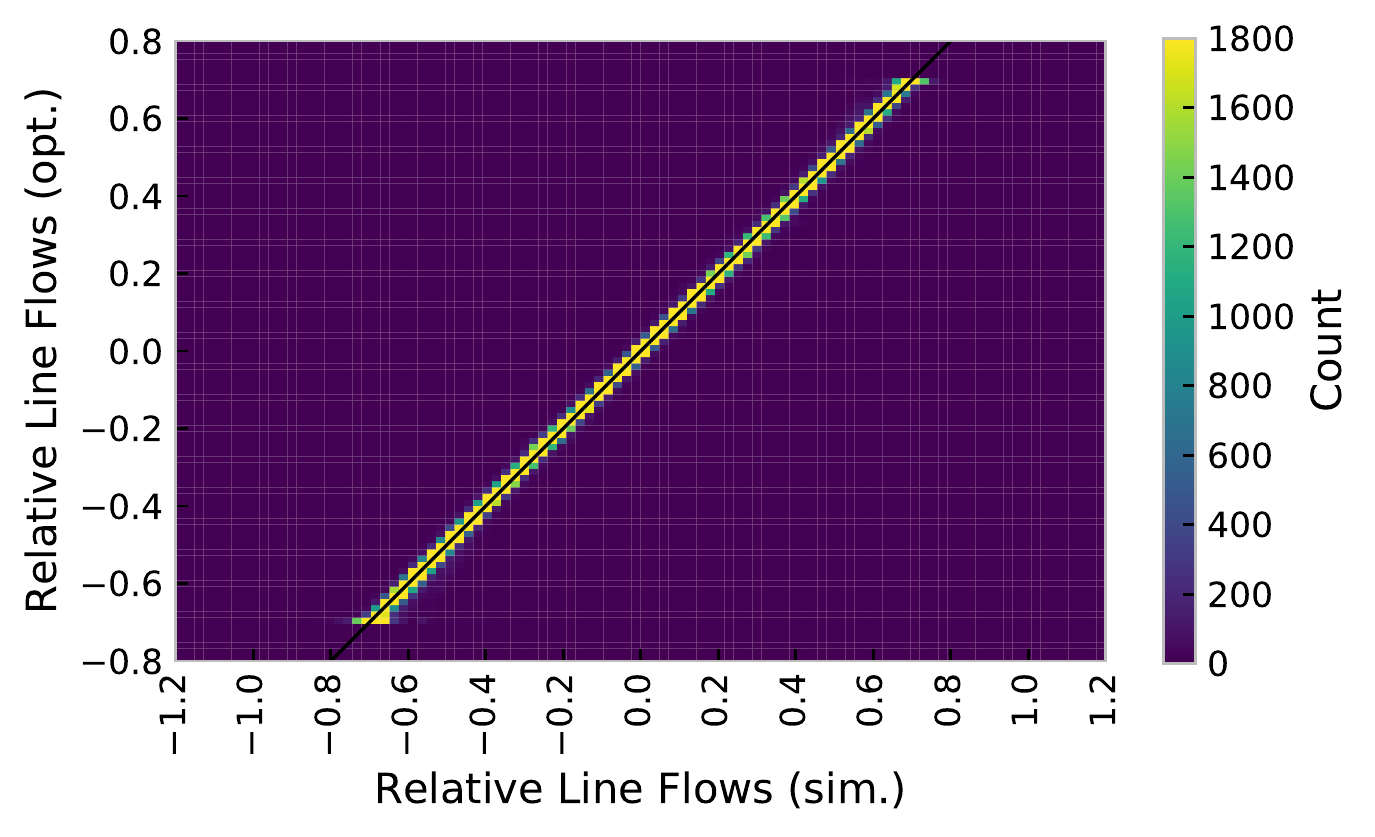}
	\end{tabular}
	\caption{Two-dimensional histograms comparing simulated flows (AC power flow) and
			 optimised flows of the indicated flow models.
			 Relative line flows are shown as $p_{\ell,t} / P_\ell$.
			 The count refers to a tuple $(\ell,t)$ of line and snapshot.
			 The black line indicates perfect alignment of simulated and optimised flows.}
	\label{fig:flow-comparison}
\end{figure}

\begin{figure}
	\centering
	\small
	\begin{tabular}{ccc}
		Transport & Lossy Transport & Lossless \\
		\includegraphics[width=0.32\textwidth]{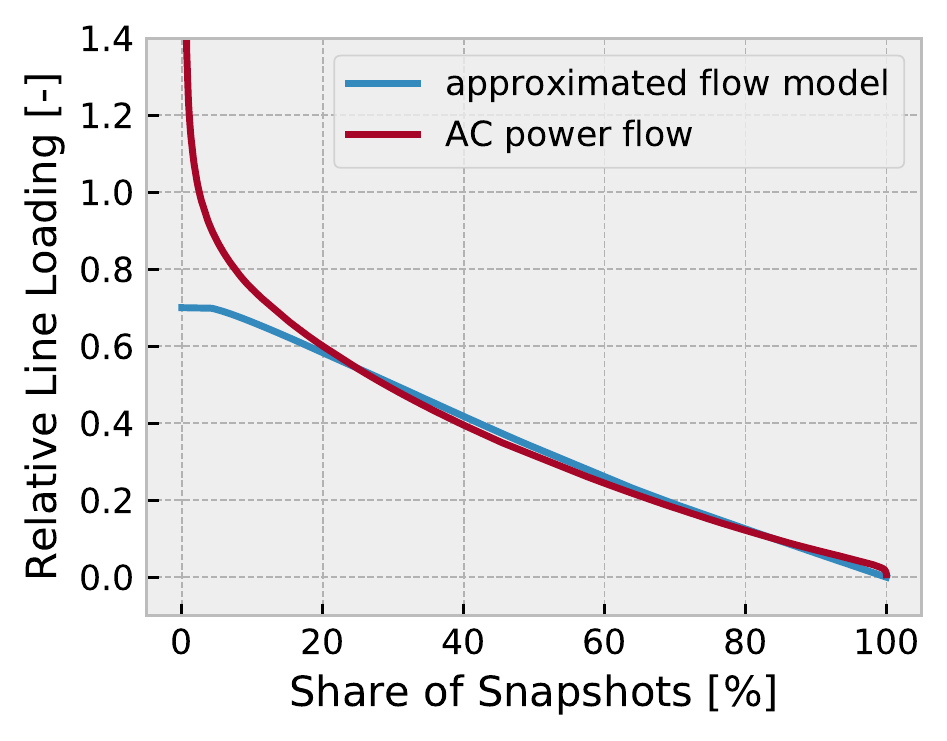} &
		\includegraphics[width=0.32\textwidth]{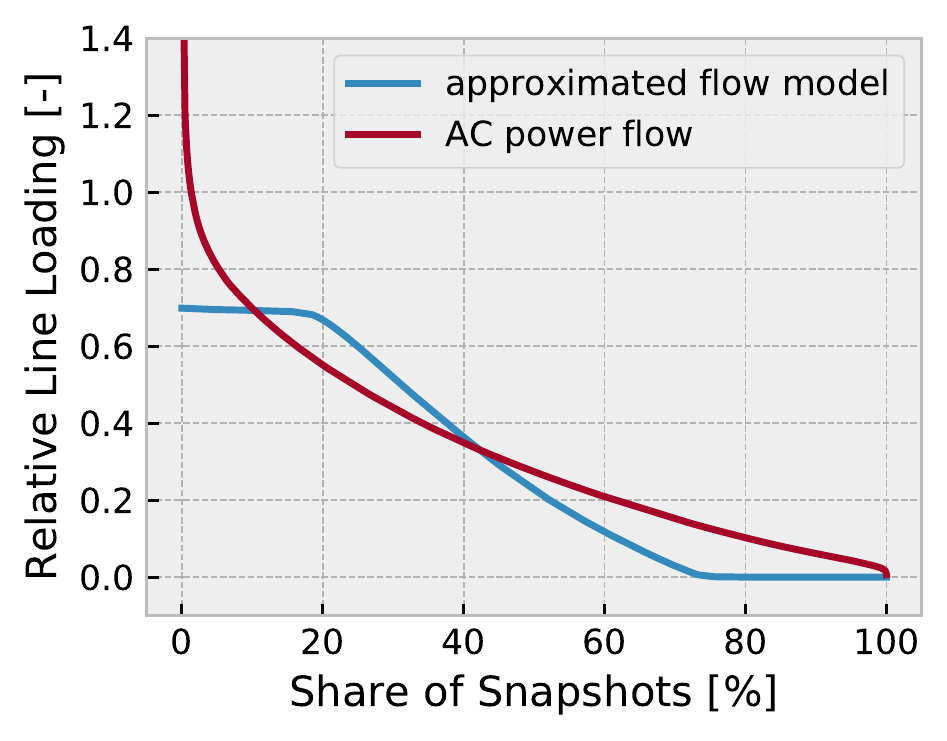} &
		\includegraphics[width=0.32\textwidth]{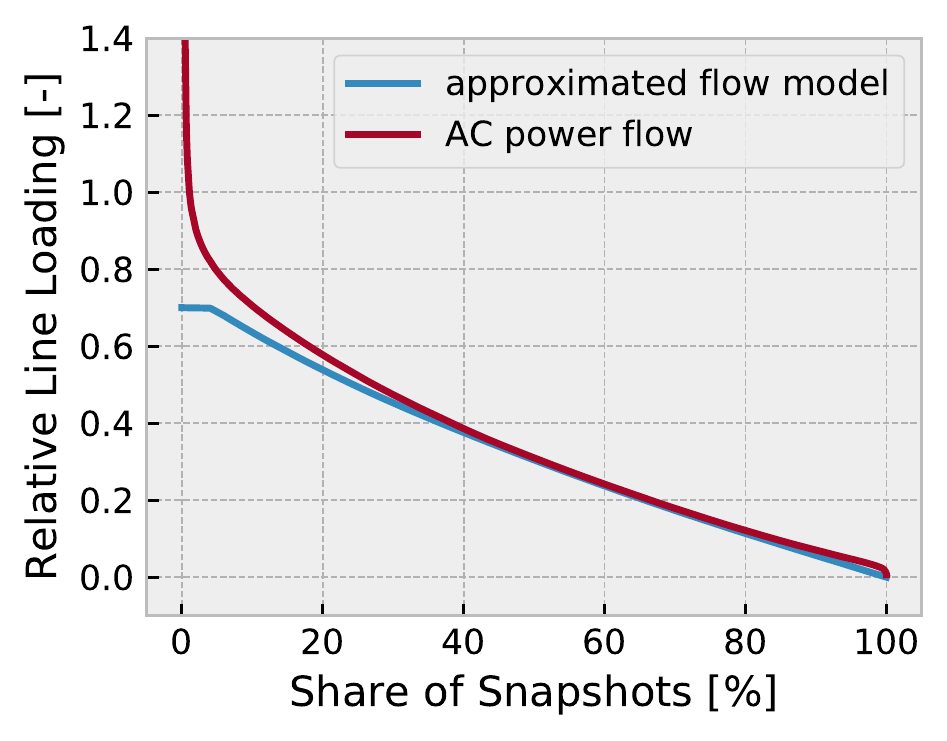} \\
		Iterative Lossless & Lossy & Iterative Lossy \\
		\includegraphics[width=0.32\textwidth]{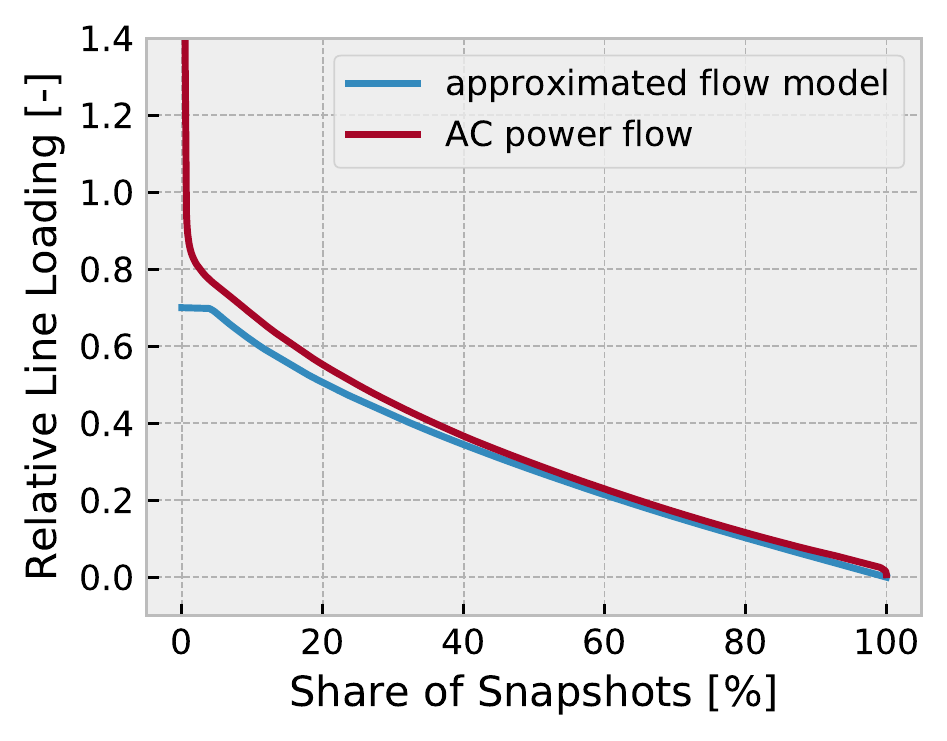} &
		\includegraphics[width=0.32\textwidth]{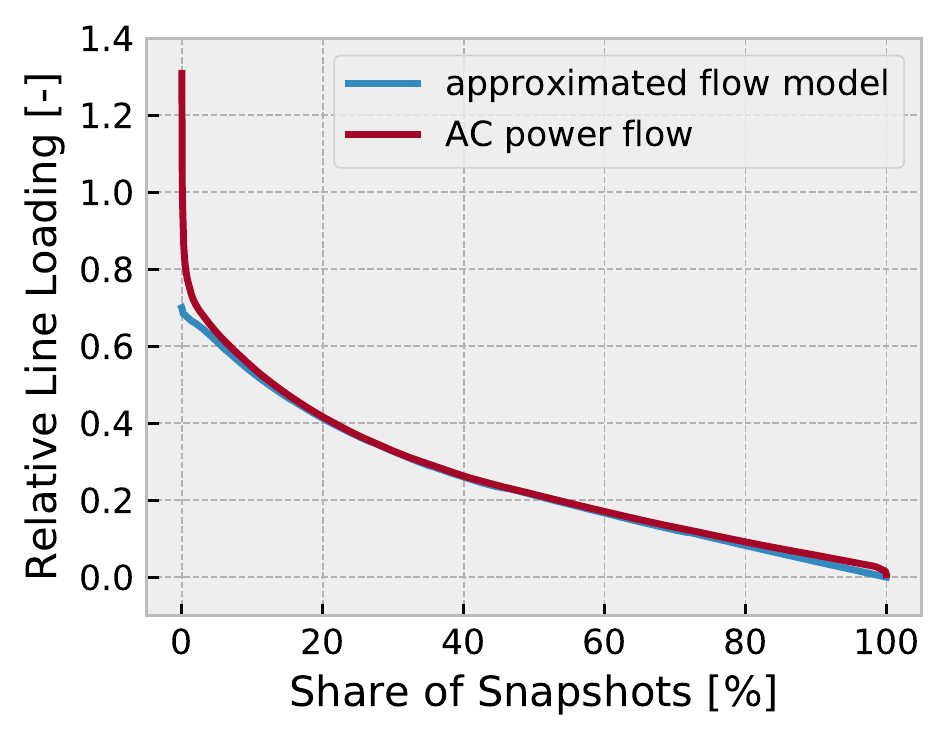} &
		\includegraphics[width=0.32\textwidth]{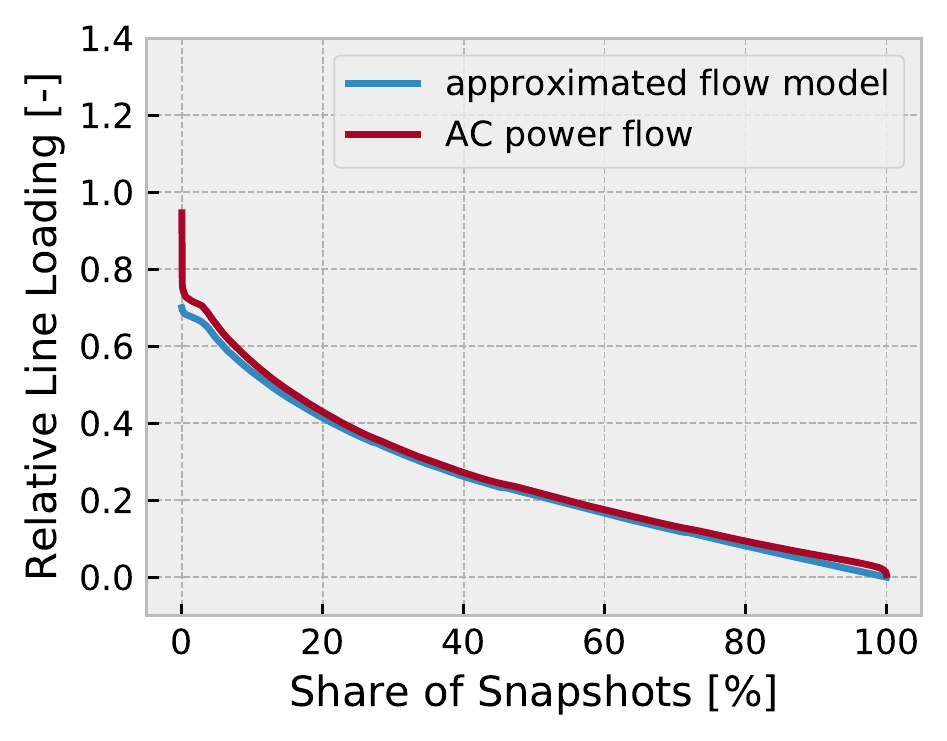}
	\end{tabular}
	\caption{Flow duration curves of simulated flows (AC power flow)
			and optimised flows for the indicated flow models.
			Relative line loading is shown as $p_{\ell,t} / P_\ell$.
			The count refers to a tuple $(\ell,t)$ of line and snapshot.}
			
	\label{fig:flow-duration-curve}
\end{figure}

\begin{figure}
	\centering
	\includegraphics[width=0.7\textwidth]{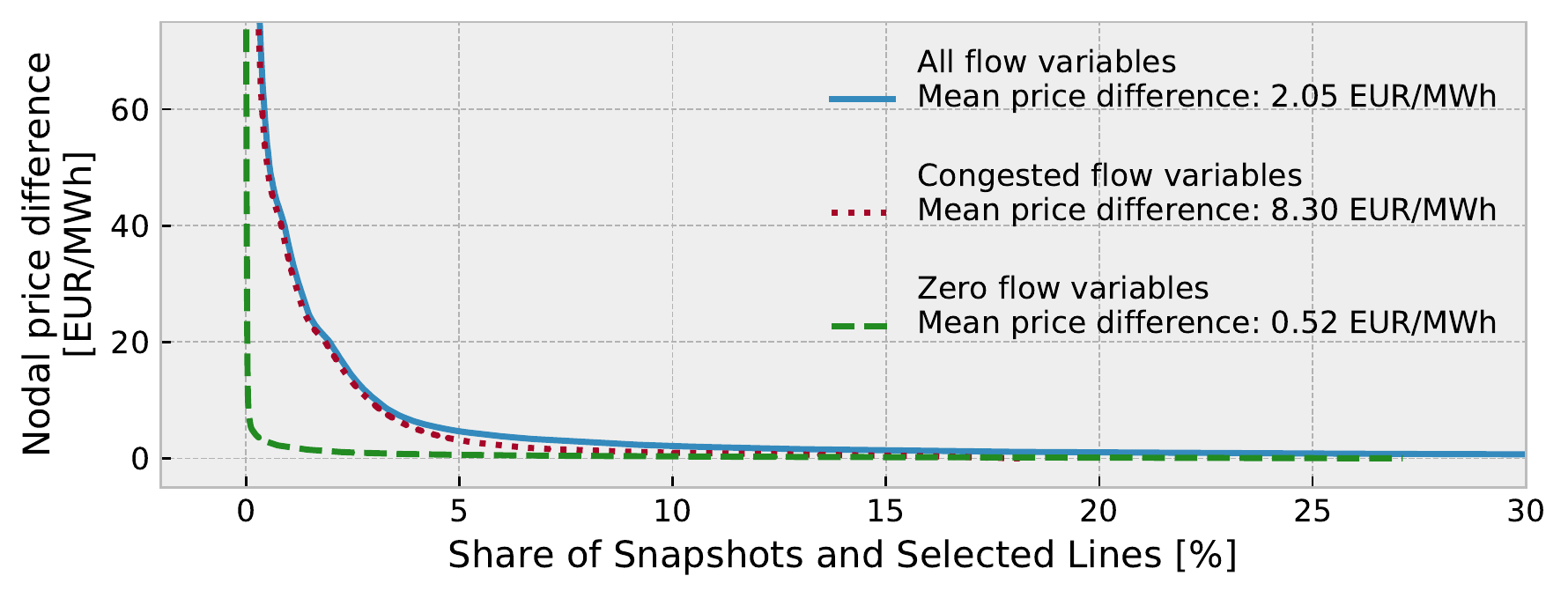}
	\caption{Duration curves (lines and snapshots) of nodal price differences for
			 lines experiencing no flow, congested lines, and all lines.}
	\label{fig:lmpdifference}
\end{figure}

\begin{figure}
	\centering
	\small
	\includegraphics[width=0.55\textwidth]{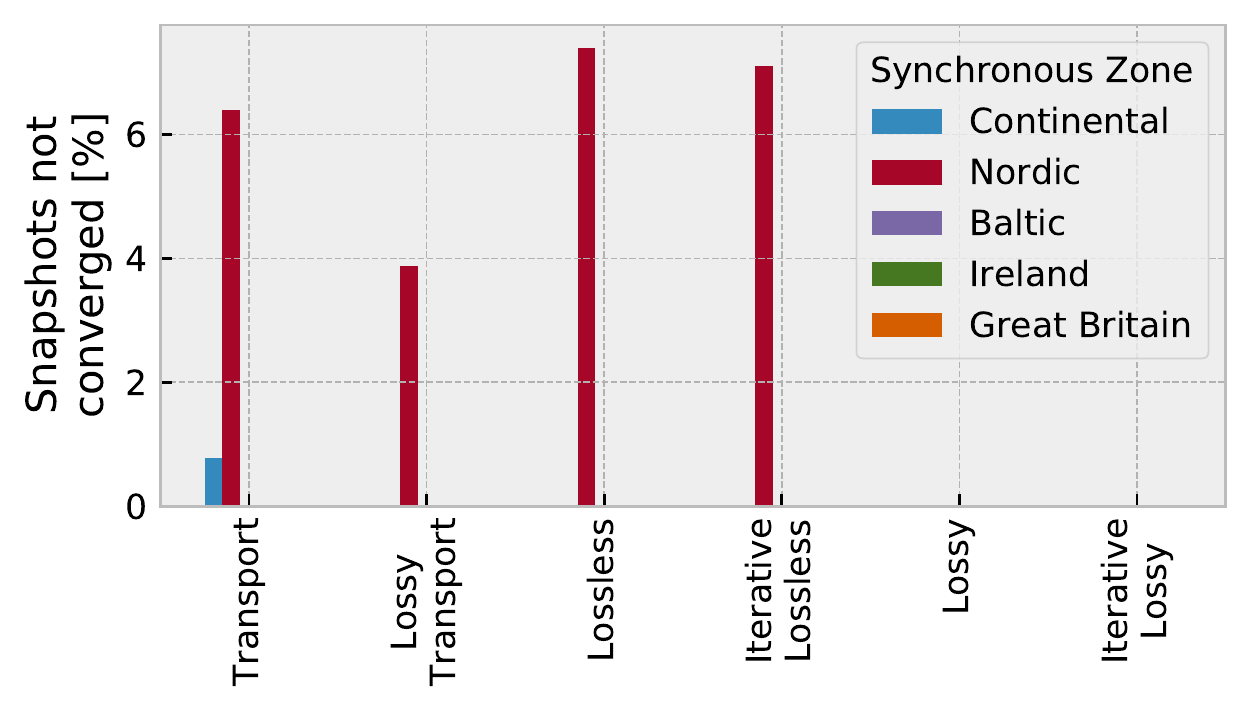}
	\caption{Share of snapshots where the Newton-Raphson algorithm
			 for solving the AC power flow equations did not converge
			 distinguished by colour-coded synchronous zone.}
	\label{fig:pf-convergence}
\end{figure}

\begin{table}
	\centering
	\footnotesize
	\begin{tabular}{ll|rrrrrr}
		\toprule
		 &  &  & Lossy &  & Iterative & & Iterative \\
		 Indicator & Unit & Transport & Transport & Lossless & Lossless & Lossy & Lossy \\ \midrule
		 Root Mean Squared (RMSE) & MW & 1468 & 1059 & 790 & 679 & 298 & 60 \\
		 Mean Absolute (MAE) & MW & 775 & 707 & 269 & 207 & 194 & 35 \\
		 Pearson Correlation (R) & -- & 0.91 & 0.94 & 0.97 & 0.98 & 0.99 & 0.998 \\
		 Coef. of Determination (R$^2$) & -- & 0.83 & 0.89 & 0.94 & 0.95 & 0.98 & 0.996 \\
		\bottomrule
	\end{tabular}
	\caption{Flow errors compared accross covered flow models.}
	\label{tab:flow-errors}
\end{table}

Figure \ref{fig:flow-comparison} compares line flows from optimisation
to simulated line flows from AC power flow for each of the flow models
in a two-dimensional histogram.
Figure \ref{fig:flow-duration-curve} displays the same information
from a different perspective as duration curves of relative line
loading for both simulated and optimised flows
(figure inspired by Brown et al.~\cite{ackermann_optimising_2016}).
Table \ref{tab:flow-errors} quantifies the alignment of optimised and simulated flows
with some standard absolute and relative measures of error
that are frequently encountered in the literature (cf.~\cite{Coffrin2012}):
root mean squared error (RMSE), mean average error (MAE),
Pearson correlation coefficient ($R$), coefficient of determination ($R^2$).

First and foremost, the results reveal that the 
iterative lossy model matches simulated flows almost perfectly.
Other formulations show deficiencies particularly under high loading conditions,
but generally get the direction of flow right.
The errors become significantly less pronounced and produce less undesired line overloading,
the more physical characteristics of power flow are considered during optimisation.
Limiting the utilisation of line capacities to 70\% prevents abundant overloading.
Remarkably, a high Pearson correlation coefficient of 0.91 is already achieved with the transport model,
indicating that despite its simplicity the model can capture the dominant flow patterns
we observe in the ex-post AC power flow simulation.

Lines with zero flow occur strikingly frequently in the lossy transport model,
causing high deviations from the simulated flows.
This can be explained with the aid of Figure \ref{fig:lmpdifference}.
There are many cases where prices are (almost) the same at two neighbouring buses.
In such cases, there is no strict economic need to move power between them. 
With a lossless transport model there is no penalty for moving power
between the two nodes, such that the optimisation yields a random value.
However, for the lossy transport there is an incentive to set
the flow to zero to avoid the losses, which is why exactly this phenomenon
frequently occurs when there is no price difference.
The physical flow constraints enforced by KVL make it complicated
to realise zero flow on a line. This is the reason why
we do not observe many lines with zero flow for models that enforce KVL.
Conversely, Figure \ref{fig:lmpdifference} also shows that congested lines
cause high nodal price differences.

In some cases the Newton-Raphson algorithm does not converge.
Typical causes can be high voltage angle differences, voltage drops, and reactive power flows.
The power flow simulation is run separately for each snapshot and each synchronous zone,
so we can check individually what prevalent network characteristics,
in combination with the underlying flow models, cause the failure to converge.
The resulting share of snapshots not converged for each synchronous zone
is presented in Figure \ref{fig:pf-convergence}.
Almost exclusively, difficulties are observed in the Nordic synchronous
zone which possesses many long (aggregated) lines,
which lead to high voltage angle differences.
With regard to the whole European system, the number of snapshots where no
convergence is reached is low.
We observe better convergence rates for more detailed flow models
and the issue is found to become less problematic as the
spatial resolution of the transmission network is increased.

Given that high voltage angle differences diminish
the accuracy of the linear power flow approximation,
a maximum of up to $\pm 30^\circ$ is commonly tolerated in the literature
\cite{doerfler2013,dvijotham2016,purchala2005}.
This domain links to the range beyond which the relative error of the
small-angle approximation of the sine exceeds 5\%.
Since the cosine approximation is a second order Taylor series expansion,
unlike the first order sine approximation,
it does not reduce the acceptable range of angle differences further
(cf.~Table \ref{tab:relerror}).
We observe that across all flow models a majority of voltage angle differences
lies within an uncritically low range
where the sine approximation is quite precise (cf.~Figure \ref{fig:vangdiff}).
The share of voltage angle differences outside $\pm 30^\circ$ reduces
considerably with more physically accurate grid modelling
(5\% for transport model versus 1.5\% for lossy model, cf.~Table \ref{tab:misc}).

\subsection{Computational Performance}
\label{sec:res-benchmark}

\begin{figure}
	\centering
	\footnotesize
	\begin{tabular}{cc}
		Peak Memory Consumption & Solving Time \\
		\includegraphics[width=0.42\textwidth]{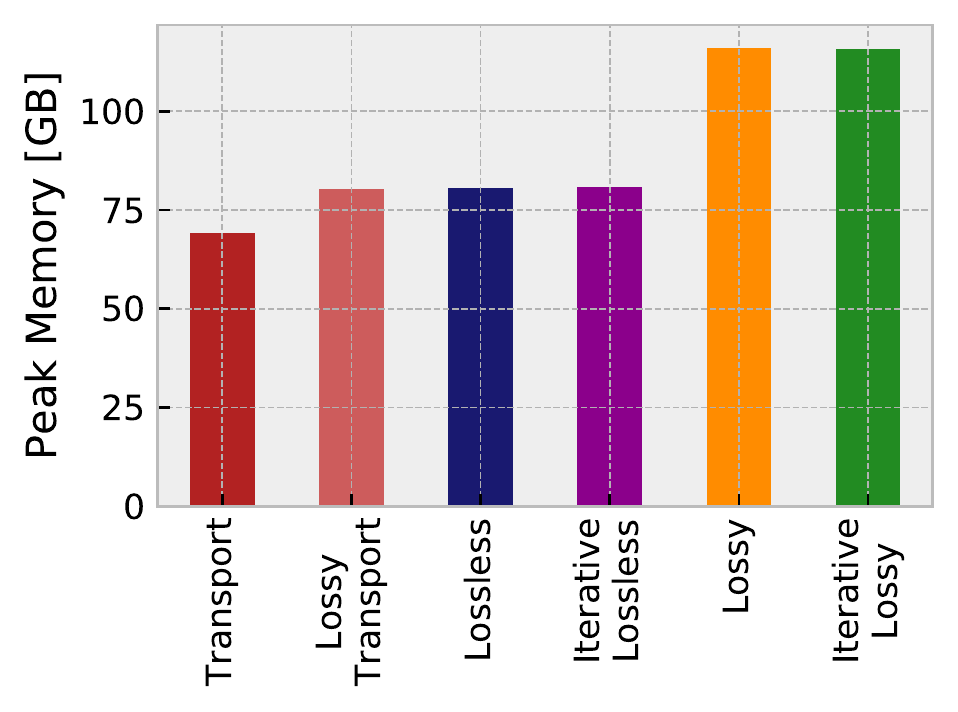} &
		\includegraphics[width=0.42\textwidth]{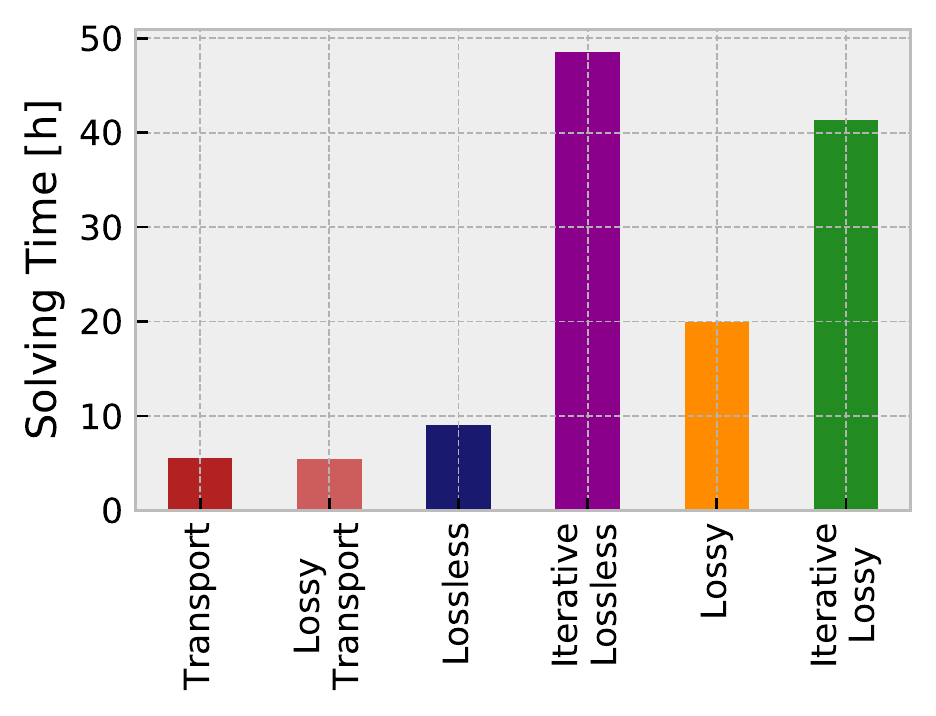}
	\end{tabular}
	\caption{Comparison of computational performance in terms of
			 peak memory consumption and solving time.}
	\label{fig:performance}
\end{figure}

The computational performance of the different flow models,
both in terms of memory and computation time, 
is shown in Figure \ref{fig:performance}.
More variables and constraints leads to higher peak memory consumption.
The spectrum ranges from 70 GB to 130 GB (around factor 2). Particularly the
loss approximation raises memory requirements significantly in relation
to added KVL constraints or constant efficiencies, also depending on the
number of tangents used for the convex envelope.
Solving times range between 5 hours and 50 hours (factor 10).
Lossy and lossless transport model are solved the fastest by far.
The lossless linearised power flow model requires almost twice the time.
Iteration has the biggest impact on solving times,
multiplying with the number of iterations.
Finally, we notice that the lossy formulations are more prone
to numerical issues, which could be circumvented by increasing
the numeric accuracy parameter of the solver at the cost of computational speed.

\subsection{Critical Appraisal}
\label{sec:critical}

The disregard of voltages and reactive power flows during optimisation
ranks among the severest shortcomings of the presented flow models.
The cost and required capacities for reactive power control are not assessed.
The confinement to linear formulations may also be considered as a weakness
in view of recent developments in convex second-order cone solvers.

Additionally, we consider the high-voltage transmission network only
and do not assess the performance of flow models in low-voltage distribution grids.
This is especially relevant in view of further closing the granularity gap.
Furthermore, losses on the distribution level are not directly modelled but taken into account
only through the electricity demand. Typically, the scale of losses is higher than at the transmission level,
as for instance the German Federal Network Agency reports \cite{bnetza2019}. In 2019,
losses at the transmission level amounted to 1.2\%, whereas losses at the distribution
level were as high as 3\%.
Moreover, the relations between ambient temperature, dynamic line rating and losses are not addressed.
Higher ambient temperatures reduce the amount of power a transmission line
can transmit safely but simultaneously increase the resistance, affecting the losses.

Although the clustered transmission system is of course also
simplified due to computational constraints,
we could observe consistent results for spatial aggregation to 100, 200 and 250 nodes.
However, the extent of network clustering also affects the length of modelled lines and we note that
for very long lines with voltage angle differences beyond $\pm 30^\circ$ in highly aggregated grid models the
standard equivalent $\Pi$ model may not be suitable \cite{kundur1994}.

\section{Conclusion}
\label{sec:conclusion}

In the present contribution we discuss best practices for incorporating
two inherently nonlinear phenomena, power flow and transmission losses,
into linear capacity expansion problems that co-optimise investments
in generation, storage and transmission infrastructure.

High model fidelity comes at the cost of high computational burden.
Given the cross-disciplinary nature of energy system modelling and
differences in complexity, the selection of a suitable representation
of power flows depends on the application,
the user's availability of computational resources,
and the level of spatial detail considered.
A highly aggregated network will not benefit from detailed power flow modelling,
whereas modelling losses is critical in the presence of continent-spanning power
transmission at sub-national detail.
The present paper provides a detailed comparison to facilitate this choice.

We find that already as little as three tangents are sufficient to
accurately approximate the quadratic losses,
which in turn are an approximation of the trigonometric losses.
We do not observe
excessive fictitious losses despite the broad convex relaxation.
However, we conclude that accounting for changing impedances as lines are expanded is essential.
Otherwise, losses will be overestimated.

The literature predominantly employs the lossless transport model in design studies, which
can already capture the main features of a cost-efficient system, but
is too inexact for subsequent nonlinear power flow calculations.
However, a representation of power flows that considers transmission losses
as well as both Kirchhoff laws allows us to bridge between techno-economic models
and more detailed electrotechnical models.

\section*{Acknowledgement}

F.N. and T.B. gratefully acknowledge funding from the Helmholtz
Association under grant no. VH-NG-1352.
The responsibility for the contents lies with the authors.
\doclicenseLongText
\doclicenseIcon

\section*{CRediT Author Statement}

\textbf{Fabian Neumann:} Conceptualization, Methodology, Investigation, Software, Validation, Formal analysis, Visualization, Writing -- Original Draft, Writing -- Review \& Editing
\textbf{Veit Hagenmeyer:} Writing -- Review \& Editing, Project administration, Funding acquisition
\textbf{Tom Brown:} Conceptualization, Writing -- Review \& Editing, Supervision, Project administration, Funding acquisition

\section*{Data Availability}

A dataset of the results including networks and graphics is available abstract \href{https://doi.org/10.5281/zenodo.3968297}{doi.org/10.5281/zenodo.3968297}. 
We also refer to the documentation of PyPSA (\href{https://pypsa.readthedocs.io}{pypsa.readthedocs.io}),
PyPSA-Eur (\href{https://pypsa-eur.readthedocs.io}{pypsa-eur.readthedocs.io}) and the source code 
repository
(\href{https://github.com/fneum/power-flow-models}{github.com/fneum/power-flow-models}).

\addcontentsline{toc}{section}{References}
\bibliography{library}

\begin{appendix}
	\section{AC Power Flow Problem Solved with Newton-Raphson}
\label{sec:app-powerflow}

\subsection{Without Distributed Slack}

Given nodal power imbalances $S_n$ at any given snapshot for each bus $n$ the AC power flow equations
are given by
\begin{equation}
	S_n = P_n + i Q_n = V_n I_n^\ast = V_n \left(\sum_m Y_{nm} V_m\right)^\ast,
\end{equation}
where $V_n=\abs{V_n}e^{i\theta_n}$ is the complex voltage, whose rotating angle is taken relative to the slack bus and
$Y_{nm}$ is the bus admittance matrix, based on the branch impedances and shunt admittances (including those attached to buses).

For the slack bus $n=0$ it is assumed $\abs{V_0}$ is given and that $\theta_0=0$; $P$ and $Q$ are to be found.
For the PV buses, $P$ and $\abs{V}$ are given; $Q$ and $\theta$ are to be found.
For the PQ buses, $P$ and $Q$ are given; $\abs{V}$ and $\theta$ are to be found.

Considering PV and PQ as sets of buses, then there are $\abs{\text{PV}}+2\abs{\text{PQ}}$ real-valued equations to solve:
\begin{align}
	\text{Re} \left[ V_n \left(\sum_m Y_{nm} V_m\right)^\ast\right] - P_n &= 0 \quad\forall\:\text{PV} \cup \text{PQ} \\
	\text{Im} \left[ V_n \left(\sum_m Y_{nm} V_m\right)^\ast\right] - Q_n &= 0 \quad\forall\:\text{PQ}
\end{align}
We need to find $\theta_n$ for all PV and PQ buses and $\abs{V_n}$ for all PQ buses.

These equations $f(x)=0$ are solved using the Newton-Raphson method, with the Jacobian
\begin{equation}
	\frac{\partial f}{\partial x} = \left(\begin{matrix}
		\frac{\partial P}{\partial \theta} &
		\frac{\partial P}{\partial \abs{V}} \\
		\frac{\partial Q}{\partial \theta} &
		\frac{\partial Q}{\partial \abs{V}}
	\end{matrix}\right)
\end{equation}
and the initial guesses $\theta_n=0$ and $\abs{V_n}=1$ for unknown quantities.
For more details see for example Grainer and Stevenson~\cite{grainer1994}.
The total active slack power, which balances remaining mismatches of
power generation and demand resulting from the AC power flow equations,
is fully allocated to the slack bus.
This can be a crude assumption,
particularly for large networks with a high
penetration of renewables.

\subsection{With Distributed Slack}

A better alternative is to distribute the total active slack power across
all generators in proportion to their capacities (or another distribution scheme)
\cite{Milano2008}.
The active power flow equations are altered to
\begin{equation}
	\text{Re} \left[ V_n \left(\sum_m Y_{nm} V_m\right)^\ast\right]
	- P_n - P_{\text{slack}} \gamma_n = 0 \quad\forall\:\text{PV} \cup \text{PQ} \cup \text{slack}
\end{equation}
where $P_{\text{slack}}$ is the total slack power and $\gamma_n$ is the share of bus $n$ of the total
generation capacity, which is used as distribution key.
We add an additional active power balance equation for the slack bus
since it is now part of the distribution scheme.

The distributed slack approach extends the Jacobian by an additional row
for the derivatives of the slack bus active power balance and by an
additional column for the partial derivatives with respect to $\gamma$
\begin{equation}
	\frac{\partial f}{\partial x} = \left(\begin{matrix}
		\frac{\partial P_0}{\partial \theta} &
		\frac{\partial P_0}{\partial \abs{V}} &
		\frac{\partial P_0}{\partial \gamma} \\
		\frac{\partial P}{\partial \theta} &
		\frac{\partial P}{\partial \abs{V}} &
		\frac{\partial P}{\partial \gamma} \\
		\frac{\partial Q}{\partial \theta} &
		\frac{\partial Q}{\partial \abs{V}} &
		\frac{\partial Q}{\partial \gamma}
	\end{matrix}\right).
\end{equation}
If $\gamma_n=0$ for all buses but the slack bus, this is equivalent to a single slack bus model. 

\section{Relations between Electrical Line Parameters}
\label{sec:app-impedance}

Following e.g.~\cite{grainer1994}, the complex per-unit impedance $z_\ell = r_\ell + i x_\ell$ is composed of ohmic resistance $r_\ell$ and reactance $x_\ell$
Likewise, the admittance $y_\ell = g_\ell + i b_\ell$ is composed of conductance $g_\ell$ and susceptance $b_\ell$
Impedance and admittance are reciprocals ($y_\ell=z_\ell^{-1}$), hence we obtain the relations
\begin{align}
	g_\ell + i b_\ell & = \frac{1}{r_\ell+i x_\ell},                                  \\
	g_\ell + i b_\ell & = \frac{r_\ell-i x_\ell}{(r_\ell+i x_\ell)(r_\ell-i x_\ell)}, \\
	g_\ell + i b_\ell & = \frac{r_\ell-i x_\ell}{r_\ell^2+x_\ell^2}.
\end{align}
By splitting real and imaginary parts we can express conductance and susceptance in terms of impedance and reactance:
\begin{align}
	g_\ell & = \text{Re}\left[\frac{r_\ell-i x_\ell}{r_\ell^2+x_\ell^2}\right] = \frac{r_\ell}{r_\ell^2+x_\ell^2}, \\
	b_\ell & = \text{Im}\left[\frac{r_\ell-i x_\ell}{r_\ell^2+x_\ell^2}\right] = \frac{x_\ell}{r_\ell^2+x_\ell^2}.
\end{align}

\section{Additional Figures and Tables}
\label{sec:addfig}

\begin{figure}
	\centering
	\small
	\begin{tabular}{ccc}
		Transport & Lossy Transport & Lossless\\
		\includegraphics[width=0.32\textwidth]{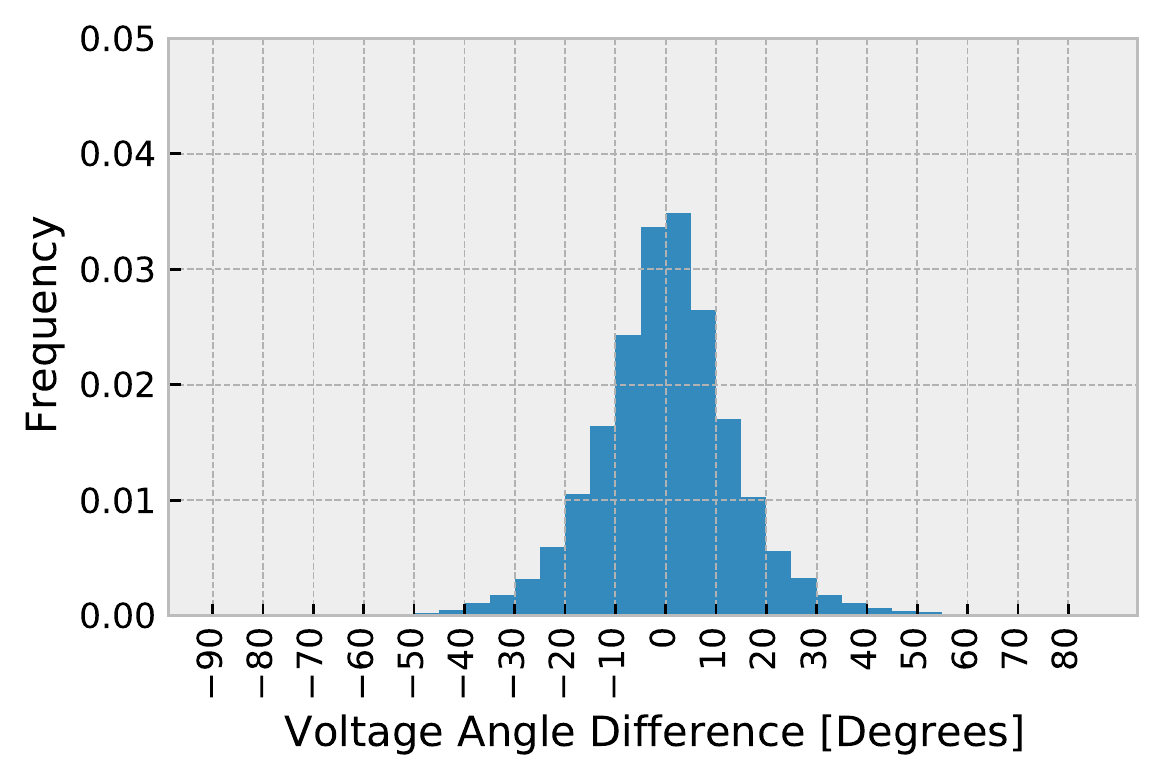} &
		\includegraphics[width=0.32\textwidth]{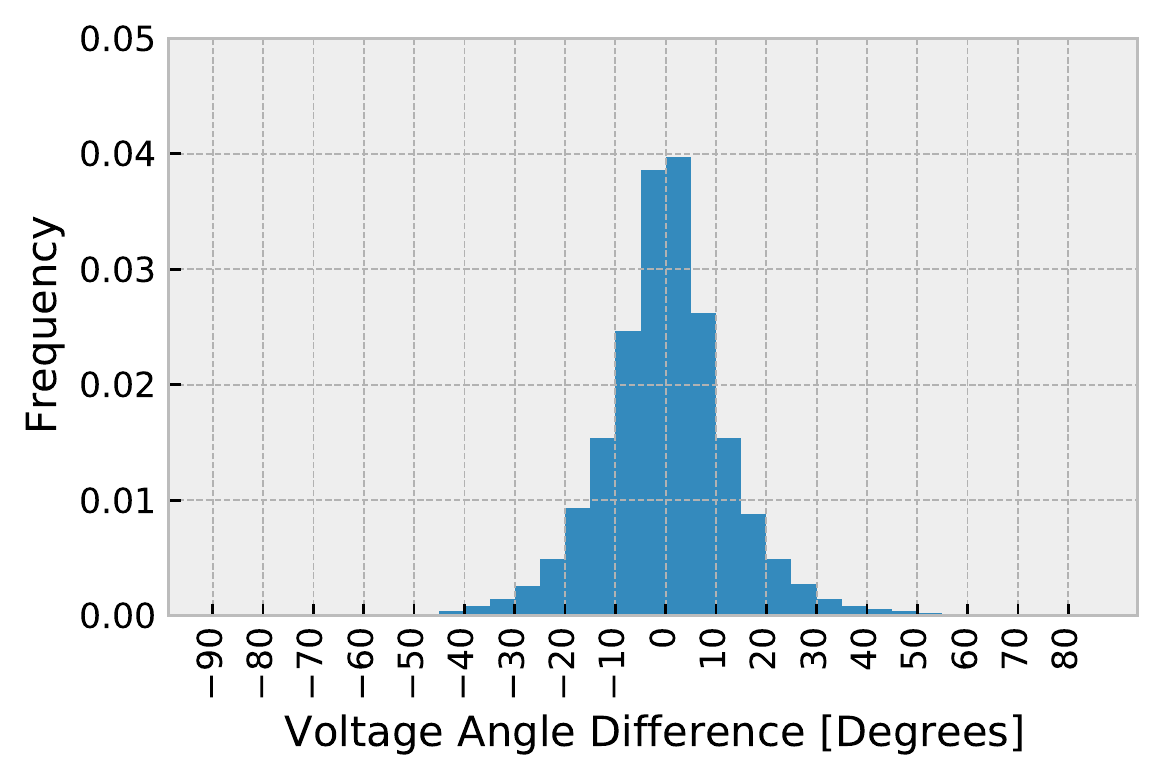} &
		\includegraphics[width=0.32\textwidth]{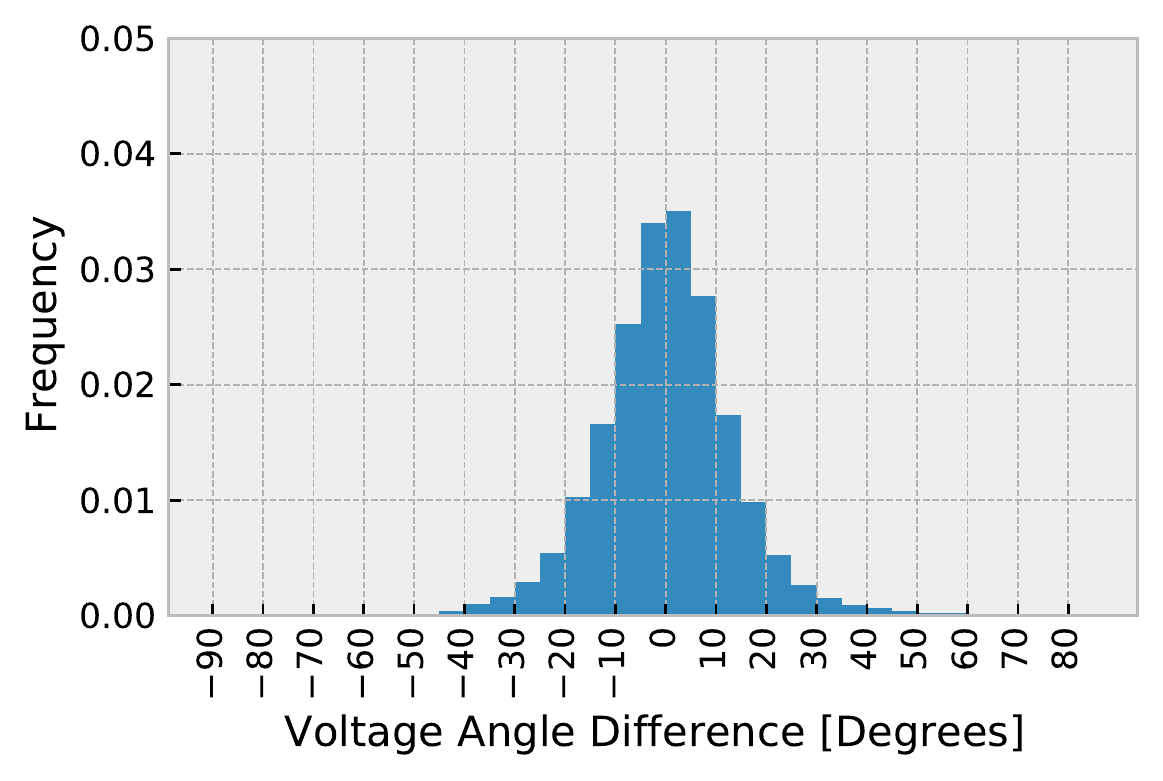} \\
		Iterative Lossless & Lossy & Iterative Lossy \\
		\includegraphics[width=0.32\textwidth]{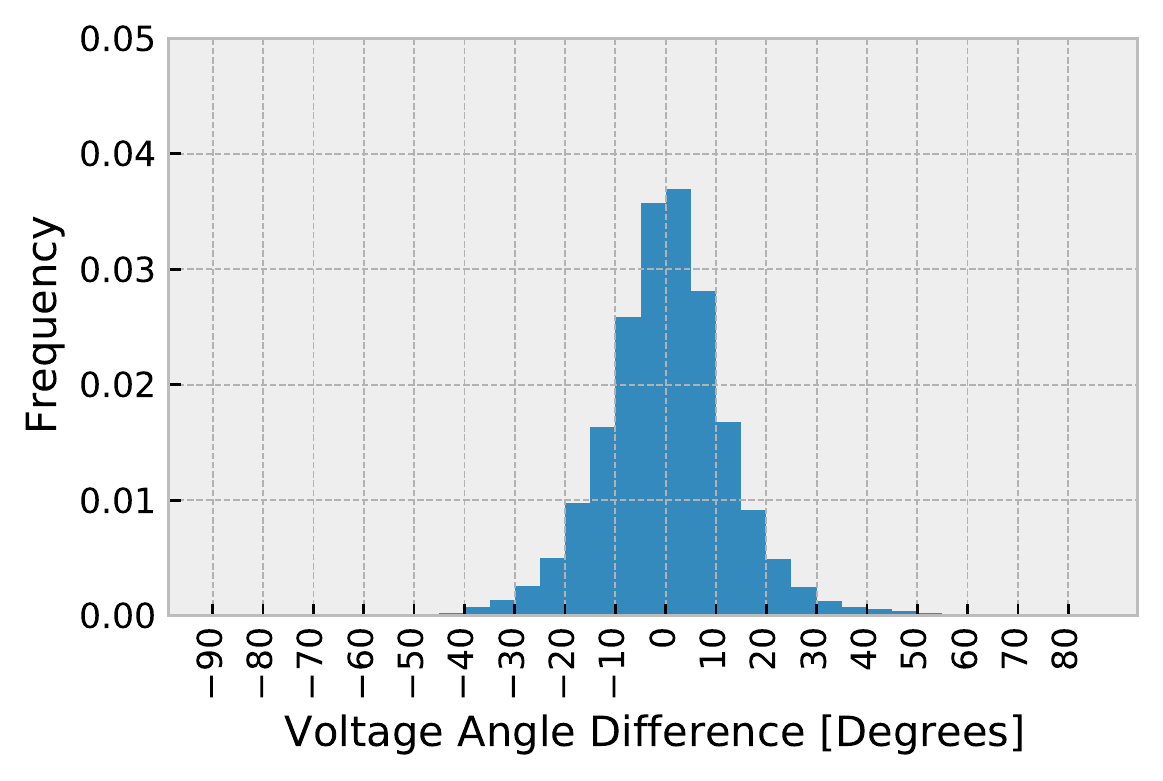} &
		\includegraphics[width=0.32\textwidth]{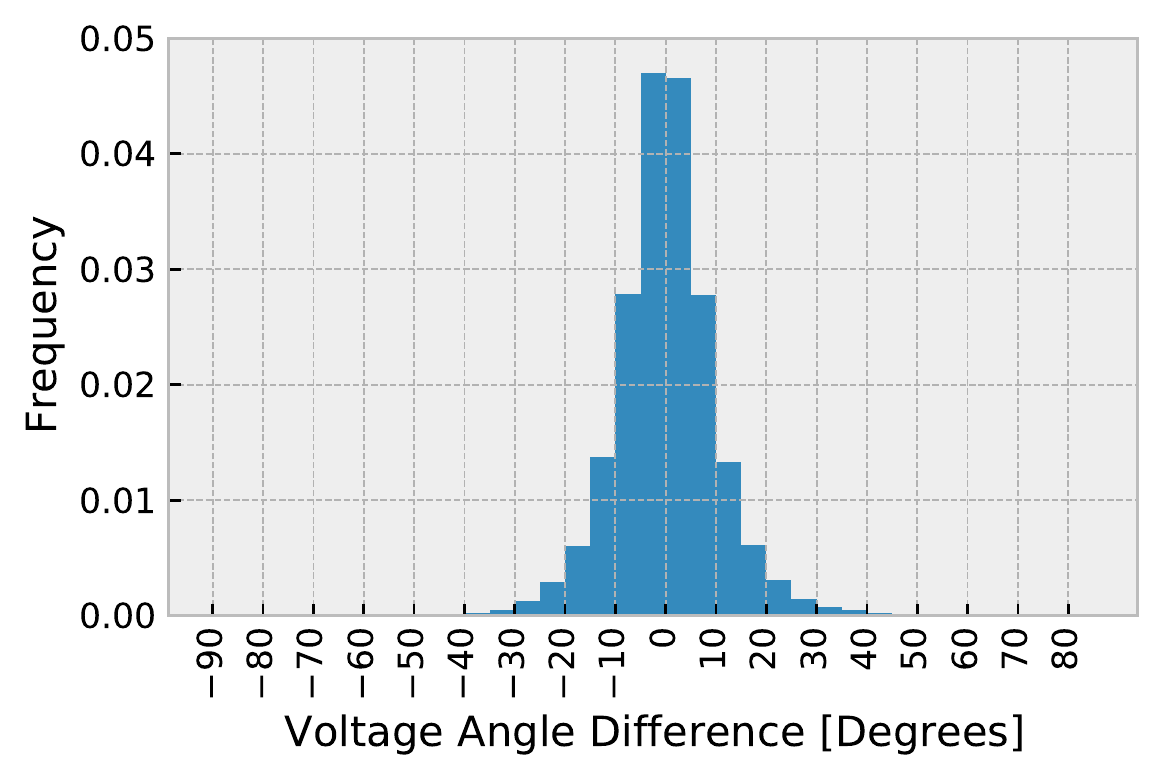} &
		\includegraphics[width=0.32\textwidth]{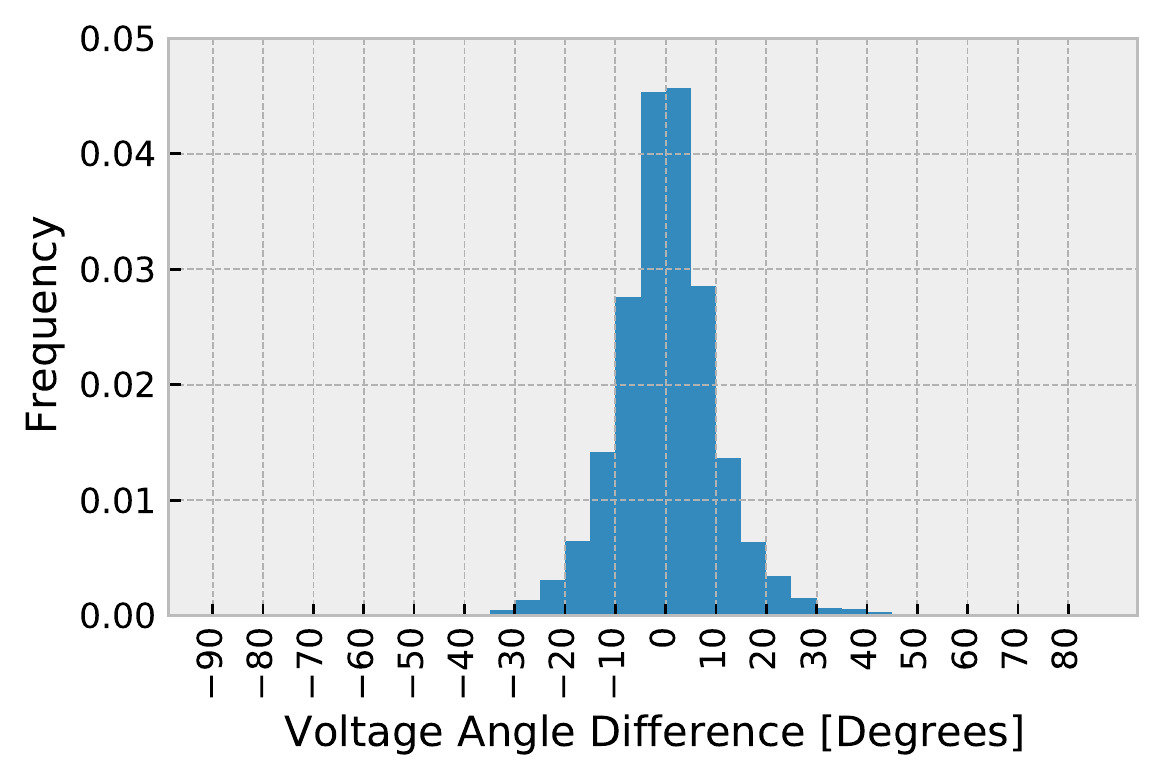}
	\end{tabular}
	\caption{Distribution of voltage angle differences for the indicated flow models.}
	\label{fig:vangdiff}
\end{figure}

\begin{figure}
	\centering
	\small
	\includegraphics[width=0.5\textwidth]{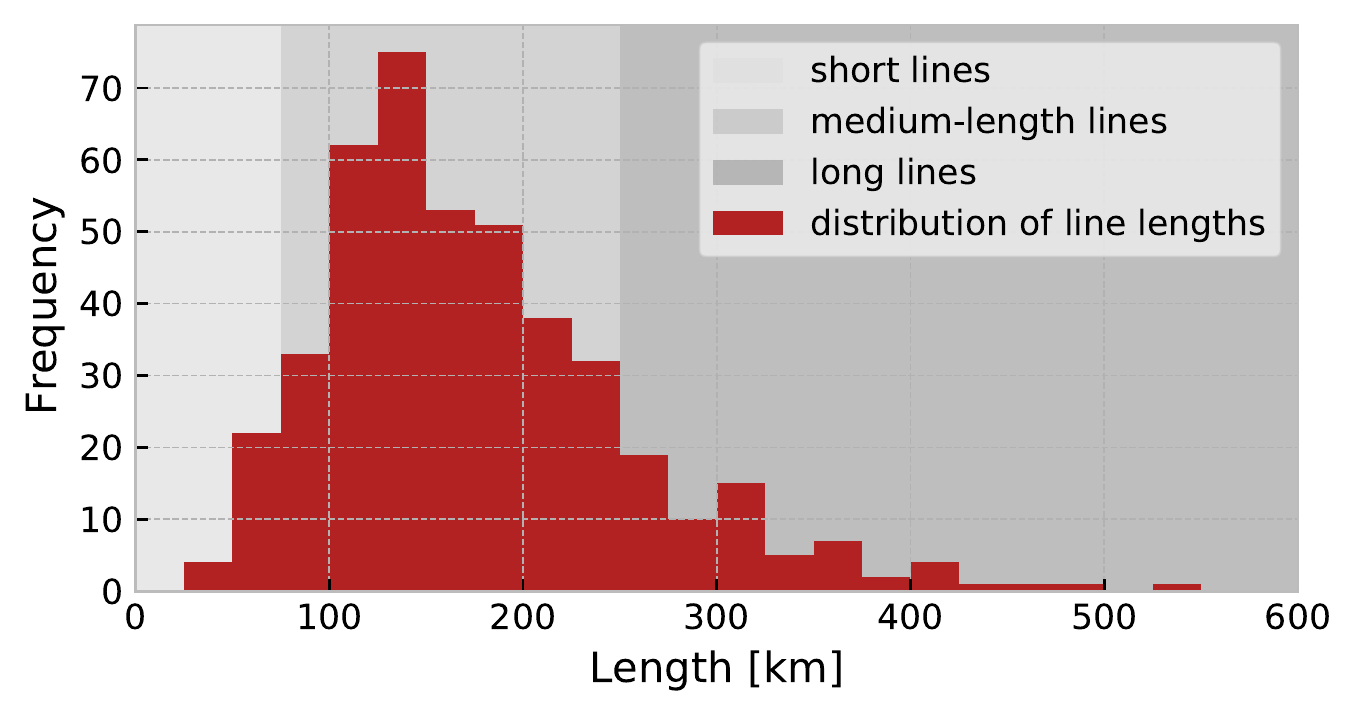}
	\caption{Distribution of line lengths in 250 node networks
	by line classification according to Kundur \cite{kundur1994}.
	Of all lines, 15\% fall into the category of long lines, while 6\% are classified as short lines.}
	\label{fig:linelengths}
\end{figure}

Labels for used mathematical symbols are included in Table \ref{tab:nomenclature}.
An overview of applied assumptions in selected equations of the present paper is given in Table \ref{tab:assumptions}.
Histograms of voltage angle differences for the evaluated flow models are plotted in Figure \ref{fig:vangdiff}.
Relative errors of first and second order small-angle approximations of trigonometric functions are listed in Table \ref{tab:relerror}.
Information on the distribution of line lengths in the considered network are displayed in Figure \ref{fig:linelengths}.
Optimised capacities and energy generation by carrier are summarised in Table \ref{tab:capacity} and Table \ref{tab:energy}.

\begin{table}
	\centering
	\begin{small}
		\begin{tabular}{@{}ll@{}}
			\toprule
			\textbf{Symbol}         & \textbf{Description}                    \\ \midrule
			$\cN$                   & set of buses                            \\
			$\cL$                   & set of lines                            \\
			$\mathcal{C}$           & set of cycles in cycle basis            \\
			$g_\ell$                & conductance                             \\
			$b_\ell$                & susceptance                             \\
			$r_\ell$                & resistance                              \\
			$x_\ell$                & reactance                               \\
			$z_\ell$                & impedance                               \\
			$y_\ell$                & admittance                              \\
			$\abs{V_i}$             & voltage magnitude                       \\
			$\theta_i$              & voltage angle                           \\
			$p_{\ell(i,j)}=p_\ell$  & active power flow                       \\
			$q_{\ell(i,j)}= q_\ell$ & reactive power flow                     \\
			$\psi_\ell$            & power loss                              \\
			$K_{i\ell}$             & incidence matrix                        \\
			$C_{\ell c}$            & cycle incidence matrix                  \\
			$p_i$                   & nodal power injection                   \\
			$P_\ell$                & line capacity                           \\
			$\overline{P}_\ell$     & maximum line capacity                   \\
			$\overline{p}_\ell$     & maximum per-unit flow                   \\
			$m_k$                   & slope of loss tangent for interval $k$  \\
			$a_k$                   & offset of loss tangent for interval $k$ \\
			$n$                     & number of intervals for loss tangents   \\
			\bottomrule
		\end{tabular}
	\end{small}
	\label{tab:nomenclature}
	\caption{Nomenclature}
\end{table}

\begin{table}
\begin{small}
	\begin{tabular}{@{}lrcccccc@{}}
		\toprule
		\hfill Assumptions                                          &                              & $\abs{V_i} \approx 1$ & $b_\ell \gg g_\ell$ & $x_\ell \gg r_\ell$ & $\sin(\theta_i - \theta_j)$   & $\cos(\theta_i - \theta_j)$                     & $q_\ell = 0$ \\
		Formulas                                                    &                              &                       &                     &                     & $\approx \theta_i - \theta_j$ & $\approx 1 - \frac{(\theta_i - \theta_j)^2}{2}$ &              \\ \midrule
		$g_\ell \approx r_\ell x_\ell^{-2}$                         & \eqref{eq:g-approx}          &                       &                     & x                   &                               &                                                 &              \\
		$b_\ell \approx x_\ell^{-1}$                                & \eqref{eq:b-approx}          &                       &                     & x                   &                               &                                                 &              \\ \midrule
		$p_\ell = b_\ell (\theta_i - \theta_j)$                     & \eqref{eq:angle-kvl-b}       & x                     & x                   &                     & x                             &                                                 & x            \\
		$\sum_\ell C_{\ell c} p_\ell b_\ell^{-1}=0$                 &                              & x                     & x                   &                     & x                             &                                                 & x            \\ \midrule
		$p_\ell = x_\ell^{-1} (\theta_i - \theta_j)$                & \eqref{eq:angle-kvl-x}       & x                     & x                   & x                   & x                             &                                                 & x            \\
		$\sum_\ell C_{\ell c} p_\ell x_\ell=0$                      & \eqref{eq:cycle-kvl-x}       & x                     & x                   & x                   & x                             &                                                 & x            \\ \midrule
		$\psi_\ell = 2 g_\ell [1-\cos(\theta_i-\theta_j)]$         & \eqref{eq:cosine-loss}       & x                     &                     &                     &                               &                                                 & x            \\
		$\psi_\ell = 2 g_\ell [1-\cos(p_\ell b_\ell^{-1})]$        &                              & x                     & x                   &                     & x                             &                                                 & x            \\
		$\psi_\ell = 2 r_\ell x_\ell^{-2} [1-\cos(p_\ell x_\ell)]$ & \eqref{eq:cosine-loss-power} & x                     & x                   & x                   & x                             &                                                 & x            \\
		$\psi_\ell = r_\ell p_\ell^2$                              & \eqref{eq:ri2}               & x                     & x                   & x                   & x                             & x                                               & x            \\ \bottomrule
	\end{tabular}
\end{small}
\caption{Overview of applied assumptions in respective equations.}
\label{tab:assumptions}
\end{table} 

\begin{table}
	\centering
	\footnotesize
	\begin{tabular}{l|rr}
		\toprule
		 $\theta_i-\theta_j$ & $\sin(\theta_i - \theta_j) \approx \theta_i - \theta_j$ & $\cos(\theta_i-\theta_j) \approx 1 - \frac{(\theta_i - \theta_j)^2}{2}$ \\ \midrule
		$\pm 10^\circ$ & 0.5\% & 0.0\% \\
		$\pm 20^\circ$ & 2.1\% & 0.1\% \\
		$\pm 30^\circ$ & 4.7\% & 0.4\% \\
		$\pm 40^\circ$ & 8.6\% & 1.3\% \\
		 \bottomrule 
	\end{tabular}
	\caption{Relative error of small-angle approximations by angle.}
	\label{tab:relerror}
\end{table}

\begin{table}
	\centering
	\footnotesize
	\begin{tabular}{ll|rrrrrr}
		\toprule
		 &  &  & Lossy &  & Iterative & & Iterative \\
		 Carrier & Unit & Transport & Transport & Lossless & Lossless & Lossy & Lossy \\ \midrule
		 Offshore Wind (AC)  & GW & 139    & 137         & 139     & 137     & 141       & 135     \\
		 Offshore Wind (DC)  & GW & 209    & 215         & 207     & 208     & 217       & 212     \\
		 Onshore Wind        & GW & 328    & 346         & 343     & 347     & 401       & 393     \\
		 Run of River        & GW & 34     & 34          & 34      & 34      & 34        & 34      \\
		 Solar               & GW & 431    & 461         & 440     & 456     & 535       & 517     \\
		 Pumped Hydro        & GW & 55     & 55          & 55      & 55      & 55        & 55      \\
		 Hydro Dam           & GW & 100     & 100          & 100      & 100      & 100          & 100     \\
		 Hydrogen Storage    & GW & 121    & 128         & 131     & 127     & 150       & 146     \\
		 Battery Storage     & GW & 47     & 45          & 44      & 46      & 42          & 43      \\
		 HVDC Links          & TWkm & 48     & 45          & 60      & 63      & 69         & 67      \\
		 HVAC Lines          & TWkm & 167    & 160         & 146     & 143     & 91        & 103    \\
		\bottomrule 
	\end{tabular}
	\caption{Optimised capacities by technology for different flow models.}
	\label{tab:capacity}
\end{table}

\begin{table}
	\centering
	\footnotesize
	\begin{tabular}{ll|rrrrrr}
		\toprule
		 &  &  & Lossy &  & Iterative & & Iterative \\
		 Carrier & Unit & Transport & Transport & Lossless & Lossless & Lossy & Lossy \\ \midrule
		 Offshore Wind (AC) & TWh & 525    & 515         & 526     & 515     & 509    & 504    \\
		 Offshore Wind (DC) & TWh & 935    & 967         & 922     & 927     & 983    & 953    \\
		 Onshore Wind & TWh & 875    & 908         & 907     & 901     & 988    & 980    \\
		 Run of River        & TWh & 139    & 140         & 139     & 138     & 140    & 140    \\
		 Solar      & TWh & 510    & 539         & 508     & 521     & 586    & 576    \\
		 Hydro Inflow     & TWh & 387    & 387         & 387     & 387     & 387    & 387  \\
		\bottomrule 
	\end{tabular}
	\caption{Energy by carrier for different flow models.}
	\label{tab:energy}
\end{table}
\end{appendix}

\end{document}